\documentclass[11pt,a4paper]{article}
\pdfoutput=1
\usepackage{jheppub,yfonts}

\usepackage[utf8]{inputenc}
\usepackage{amsmath}
\usepackage{amssymb}
\usepackage{amsthm}
\usepackage{mathrsfs}
\usepackage{graphicx}
\usepackage{fancyhdr}
\usepackage{array}
\usepackage{latexsym}
\usepackage[all]{xy}
\usepackage{eufrak}
\usepackage{euscript}
\usepackage{enumerate}
\usepackage{dsfont}
\usepackage{slashed}
\usepackage{hyperref}
\usepackage{caption}

\newcommand{\be}{\begin{equation}}
\newcommand{\ee}{\end{equation}}
\newcommand{\bea}{\begin{eqnarray}}
\newcommand{\eea}{\end{eqnarray}}

\def \v {\vec}

\def\lsim{\mathop{\hbox{${\lower 3.8pt\hbox{$<$}}\atop{\raise 0.2pt\hbox{$\sim$}}$}}}
\def\gsim{\mathop{\hbox{${\lower 3.8pt\hbox{$>$}}\atop{\raise 0.2pt\hbox{$\sim$}}$}}}

\title{Energy loss, equilibration, and thermodynamics of a baryon rich strongly coupled quark-gluon plasma}

\author[a]{Romulo Rougemont,}
\author[b]{Andrej Ficnar,}
\author[a,c]{Stefano I. Finazzo,}
\author[a,d]{and Jorge Noronha}

\affiliation[a]{Instituto de F\'{i}sica, Universidade de S\~{a}o Paulo, Rua do Mat\~{a}o, 1371, Butant\~{a}, CEP 05508-090, S\~{a}o Paulo, SP, Brazil}
\affiliation[b]{Rudolf Peierls Centre for Theoretical Physics, University of Oxford, 1 Keble Road, Oxford OX1 3NP, United Kingdom}
\affiliation[c]{Instituto de F\'{i}sica Te\'orica, Universidade do Estado de S\~{a}o Paulo, Rua Dr. Bento T. Ferraz, 271, CEP 01140-070, S\~{a}o Paulo, SP, Brazil}
\affiliation[d]{Department of Physics, Columbia University, 538 West 120th Street, New York, NY 10027, USA}

\date{\today}

\abstract{Lattice data for the QCD equation of state and the baryon susceptibility near the crossover phase transition (at zero baryon density) are used to determine the input parameters of a 5-dimensional Einstein-Maxwell-Dilaton holographic model that provides a consistent holographic framework to study both equilibrium and out-of-equilibrium properties of a hot and {\it baryon rich} strongly coupled quark-gluon plasma (QGP). We compare our holographic equation of state computed at nonzero baryon chemical potential, $\mu_B$, with recent lattice calculations and find quantitative agreement for the pressure and the speed of sound for $\mu_B \leq 400$ MeV. This holographic model is used to obtain holographic predictions for the temperature and $\mu_B$ dependence of the drag force and the Langevin diffusion coefficients associated with heavy quark jet propagation as well as the jet quenching parameter $\hat{q}$ and the shooting string energy loss of light quarks in the baryon dense plasma. We find that the energy loss of heavy and light quarks generally displays a nontrivial, fast-varying behavior as a function of the temperature near the crossover. Moreover, energy loss is also found to generally increase due to nonzero baryon density effects even though this strongly coupled liquid cannot be described in terms of well defined quasiparticle excitations. Furthermore, to get a glimpse of how thermalization occurs in a hot and baryon dense QGP, we study how the lowest quasinormal mode of an external massless scalar disturbance in the bulk is affected by a nonzero baryon charge. We find that the equilibration time associated with the lowest quasinormal mode decreases in a dense medium.}
   

\emailAdd{romulo@if.usp.br, andrej.ficnar@physics.ox.ac.uk, stefano@ift.unesp.br, noronha@if.usp.br}
 
\begin{document}

\maketitle

\setlength{\parskip}{8pt}


\section{Introduction}
\label{introduction}

\hspace{5 mm} In high energy heavy ion collisions \cite{QGP1,QGP2,QGP3,QGP4,Aad:2013xma} where nonzero baryon density effects are small, it is currently understood that the quark-gluon plasma (QGP) formed in these reactions behaves as a nearly perfect liquid \cite{QGP} where viscous effects are surprisingly small. For instance, current estimates based on hydrodynamical modeling (which does not include dynamical effects from a nonzero baryonic charge) find $\eta/s$ between $0.095$ \cite{Ryu:2015vwa} and $0.2$ \cite{Gale:2012rq,reviewQGP1}, where $\eta$ is the shear viscosity and $s$ is the entropy density of the plasma. These values are very close to the uncertainty principle estimate derived three decades ago in \cite{Danielewicz:1984ww} and, more importantly, to the celebrated $1/4\pi$ result found in strongly coupled holographic non-Abelian plasmas with large number of colors \cite{Kovtun:2004de,Buchel:2003tz}. Such a small value for this ratio is an order of magnitude below the result of calculations in transport models involving ordinary hadrons \cite{Demir:2008tr} or in QCD at weak coupling \cite{Arnold:2000dr}. Alternative physical mechanisms involving the effects of additional heavy resonances near the phase transition \cite{NoronhaHostler:2008ju,NoronhaHostler:2012ug} or Polyakov loop degrees of freedom \cite{Hidaka:2009ma} have been devised to obtain a small $\eta/s$ near the QCD crossover transition \cite{Aoki:2006we}, though it remains unclear what is the main non-perturbative element that is responsible for the putative near perfect fluidity of the QGP. Much less is known about the perfect fluid nature of the QGP when not only the temperature $T$ but also the baryon chemical potential $\mu_B$ is large. 

In the last years, the RHIC low energy scan has been used to study heavy ion collisions at lower energies with the intention to probe the QCD phase diagram when $\mu_B \lesssim T$ while also (hopefully) finding signals that could confirm the existence of a critical point \cite{Stephanov:1998dy,Stephanov:1999zu}. In general, the phase diagram of QCD may be studied as a function of many other variables besides $T$ and $\mu_B$ such as the other chemical potentials associated with conserved charges, external electric and magnetic fields, quark masses, number of colors $N_c$ and flavors $N_f$, and others (for reviews see, for instance, \cite{phasediagram1,phasediagram2,phasediagram5}). An important question that requires further investigation is whether the baryon rich QGP studied experimentally via the RHIC beam energy scan (and in a few years by the future FAIR facility at GSI) is also a strongly coupled liquid that displays nearly perfect fluidity. This seems to be the case given the large value of elliptic flow observed by the low energy scan at RHIC \cite{Adamczyk:2012ku}, which may indicate that even the hadron gas formed in the last stages of the collisions can exhibit strong collective behavior at large baryon densities \cite{Denicol:2013nua}. Therefore, one may say that there is now experimental evidence that the significant collective behavior observed in both high and low energies heavy ion collisions is driven by the formation of a strongly coupled, nearly inviscid QGP. 

Even though lattice QCD is currently the main non-perturbative tool to understand strongly interacting QCD physics, when $\mu_B \neq 0$ lattice approaches suffer from the famous sign problem of the fermion determinant, which makes the Monte Carlo-based numerical simulations considerably more challenging. While alternative techniques have been developed to circumvent this issue (see, for instance, \cite{Aarts:2011ax,Fodor:2001au,Gavai:2003mf,Allton:2003vx,deForcrand:2003hx,deForcrand:2002ci,D'Elia:2002gd,fodorreview,Philipsen:2012nu}) making it possible at least in principle to investigate small to moderate chemical potentials on the lattice, it is certainly desirable to have at one's disposal complementary theoretical tools to study strongly coupled QCD phenomena at nonzero temperature and density both in equilibrium as well as out of equilibrium.

The gauge/gravity holographic duality \cite{adscft1,adscft2,adscft3} may be employed to pursue this goal. The duality has provided valuable insights in the study of real time non-equilibrium dynamical phenomena allowing, for instance, the calculation of many transport coefficients\footnote{See \cite{transport} for a recent calculation of a large set of first and second order hydrodynamic transport coefficients in a nonconformal holographic model for QCD at $\mu_B=0$; see also Ref.\ \cite{conductivity} for the computation of the electric conductivity in the same holographic model. For recent reviews on applications of holography in the study of the strongly coupled QGP at $\mu_B=0$, see \cite{solana,adams}.} of the QGP near the crossover transition. We remark that real time non-equilibrium phenomena (even when $\mu_B=0$) are very difficult to investigate using lattice techniques, which are naturally defined in Euclidean imaginary time.

A 5-dimensional bottom-up Einstein-Maxwell-Dilaton holographic model was proposed by DeWolfe, Gubser, and Rosen in \cite{gubser1} as a phenomenological gravity dual for the QGP at finite temperature and baryon chemical potential in the strongly coupled regime. This model contains three fields: the bulk metric $g_{\mu\nu}$, dual to the stress-energy tensor of the boundary quantum field theory, an Abelian vector field $A_\mu$ whose temporal component's boundary value plays the role of the baryon chemical potential, and a real scalar (dilaton) field $\phi$, with a nontrivial profile in the bulk that is responsible for the dynamical breaking of conformal symmetry in the gauge theory, emulating the effects of a dynamically generated infrared scale such as $\Lambda_{\textrm{QCD}}$. In Ref.\ \cite{gubser1}, an estimate was obtained for the QCD critical point in the $(T,\mu_B)$ plane and the corresponding critical exponents were calculated. Moreover, in Ref.\ \cite{gubser2}, the same holographic setup was used to calculate the bulk viscosity and the baryon charge conductivity for the corresponding hot and dense holographic plasma\footnote{See also Ref.\ \cite{veneziano} for a bottom-up holographic dual at finite temperature and chemical potential in the Veneziano limit of $N_f, N_c\rightarrow\infty$ with finite ratio $x_f=N_f/N_c$.}. We also point out that recently in Ref.\ \cite{EMD+B} an anisotropic Einstein-Maxwell-Dilaton holographic model was constructed to describe a holographic plasma under the influence of a strong external magnetic field and the corresponding holographic equation of state and crossover temperature dependence on the external magnetic field were shown to be in good quantitative agreement with recent lattice data \cite{Bali:2014kia} for values of the magnetic field relevant for noncentral ultrarelativistic heavy ion collisions.

These bottom-up models, though not directly derivable from top-down string theory constructions, rely on the conjecture regarding the validity of the holographic dictionary under more general circumstances in a way that closely resembles the majority of the ongoing efforts to apply holographic approaches to strongly interacting condensed matter systems \cite{Hartnoll:2009sz}. The key idea here is that these 5-dimensional, holographic effective models for the strong coupling limit of many-body QCD phenomena are able to describe the nearly perfect fluidity properties of strongly coupled non-Abelian gauge theories through the physics of black holes. Given that nearly perfect fluidity is a fundamental property of holography \cite{Kovtun:2004de,Buchel:2003tz} and that these bottom-up models can be consistently tuned to describe the known equilibrium properties of the QGP calculated on the lattice \cite{GN}, their current application in the calculation of non-equilibrium phenomena may not only be natural but also necessary in order to obtain quantitative predictions near the crossover transition that can be later used in heavy ion phenomenology, such as done in \cite{transport}.

In the present work, we use the general holographic setup proposed in \cite{gubser1,gubser2} to obtain some equilibrium and out-of-equilibrium properties of a hot and {\it baryon rich} strongly coupled QGP. We compare our holographic equation of state computed at nonzero baryon chemical potential, $\mu_B$, with recent lattice calculations \cite{fodor1} and find surprising quantitative agreement for the pressure and the speed of sound for $\mu_B \leq 400$ MeV. Moreover, we find that the crossover temperature decreases with $\mu_B$ in our model in a way that quantitatively matches the behavior found on the lattice \cite{Bonati:2015bha,Bellwied:2015rza}. This shows that this holographic construction provides a consistent framework to study both finite temperature and nonzero baryon density effects at strong coupling. We obtain holographic predictions for the temperature and $\mu_B$ dependence of the drag force and the Langevin diffusion coefficients associated with heavy quark jets as well as the jet quenching parameter $\hat{q}$ and the shooting string energy loss of light quarks in this baryon dense holographic plasma. We find that all these quantities generally display a nontrivial, fast-varying behavior as a function of the temperature near the crossover. We also find that the energy loss of heavy and light quarks in the plasma generally increases in the dense QGP as one increases $\mu_B$. Thus, even though this strongly coupled liquid cannot be described in terms of well defined quasiparticle excitations, the general expectation that jet probes should lose more energy as they plow through a baryon dense medium remains valid at strong coupling. Moreover, we initiate a study on how thermalization occurs in a hot and baryon dense QGP by computing the lowest quasinormal mode of an external massless scalar disturbance in the bulk (which gives information about the linearized approach to equilibrium of different hydrodynamic channels in the spatially uniform limit) taking into account a nonzero $\mu_B$. We find that the equilibration time associated with this lowest quasinormal mode decreases in a baryon dense medium, which may be relevant for the phenomenology of low energy collisions currently under investigation at RHIC.

This paper is organized as follows. In Section \ref{sec2} we review in detail the bottom-up holographic setup proposed in Refs.\ \cite{gubser1,gubser2}. In Section \ref{sec3-new} we fix the unknown functions and parameters of the model using $(2+1)$-flavor lattice QCD data for the equation of state and baryon susceptibility at $\mu_B=0$ from Refs.\ \cite{fodor1,fodor2}, compute the equilibrium properties of the plasma when $\mu_B \neq 0$, and finally compare it with lattice results at nonzero baryon density \cite{fodor1}. In Section \ref{sec4} we compute the energy loss of heavy quarks and light quarks in this model. Our investigation regarding the equilibration of the plasma at nonzero $\mu_B$ is done in Section \ref{QNM}. Some rather technical discussions on the structure of our holographic model are deferred to Appendix \ref{apa}. We finish this paper in Section \ref{conclusion} with a discussion of our main results and we also point out some related studies to be pursued in the future.

\section{The holographic model}
\label{sec2}

\hspace{5 mm} We begin this section by reviewing the main points in the derivation of the classical backgrounds proposed in the approach of Refs. \cite{gubser1,gubser2} to holographically model QCD thermodynamics at finite temperature and baryon chemical potential. The slight modifications to this approach which shall be implemented here, regarding the determination of the free functions and parameters of the holographic model, will be discussed in Section \ref{sec2.4}.

The Einstein-Maxwell-Dilaton holographic model of \cite{gubser1,gubser2} is given by the following action\footnote{Throughout this paper we use a mostly plus metric signature and natural units where $c=\hbar=k_B=1$.}
\begin{align}
S&=\frac{1}{16\pi G_5}\int_{\mathcal{M}_5}d^5x\sqrt{-g}\left[\mathcal{R}-\frac{1}{2}(\partial_\mu\phi)^2-V(\phi) -\frac{f(\phi)}{4}F_{\mu\nu}^2\right] +S_{\textrm{GHY}}+S_{\textrm{CT}},
\label{2.1}
\end{align}
where $S_{\textrm{GHY}}$ is the Gibbons-Hawking-York boundary term \cite{ghy1,ghy2} necessary to establish a well-defined variational problem with Dirichlet boundary condition for the metric field, and $S_{\textrm{CT}}$ is the counterterm action needed to regularize the ultraviolet divergences of the on-shell action, which is done through the holographic renormalization procedure \cite{ren1,ren2,ren3,ren4,ren5}.\footnote{One could also consider adding to the action an Abelian Chern-Simons term of the form $A\wedge F\wedge F$. However, as discussed in \cite{gubser1}, this term does not contribute to the classical equations of motion and it vanishes on-shell for the solutions considered here, which have only electric but no magnetic charge.} Since these two boundary terms only contribute to the on-shell action, which will not be needed in this work, we do not write down their explicit form. The holographic model in \eqref{2.1} contains two unknown functions: the dilaton potential $V(\phi)$ and the Maxwell-dilaton gauge coupling, $f(\phi)$. These functions, together with the value of the 5-dimensional gravitational constant $G_5$ and the temperature scale in MeV, will be fixed in Section \ref{sec2.4} using the $\mu_B=0$ lattice data.

In order to holographically model a field theory at finite temperature and nonzero chemical potential, we take the following charged (and spatially isotropic) black brane Ansatz for the bulk gravity fields
\begin{align}
ds^2=e^{2A(r)}\left[-h(r)dt^2+d\v{x}^2\right]+\frac{e^{2B(r)}dr^2}{h(r)},\,\,\,\phi=\phi(r),\,\,\,A=A_\mu dx^\mu=\Phi(r)dt,
\label{2.2}
\end{align}
where the radial location of the black hole horizon, $r_H$, is given by the largest simple root of the equation $h(r_H)=0$ (we consider coordinates where the boundary of the asymptotically $AdS_5$ space is located at $r\rightarrow\infty$). Also, for simplicity, we set to unity the AdS radius.

With Ansatz \eqref{2.2}, the equation of motion for the dilaton field following from the action \eqref{2.1} is
\begin{equation}\label{2.4}
\begin{aligned}
& \phi''(r)+\left(\frac{h'(r)}{h(r)}+4A'(r)-B'(r)\right)\phi'(r)\cr 
&\qquad\,\,-\frac{e^{2B(r)}}{h(r)}\left( \frac{\partial V(\phi)}{\partial\phi}-\frac{e^{-2[A(r)+B(r)]}\Phi'(r)^2}{2}\frac{\partial f(\phi)}{\partial\phi} \right)=0,
\end{aligned}
\end{equation}
where the prime denotes derivative with respect to the radial direction. The relevant equation of motion for the Maxwell field is
\begin{align}
\Phi''(r)+\left(2A'(r)-B'(r)+\frac{d\left[\ln\left(f(\phi)\right)\right]}{d\phi}\phi'(r)\right)\Phi'(r)=0.
\label{2.6}
\end{align}

Combining the independent components of Einstein equations, we arrive at the following set of equations of motion for the metric
\begin{eqnarray}\label{2.15}
A''(r)-A'(r)B'(r)+\frac{\phi'(r)^2}{6}&=&0,\\ \label{2.17}
h''(r)+[4A'(r)-B'(r)]h'(r)-e^{-2A(r)}f(\phi)\Phi'(r)^2&=&0, \\ \label{2.16}
h(r)[24A'(r)^2-\phi'(r)^2]+6A'(r)h'(r)+2e^{2B(r)}V(\phi)+e^{-2A(r)}f(\phi)\Phi'(r)^2&=&0.
\end{eqnarray}
We will use the set of equations of motion \eqref{2.4}-\eqref{2.17} and the constraint \eqref{2.16} to numerically solve for the unknown functions $A(r)$, $h(r)$, $\phi(r)$, and $\Phi(r)$. Due to reparametrization invariance of the radial coordinate, $B(r)$ may be conveniently chosen in order to simplify the calculations. We will specify a convenient metric gauge in Sections \ref{sec2.1} and \ref{sec2.2}.

There are two conserved charges related to the equations of motions: the Gauss charge $Q_G$, and the Noether charge $Q_N$, which are given by
\begin{equation}\label{2.18}
\begin{aligned}
Q_G(r)&=f(\phi)e^{2A(r)-B(r)}\Phi'(r), \cr
Q_N(r)&=e^{2A(r)-B(r)}[e^{2A(r)}h'(r)-f(\phi)\Phi(r)\Phi'(r)].
\end{aligned}
\end{equation}
The equation of motion for the Maxwell field, Eq.\ \eqref{2.6}, may be written as $dQ_G/dr=0$, while the equation of motion \eqref{2.17} for the blackening function, $h(r)$, may be written as $dQ_N/dr=0$. Since these are conserved charges in the radial direction, one may evaluate \eqref{2.18} at any value of $r$.

We should also remark the limitations of the holographic model discussed here. This model cannot describe phenomena directly related to the breaking of chiral symmetry and its restoration at finite temperature and density, for which one would need to include flavored branes in the bulk \cite{veneziano}. Also, the holographic setup discussed here cannot describe hadron thermodynamics: the pressure of the holographic plasma in the deconfined phase goes like $\sim N_c^2$, while the pressure in the confined hadronic phase goes like $\sim N_c^0$, requiring quantum string corrections in the bulk in order to be properly taken into account\footnote{Note also that this model does not display confinement in the strict sense of an area law for the Wilson loop at zero temperature and zero density, as expected for a boundary non-Abelian gauge theory with dynamical fermions. By suppressing the fermion dynamics one obtains a pure glue theory which does confine probe charges according to the Wilson criterion, which may be achieved by adequately choosing the functional form of the dilaton potential (see, for instance, \cite{ihqcd-1,ihqcd-2,Gubser:2008yx,Finazzo:2014zga}).}. Furthermore, our holographic dual defined in the classical gravity limit asymptotes to a strongly coupled ultraviolet fixed point where conformal invariance is restored and, therefore, the gauge theory investigated here is not asymptotically free at very high temperatures or densities (as it is the case in QCD). Therefore, this model may be applicable in the description of the strongly coupled QGP with $T \sim 150 - 400$ MeV and, as the results in Section \ref{sec3} will show, $\mu_B =0-400$ MeV,\footnote{This defines some kind of ``holographic Goldilocks regime": the plasma cannot be too hot or too cold for the approach described here to be useful.} which is perhaps the regime in which there is still great theoretical uncertainty in QGP modeling.

\subsection{Thermodynamics}
\label{sec2.1}

\hspace{5 mm} For convenience, we now choose the gauge $\tilde{B}(\tilde{r})=0$ with the $AdS_5$ boundary located at $\tilde{r}\rightarrow\infty$ and the horizon at $\tilde{r}=\tilde{r}_H$.\footnote{The tilde is used here to denote the \textit{standard coordinates} associated with the condition $\tilde{B}(\tilde{r})=0$ where the blackening function goes to unity at the boundary. In these coordinates, we will obtain standard expressions for thermodynamical quantities such as the temperature and entropy density. However, in order to numerically solve the equations of motion for the bulk fields, we will need to rescale these coordinates and we reserve the variables without the tilde to denote the corresponding rescaled ``numerical'' coordinates, which we shall specify in Section \ref{sec2.2}.} In this gauge, we have
\begin{align}
d\tilde{s}^2=e^{2\tilde{A}(\tilde{r})}\left[-\tilde{h}(\tilde{r})d\tilde{t}^2+d\v{\tilde{x}}^2\right]+ \frac{d\tilde{r}^2}{\tilde{h}(\tilde{r})},\,\,\,\tilde{\phi}=\tilde{\phi}(\tilde{r}),\,\,\, \tilde{A}=\tilde{A}_\mu d\tilde{x}^\mu=\tilde{\Phi}(\tilde{r})d\tilde{t},
\label{2.19}
\end{align}
and the near-boundary, far from the horizon asymptotics for the bulk fields are given by \cite{gubser1}
\begin{align}
\tilde{A}(\tilde{r})&=\tilde{r}+\mathcal{O}\left(e^{-2\nu\tilde{r}}\right),\nonumber\\
\tilde{h}(\tilde{r})&=1+\mathcal{O}\left(e^{-4\tilde{r}}\right),\nonumber\\
\tilde{\phi}(\tilde{r})&=e^{-\nu\tilde{r}}+\mathcal{O}\left(e^{-2\nu\tilde{r}}\right),\nonumber\\
\tilde{\Phi}(\tilde{r})&=\tilde{\Phi}_0^{\textrm{far}}+\tilde{\Phi}_2^{\textrm{far}}e^{-2\tilde{r}}+ \mathcal{O}\left(e^{-(2+\nu)\tilde{r}}\right).
\label{2.20}
\end{align}
Here $\nu\equiv d-\Delta$, where $d=4$ is the number of dimensions at the boundary and $\Delta=(d+\sqrt{d^2+4m^2})/2$ is the scaling dimension of the boundary gauge theory operator dual to the dilaton field, with $m$ being the effective dilaton mass obtained from the expansion of the dilaton potential close to the boundary. For the potential we consider here (which will be specified in Section \ref{sec2.4}), $\Delta \approx 3$, and the bulk dilaton field is dual to a relevant operator in the boundary field theory that creates a renormalization group flow from an ultraviolet fixed point towards a nonconformal state in the infrared.  

The temperature in the dual field theory is given by the Hawking temperature of the black hole solution\footnote{We use a ``hat" to designate thermodynamical observables expressed in terms of the quantities on the gravity side. The counterparts of these observables without the hat will be used to refer to thermodynamical quantities expressed in powers of MeV, as will be discussed in Section \ref{sec2.4}.}
\begin{align}
\hat{T}=\frac{\sqrt{-g'_{\tilde{t}\tilde{t}} g^{\tilde{r}\tilde{r}}\,'}}{4\pi}\biggr|_{\tilde{r}=\tilde{r}_H}= \frac{e^{\tilde{A}(\tilde{r}_H)}}{4\pi}|\tilde{h}'(\tilde{r}_H)|.
\label{2.21}
\end{align}
The entropy density is given by the Bekenstein-Hawking formula \cite{bek1,bek2}
\begin{align}
\hat{s}=\frac{A_H/4G_5}{V}=\frac{2\pi}{\kappa^2}e^{3\tilde{A}(\tilde{r}_H)},
\label{2.22}
\end{align}
where $\kappa^2\equiv 8\pi G_5$. The chemical potential is given by the boundary value of the Maxwell field
\begin{align}
\hat{\mu}=\lim_{\tilde{r}\rightarrow\infty}\tilde{\Phi}(\tilde{r})=\tilde{\Phi}_0^{\textrm{far}},
\label{2.23}
\end{align}
while the charge density is associated with the boundary value of the radial momentum conjugate to the gauge field\footnote{In Eq.\ \eqref{2.24} the metric determinant is included in the definition of the Lagrangian density, i.e. $S=\int_{\mathcal{M}_5} d^5\tilde{x}\,\mathcal{L}$.}
\begin{align}
\hat{\rho}=\langle J^{\tilde{t}}\rangle=\lim_{\tilde{r}\rightarrow\infty} \frac{\partial\mathcal{L}}{\partial\left(\partial_{\tilde{r}}\tilde{\Phi}\right)}
=\frac{Q_G(\tilde{r}\rightarrow\infty)}{2\kappa^2}
=-\frac{\tilde{\Phi}_2^{\textrm{far}}}{\kappa^2},
\label{2.24}
\end{align}
where we used Eqs.\ \eqref{2.18} and \eqref{2.20}.

\subsection{Numerical procedure}
\label{sec2.2}

\hspace{5 mm} In order to numerically solve the equations of motion \eqref{2.4}-\eqref{2.17} we follow \cite{gubser1} and adopt numerical coordinates where the near-horizon Taylor expansions for the bulk fields $X=\left\{A,h,\phi,\Phi\right\}$ are given by $X(r)=\sum_{n=0}^\infty X_n(r-r_H)^n$ with
\begin{align}
r_H=0;\,\,\,h_0=0,\,\,\,h_1=1,\,\,\,A_0=0,\,\,\,\Phi_0=0.
\label{2.25}
\end{align}
The location of the horizon fixed at $r_H=0$ may be obtained by rescaling the radial coordinate. Clearly, $h_0=0$ comes from the fact that, by definition, $h(r)$ has a simple zero at the horizon while $h_1=1$ may be obtained by rescaling $t$. Also, $A_0=0$ may be obtained by rescaling $(t,\v{x})$ by a common factor. Furthermore, $\Phi_0=0$ is required in order to have a well defined gauge field $A=\Phi(r)dt$ at the horizon\footnote{As mentioned in \cite{cond}, $dt$ has infinite norm at the horizon such that if $\Phi(r_H)=\Phi_0\neq 0$ one would find that $A=\Phi(r)dt$ would be ill-defined at the horizon.}. The reason for using this set of numerical coordinates comes from the fact that in order to numerically solve the equations of motion we need to specify numerical values for each parameter in the near-horizon Taylor expansions for the bulk fields, which is possible after implementing the coordinate rescalings discussed above. With Eq.\ \eqref{2.25} at hand, all the remaining coefficients in the near-horizon Taylor expansions for the bulk fields may be determined from a two-parameter initial condition, $(\phi_0,\Phi_1)$, by solving the equations of motion order by order in these expansions.

The numerical strategy we adopt here to solve the equations of motion \eqref{2.4}-\eqref{2.17} proceeds as follows\footnote{Of course, to numerically solve the equations of motion we need to specify the dilaton potential, $V(\phi)$, and the Maxwell-dilaton gauge coupling, $f(\phi)$, which will be done in Section \ref{sec2.4}.}. In order to avoid the singular point of the equations of motion at the horizon $r_H=0$, we start our numerical integration slightly above it at $r_{\textrm{start}}=10^{-8}$, and use second order near-horizon expansions, $X(r_{\textrm{start}})=X_0+X_1r_{\textrm{start}}+X_2r_{\textrm{start}}^2+\mathcal{O}(r_{\textrm{start}}^3)$, to initialize the numerical integration going up to $r_{\textrm{max}}=2$, which we choose to be the ultraviolet cutoff of our calculations\footnote{At this value of the radial coordinate our numerical solutions already reach the ultraviolet fixed point corresponding to the $AdS_5$ geometry.}. Using Eq.\ \eqref{2.25} we then determine the remaining coefficients of the second order near-horizon expansions, $A_1$, $A_2$, $h_2$, $\phi_1$, $\phi_2$, and $\Phi_2$ as functions of the initial conditions, $(\phi_0,\Phi_1)$, by substituting the second order near-horizon expansions into the equations of motion \eqref{2.4}-\eqref{2.17} and the constraint \eqref{2.16}, setting separately to zero each power of $r_{\textrm{start}}$ in the resulting algebraic equations. Note that we have set to zero $\mathcal{O}(r_{\textrm{start}}^{-1})$ in \eqref{2.4}, $\mathcal{O}(r_{\textrm{start}}^{0})$ in \eqref{2.6}, $\mathcal{O}(r_{\textrm{start}}^{0})$ and $\mathcal{O}(r_{\textrm{start}}^{1})$ in \eqref{2.15}, $\mathcal{O}(r_{\textrm{start}}^{0})$ in \eqref{2.17}, and $\mathcal{O}(r_{\textrm{start}}^{0})$ in \eqref{2.16}. The boundary conditions specified near the horizon, which are necessary to initialize the numerical integration of the second order differential equations \eqref{2.4}-\eqref{2.17}, are then given by $X(r_{\textrm{start}})$ and $X'(r_{\textrm{start}})$.

Any asymptotically $AdS_5$ geometry must satisfy $\mathcal{R}(r_{\textrm{max}})=-20$. For each value of $\phi_0$, there is a bound on the maximum value of $\Phi_1$ above which the solutions are not asymptotically $AdS_5$. In order to determine this bound, we note that in the gauge $B(r)=0$ the equation of motion \eqref{2.15} gives $A''(r)=-\phi'(r)^2/6\le 0$ so that $A(r)$ is a concave function of the radial coordinate. Since for asymptotically $AdS_5$ geometries in the gauge $B(r)=0$ the function $A(r)$ must increase for large $r$ and, since it is a concave function of $r$, this means that $A(r)$ must increase monotonically from the horizon towards the boundary. This, in turn, means that the derivative of $A(r)$ at the horizon is positive, i.e., for asymptotically $AdS_5$ geometries one must have $A_1>0$. By plugging the near-horizon expansions into the constraint equation \eqref{2.16} and evaluating it at the horizon using the conditions \eqref{2.25} we obtain
\begin{align}
A_1=-\frac{1}{6}\left[2V(\phi_0)+f(\phi_0)\Phi_1^2\right].
\label{2.26}
\end{align}
Since the dilaton potential we will work with is negative definite and the Maxwell-dilaton coupling is positive definite, and since $A_1>0$ for asymptotically $AdS_5$ geometries, Eq.\ \eqref{2.26} gives us the bound\footnote{We remark that the dilaton potential in Ref.\ \cite{transport}, which will also be used here, obeys the bound derived in \cite{gubserbound}.} \cite{gubser1}
\begin{align}
\Phi_1<\sqrt{-\frac{2V(\phi_0)}{f(\phi_0)}}\equiv\Phi_1^{\textrm{max}}(\phi_0).
\label{2.27}
\end{align}

\subsection{Coordinate transformations}
\label{sec2.3}

\hspace{5 mm} Here we will express various relevant quantities in the standard coordinates in terms of the numerical coordinates of the gauge $B(r)=0$ discussed above. In order to do so, we first need to inspect the far-from-horizon ($r\rightarrow \infty$) asymptotics of the bulk fields in the numerical coordinates, which are known to have the following form \cite{gubser1}
\begin{align}
A(r)&=\alpha(r)+\mathcal{O}\left(e^{-2\nu\alpha(r)}\right);\,\,\,\alpha(r)= A_{-1}^{\textrm{far}}r+A_0^{\textrm{far}},\nonumber\\
h(r)&=h_0^{\textrm{far}}+\mathcal{O}\left(e^{-4\alpha(r)}\right),\nonumber\\
\phi(r)&=\phi_A e^{-\nu\alpha(r)}+\mathcal{O}\left(e^{-2\nu\alpha(r)}\right),\nonumber\\
\Phi(r)&=\Phi_0^{\textrm{far}}+\Phi_2^{\textrm{far}}e^{-2\alpha(r)}+ \mathcal{O}\left(e^{-(2+\nu)\alpha(r)}\right).
\label{2.28}
\end{align}
After evaluating the constraint \eqref{2.16} at the boundary and using \eqref{2.28}, we obtain 
\begin{align}
A_{-1}^{\textrm{far}}=\frac{1}{\sqrt{h_0^{\textrm{far}}}},
\label{2.29}
\end{align}
where we also used that $\phi\to 0$ at the boundary where the potential $V\to -12$. The Gauss charge in Eq.\ \eqref{2.18} evaluated at the horizon is given by $Q_G(r_H)=f(\phi_0)\Phi_1$, using \eqref{2.25}. When evaluated at the boundary, one finds $Q_G(r\rightarrow\infty)=-2f(0)\Phi_2^{\textrm{far}}/\sqrt{h_0^{\textrm{far}}}$ using \eqref{2.28} and \eqref{2.29}. Since this is a conserved charge in the radial direction, it follows that
\begin{align}
\Phi_2^{\textrm{far}}=-\frac{\sqrt{h_0^{\textrm{far}}}}{2f(0)}f(\phi_0)\Phi_1.
\label{2.30}
\end{align}

As we shall see in a moment, besides $\Phi_2^{\textrm{far}}$ that may be calculated using Eq.\ \eqref{2.30} once we determine $h_0^{\textrm{far}}$, we also need $\Phi_0^{\textrm{far}}$ and $\phi_A$ to compute thermodynamical quantities such as the temperature \eqref{2.21}, the entropy density \eqref{2.22}, the chemical potential \eqref{2.23}, and the charge density \eqref{2.24} directly in terms of quantities extracted from the numerical solutions for $A(r)$, $h(r)$, $\phi(r)$, and $\Phi(r)$. The numerical solutions for $h(r)$ and $\Phi(r)$ are found to quickly converge to their asymptotic values at large $r$ and, thus, we can set $h_0^{\textrm{far}}=h(r_{\textrm{max}})$ and $\Phi_0^{\textrm{far}}=\Phi(r_{\textrm{max}})$. On the other hand, we fix $\phi_A$ through the following near-boundary fit evaluated in the interval $r\in[r_{\textrm{max}}-1,r_{\textrm{max}}]$: $\phi(r)=\phi_A e^{-\nu A(r)}$.\footnote{We checked that $A(r)\approx\alpha(r)$ within this interval.}

In order to express $\hat{T}$, $\hat{\mu}$, $\hat{s}$, and $\hat{\rho}$ in terms of $h_0^{\textrm{far}}$, $\Phi_0^{\textrm{far}}$, $\Phi_2^{\textrm{far}},$ and $\phi_A$, we need to derive some relations between the standard and numerical coordinates. Taking $\tilde{\phi}(\tilde{r})=\phi(r)$, $d\tilde{s}^2=ds^2$, and $\tilde{\Phi}(\tilde{r})d\tilde{t}=\Phi(r)dt$ we obtain\footnote{As mentioned in \cite{gubser1}, the following relations are valid for $\phi_A>0$. If $\phi_A<0$ one must replace $\phi_A\mapsto|\phi_A|$.} by comparing the near-boundary asymptotics \eqref{2.20} and \eqref{2.28} for $r\rightarrow\infty$
\begin{align}
\tilde{r}&=\frac{r}{\sqrt{h_0^{\textrm{far}}}}+A_0^{\textrm{far}}-\ln(\phi_A^{1/\nu}), \label{2.31}\\
\tilde{A}(\tilde{r})&=A(r)-\ln(\phi_A^{1/\nu}), \label{2.32}\\
\v{\tilde{x}}&=\phi_A^{1/\nu}\v{x}, \label{2.33}\\
\tilde{t}&=\phi_A^{1/\nu}\sqrt{h_0^{\textrm{far}}}t, \label{2.34}\\
\tilde{h}(\tilde{r})&=\frac{h(r)}{h_0^{\textrm{far}}}, \label{2.35}\\
\tilde{\Phi}(\tilde{r})&=\frac{\Phi(r)}{\phi_A^{1/\nu}\sqrt{h_0^{\textrm{far}}}}. \label{2.36}
\end{align}
At order $\mathcal{O}(r^0)$ and $\mathcal{O}(e^{-2r})$ in Eq.\ \eqref{2.36} one finds, respectively
\begin{align}
\tilde{\Phi}_0^{\textrm{far}}=\frac{\Phi_0^{\textrm{far}}}{\phi_A^{1/\nu}\sqrt{h_0^{\textrm{far}}}},\,\,\, \tilde{\Phi}_2^{\textrm{far}}=\frac{\Phi_2^{\textrm{far}}}{\phi_A^{3/\nu}\sqrt{h_0^{\textrm{far}}}}.
\label{2.37}
\end{align}
Using the above relations and Eqs.\ \eqref{2.21}-\eqref{2.25} we can express the temperature, baryon chemical potential, entropy density, and baryon charge density as functions of the coefficients of the near-boundary asymptotics in the numerical coordinates
\begin{eqnarray}\label{2.40}
\hat{T}&=&\frac{1}{4\pi\phi_A^{1/\nu}\sqrt{h_0^{\textrm{far}}}}, \\ \label{2.41}
\hat{\mu}&=&\frac{\Phi_0^{\textrm{far}}}{\phi_A^{1/\nu}\sqrt{h_0^{\textrm{far}}}}, \\ \label{2.42}
\hat{s}&=&\frac{2\pi}{\kappa^2\phi_A^{3/\nu}}, \\ \label{2.43}
\hat{\rho}&=&-\frac{\Phi_2^{\textrm{far}}}{\kappa^2\phi_A^{3/\nu}\sqrt{h_0^{\textrm{far}}}}.
\end{eqnarray}
In the next section we will express these thermodynamical quantities in physical units (which shall be then denoted without the hat) using lattice data for the thermodynamic functions and the quark susceptibility at vanishing baryon chemical potential.

\section{Fixing the model parameters and comparison with lattice data at nonzero baryon density}
\label{sec3-new}

\hspace{5 mm} We use lattice data from \cite{fodor1,fodor2} \emph{at vanishing baryon chemical potential}\footnote{We remark that there is now agreement between different lattice collaborations when it comes to the thermodynamic properties of the QGP at $\mu_B=0$ \cite{Borsanyi:2013bia,Bazavov:2014pvz}.} to determine the unknown functions in our effective holographic model: the dilaton potential and the Maxwell-dilaton gauge coupling. Once this is done, the model is fully specified and any quantity computed at $\mu_B \neq 0$ can be interpreted as a direct prediction from the holographic setup. In fact, later in this section we will compare our results for the thermodynamic functions with $\mu_B\neq 0$ to the corresponding lattice data.

Some technical aspects regarding the form of our dilaton potential and Maxwell-dilaton gauge coupling and their relations with the thermodynamic stability of our black hole solutions will be discussed in Appendix \ref{apa}.

\subsection{Extracting $V(\phi)$ and $f(\phi)$ from $\mu_B = 0$ lattice data}
\label{sec2.4}

\hspace{5 mm} The gauge field in the bulk is associated with a conserved charge at the boundary and in our bottom-up model we ensure that this conserved charge plays the role of a baryon charge by matching the susceptibility computed in the model (at zero chemical potential) with the corresponding baryon susceptibility $\chi_2^B(T,\mu_B=0)$ computed on the lattice. A holographic formula for the susceptibility at zero chemical potential was obtained in Ref.\ \cite{gubser1} and we will review this calculation below since the final result in \eqref{2.52} is needed to fix the Maxwell-dilaton coupling function $f(\phi)$. 

The baryon susceptibility is given by the second derivative of the Helmholtz free energy density $\mathcal{F}$ with respect to the baryon chemical potential and, using standard thermodynamical relations, one can express it as \cite{gubser1}
\begin{align}
\chi_2^B=-\frac{\partial^2 \mathcal{F}}{\partial\mu_B^2}=\frac{\partial\rho}{\partial\mu_B},
\label{2.44}
\end{align}
which is evaluated at a fixed temperature. At zero chemical potential (and, consequently, zero charge density), we have
\begin{align}
\chi_2^B(\mu_B=0)=\lim_{\mu_B,\delta\rightarrow 0} \frac{\rho(\mu_B+\delta)-\rho(\mu_B)}{\delta}= \lim_{\delta\rightarrow 0}\frac{\rho(\delta)}{\delta}=\lim_{\mu_B\rightarrow 0}\frac{\rho(\mu_B)}{\mu_B}.
\label{2.45}
\end{align}
Since we want to obtain a holographic formula to compute \eqref{2.45} using black hole physics, we approximate $\Phi(r)$ as a linear perturbation in the equations of motion such that $\Phi'(r)^2\sim 0$. In this limit, the equations of motion for $A(r)$, $h(r)$, and $\phi(r)$ decouple from the equation for $\Phi(r)$ and they may be numerically solved yielding uncharged black hole geometries dual to a field theory at finite temperature and zero chemical potential \cite{GN}. The equation of motion for the linear perturbation $\Phi(r)$ may be then solved on top of these uncharged backgrounds. Using Eqs.\ \eqref{2.40}-\eqref{2.43} and \eqref{2.30} in \eqref{2.45}, one obtains\footnote{It is convenient to consider the dimensionless ratio $\chi_2^B/T^2$ since it asymptotes to a constant at high temperatures.}
\begin{align}
\frac{\chi_2^B(\mu_B=0)}{T^2}=\frac{8\pi^2(h_0^{\textrm{far}})^{3/2}}{\kappa^2 f(0)} \frac{Q_G}{\Phi_0^{\textrm{far}}}.
\label{2.46}
\end{align}
Using the near-horizon expansions \eqref{2.25} and the near-boundary asymptotics \eqref{2.28} one can express $\Phi_0^{\textrm{far}}$ as follows\begin{align}
\int_{r_H}^\infty dr\,\Phi'(r)=\Phi_0^{\textrm{far}}.
\label{2.47}
\end{align}
Now one needs to find a suitable formula for $Q_G$. Since the Gauss charge \eqref{2.18} is conserved in the radial direction, we can take it outside the following integral
\begin{align}
\int_{r_H}^\infty dr \,Q_G \,e^{-2A(r)}f^{-1}(\phi(r))=Q_G\int_{r_H}^\infty dr\, e^{-2A(r)}f^{-1}(\phi(r)).
\label{2.48}
\end{align}
From the definition of the Gauss charge \eqref{2.18} and fixing the metric gauge $B(r)=0$, we see that the left hand side of Eqs.\ \eqref{2.47} and \eqref{2.48} are equal and, thus
\begin{align}
\frac{Q_G}{\Phi_0^{\textrm{far}}}=\frac{1}{\int_{r_H}^\infty dr\, e^{-2A(r)}f^{-1}(\phi(r))}.
\label{2.49}
\end{align}
Also, using \eqref{2.40} and \eqref{2.42} we obtain
\begin{align}
\frac{s}{T^3}=\frac{128\pi^4(h_0^{\textrm{far}})^{3/2}}{\kappa^2}.
\label{2.51}
\end{align}
Finally, plugging \eqref{2.49} and \eqref{2.51} into \eqref{2.46} we arrive at
\begin{align}
\frac{\chi_2^B(\mu_B=0)}{T^2}=\frac{1}{16\pi^2} \frac{s}{T^3} \frac{1}{f(0)\int_{r_H}^\infty dr\, e^{-2A(r)}f^{-1}(\phi(r))}.
\label{2.52}
\end{align}
Note that in \eqref{2.52} there is no dependence on the linear perturbation $\Phi(r)$ so that $\chi_2^B(\mu_B=0)/T^2$ can be evaluated using the solutions corresponding to the uncharged black holes with zero chemical potential without having to explicitly solve the equation of motion for $\Phi(r)$.

In this work we have numerically generated $\sim 10^3$ uncharged black hole solutions by setting $\Phi_1=0$ and varying $\phi_0$ in the holographic setup discussed in the previous sections. Guided by the functional forms for the dilaton potential $V(\phi)$ considered in \cite{gubser1} and by the recent lattice data \cite{fodor1} for the speed of sound squared, $c_s^2$, and the normalized pressure, $p/T^4$, in $(2+1)$-flavor QCD at zero baryon chemical potential, we fixed, respectively, the dilaton potential and the gravitational constant to be \cite{transport,EMD+B}
\begin{align}
V(\phi)=-12\cosh(0.606\,\phi)+0.703\,\phi^2-0.1\,\phi^4+0.0034\,\phi^6;\,\,\,\kappa^2 = 8\pi G_5 = 12.5.
\label{2.53}
\end{align}
From this potential, we see that the effective dilaton mass is $m^2\approx-3$ (therefore, satisfying the Breitenlohner-Freedman bound \cite{Breitenlohner:1982jf,Breitenlohner:1982bm} for massive scalar fields on asympotitcally $AdS_5$ geometries) and that the scaling dimension of the relevant gauge field theory operator dual to the dilaton is $\Delta \approx 3$, as anticipated in Section \ref{sec2.1}.

\begin{figure}[h]
\begin{tabular}{cc}
\includegraphics[width=0.46\textwidth]{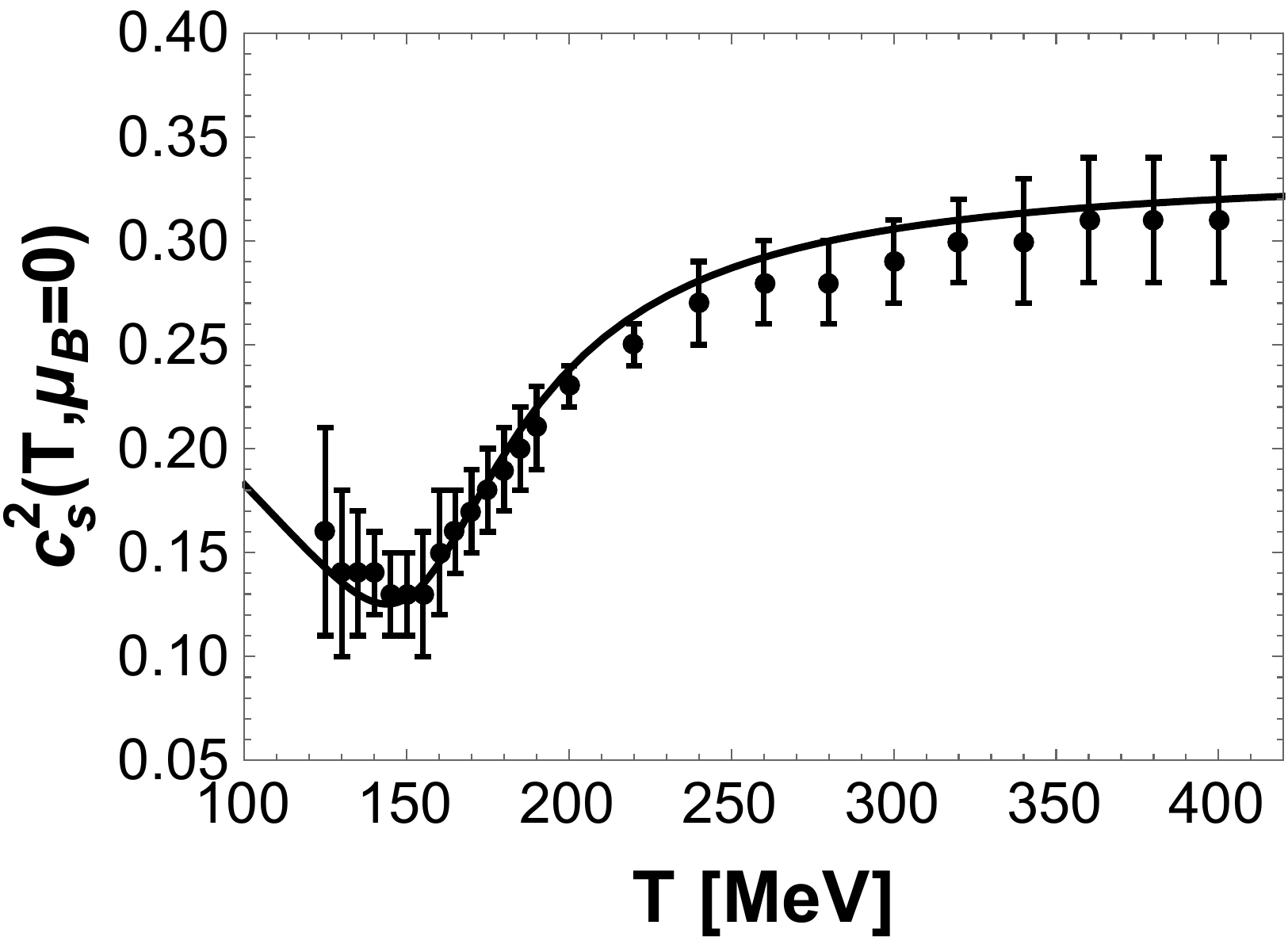} & \includegraphics[width=0.46\textwidth]{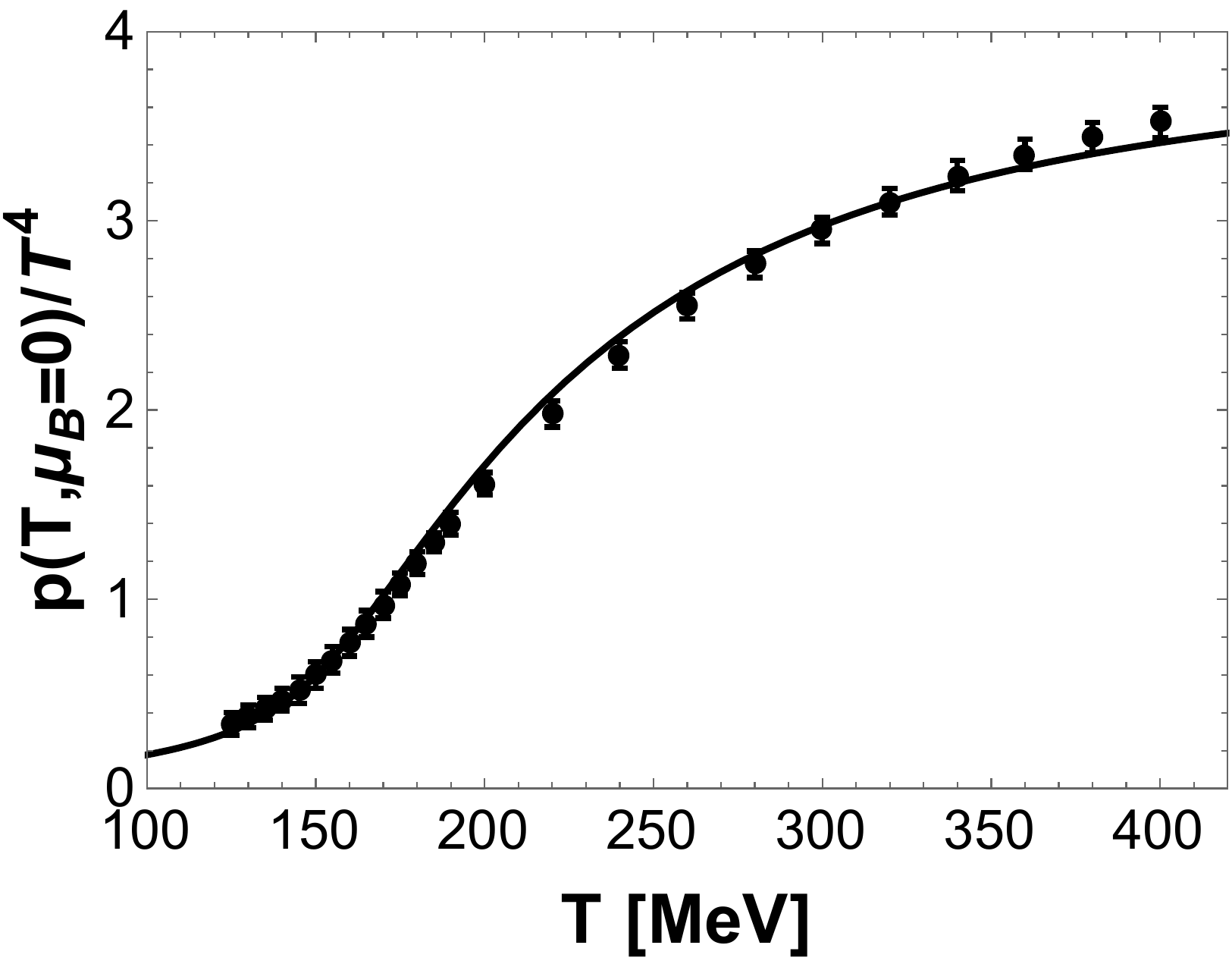} 
\end{tabular}
\begin{center}
\begin{tabular}{c}
\includegraphics[width=0.46\textwidth]{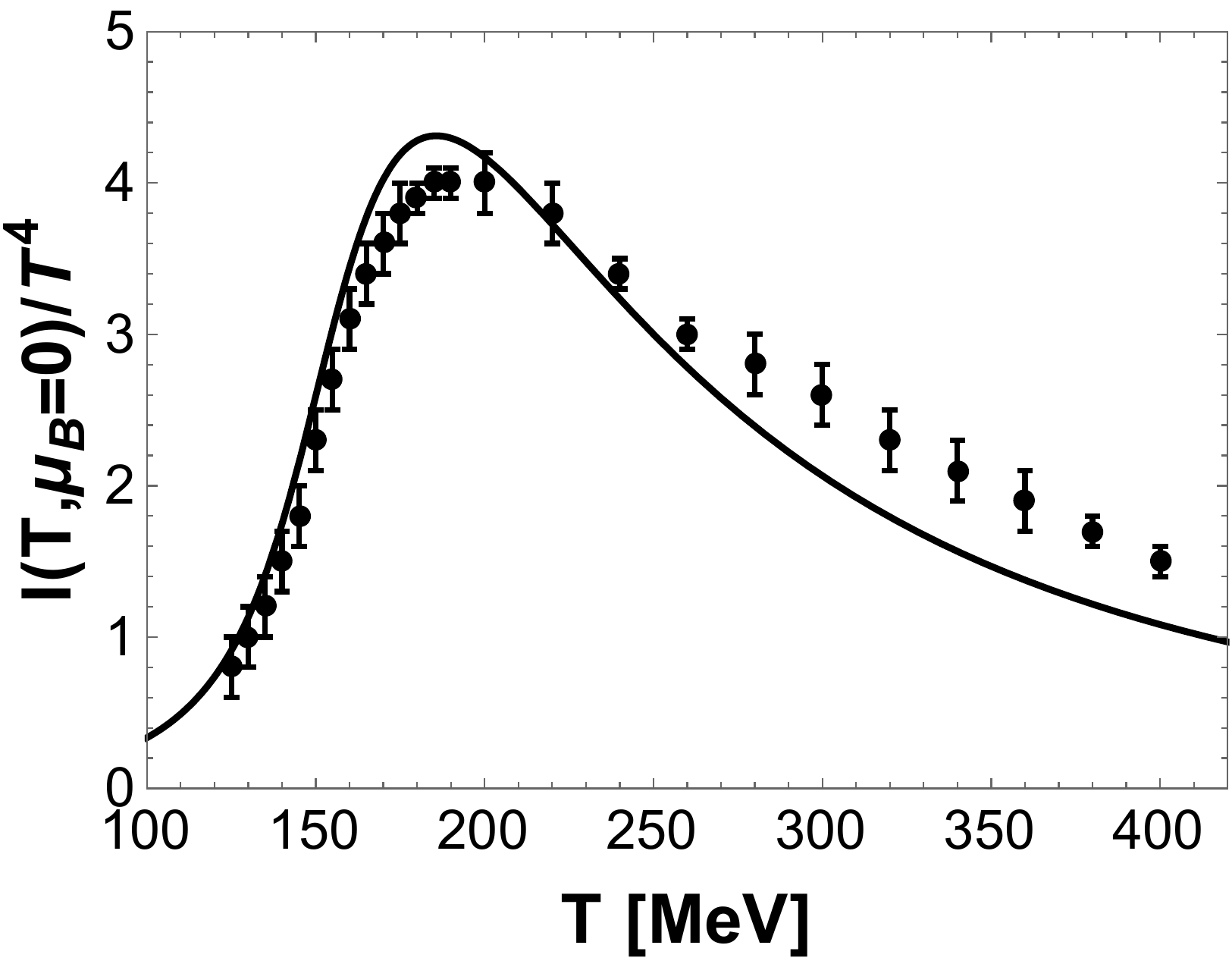} 
\end{tabular}
\end{center}
\caption{Speed of sound squared $c_s^2$, normalized pressure $p/T^4$, and normalized trace anomaly $I/T^4$ at zero baryon chemical potential as functions of the temperature. The data points correspond to lattice calculations for QCD with physical quark masses from \cite{fodor1}.}
\label{fig2}
\end{figure}

Some remarks are in order at this point. The speed of sound squared at zero baryon chemical potential was calculated using 
\begin{align}
c_s^2(T,\mu_B=0)\equiv\frac{dp}{d\epsilon}\biggr|_{\mu_B=0}=\frac{d\ln T}{d\ln s}\biggr|_{\mu_B=0},
\end{align}
where $\epsilon(T,\mu_B=0)=Ts(T,\mu_B=0)-p(T,\mu_B=0)$ is the internal energy density. Also, since we have not renormalized the Einstein-Maxwell-Dilaton action in the present work, we did not calculate the pressure directly from the renormalized free energy density. Rather, here we followed \cite{transport} and computed pressure differences with respect to a very small reference pressure $p_0$ evaluated at some low temperature $T_{\textrm{low}}$ by integrating the entropy density calculated via the Bekenstein-Hawking formula \eqref{2.22}
\begin{align}
p(T,\mu_B=0)\approx p(T,\mu_B=0)-p_{0}=\int_{T_{\textrm{low}}}^T d\bar{T}\, s(\bar{T},\mu_B=0),
\end{align}
where we took $T_{\textrm{low}}=22$ MeV \cite{transport,EMD+B}. Then, the trace anomaly was calculated at zero chemical potential using its definition $I(T,\mu_B=0)\equiv\epsilon(T,\mu_B=0)-3p(T,\mu_B=0)$. In Fig.\  \ref{fig2} we show a comparison between the model calculations for $c_s^2$, $p/T^4$, and the trace anomaly at $\mu_B = 0$ and the corresponding lattice data \cite{fodor1}. One can see that the choice of parameters in \eqref{2.53} gives an overall good description of the lattice data (though the trace anomaly is underpredicted by the model at high temperatures). 

\begin{figure}[h]
\begin{centering}
\includegraphics[scale=0.55]{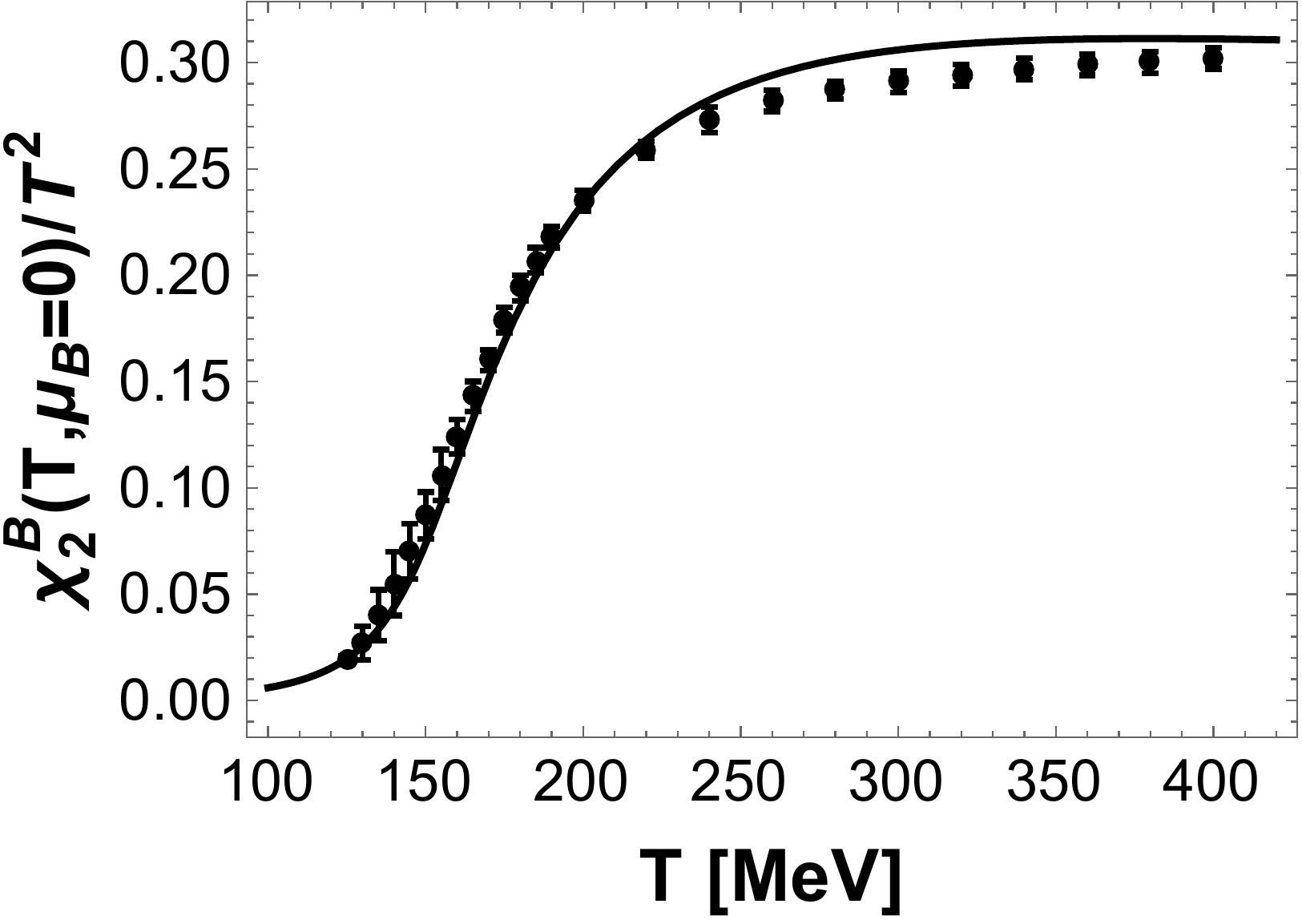}
\par\end{centering}
\caption{Holographic result for the baryon susceptibility at zero baryon chemical potential as a function of the temperature. The data points correspond to lattice calculations for QCD with physical quark masses from \cite{fodor2}.}
\label{fig6}
\end{figure}

Guided by the functional forms for the Maxwell-dilaton gauge coupling $f(\phi)$ used in \cite{gubser1} and by the recent lattice data from Ref.\ \cite{fodor2} (which is in agreement with data from other lattice groups \cite{Bazavov:2012jq}) for the baryon susceptibility of $(2+1)$-flavor QCD at zero baryon chemical potential and physical quark masses, we fixed this coupling in this work to be
\begin{align}
f(\phi)=\frac{\textrm{sech}(1.2\,\phi-0.69)}{3\,\textrm{sech}(0.69)}+\frac{2}{3}\,e^{-100\,\phi}.
\label{2.54}
\end{align}
In Fig.\ \ref{fig6} we compare the results for $\chi_2^B$ from our holographic model with potential $V(\phi)$ and coupling $f(\phi)$ given by \eqref{2.53} and \eqref{2.54} with the lattice data from Ref.\ \cite{fodor2}. Our choice for the Maxwell-dilaton coupling gives a very good description of the data in the crossover transition though the agreement becomes worse at high temperatures. 

In order to perform a meaningful comparison of our holographic results with the lattice data, one needs to express the quantities computed using the black hole physics (which are defined in units of the asymptotic AdS radius) in terms of physical units. This can be done by matching the temperature at which $c_s^2$ displays a minimum in our calculations with the corresponding temperature extracted from the lattice data\footnote{We obtained the lattice value $T_{\textrm{min.}\,c_s^2}^{\textrm{lattice}}\approx 143.8$ MeV by finding the minimum of $c_s^2$ derived using simple thermodynamical identities and the analytical fit for the trace anomaly given in Eq.\ (6.1) of Ref.\ \cite{fodor1}. This estimate is compatible with the data set available in table 4 of \cite{fodor1}.}, which gives the energy scale
\begin{align}
\Lambda\equiv \frac{T_{\textrm{min.}\,c_s^2}^{\textrm{lattice}}\,[\textrm{MeV}]}{T_{\textrm{min.}\,c_s^2}^{\textrm{BH}}}\approx \frac{143.8\,\textrm{MeV}}{0.173}\approx 831\,\textrm{MeV}.
\label{2.55}
\end{align}
We can use this to express any dimensionful quantity on the gravity side $\hat{X}$ in terms of its counterpart $X$ with mass dimension [MeV$^p$] through $X = \Lambda^p \hat{X}$.

Our procedure for fixing the free parameters of the holographic model, and our results for them discussed above, differ from those presented in Ref.\ \cite{gubser1}. Here, we fitted the lattice data at zero baryon chemical potential from \cite{fodor1,fodor2} while in Ref.\ \cite{gubser1} the data used was from an older calculation performed in \cite{BNL2007}. Also, in \cite{gubser1} four independent scaling parameters were introduced to express $T$, $\mu$, $s$, and $\rho$ in physical units while here we employed a single scaling parameter given by \eqref{2.55}. 

With the dilaton potential and the Maxwell-dilaton gauge coupling fixed by lattice data at $\mu_B=0$, the holographic model is now fully specified and the numerical charged black hole geometries obtained from it may be used to derive holographic predictions for the physics of the dual field theory at nonzero baryon chemical potential. In the next section, we are going to compare our results for the thermodynamics in the presence of nonzero baryon density with the lattice data \cite{fodor1} for baryon chemical potentials up to $\mu_B=400$ MeV.

We close this section by remarking that the forms of the dilaton potential $V(\phi)$ and Maxwell-Dilaton gauge coupling $f(\phi)$ fixed in Eqs. \eqref{2.53} and \eqref{2.54}, respectively, were only partially constrained by lattice data at $\mu_B=0$ in the range of temperatures $T\sim 100 - 400$ MeV. In the deep infrared, for large values of $\phi$ (which generically translate into small values of $T$), one could in principle deform $V(\phi)$ in infinitely many different ways such as to obtain very different results for low $T$ physics, while still keeping basically the same agreement with lattice data in the range of temperatures spanned by $T\sim 100 - 400$ MeV. This is the reason why we shall not consider values of $T$ outside this region in the present work. We also remark that, since $V(\phi)$ and $f(\phi)$ were fixed by lattice data at zero chemical potential, all of our results for nonzero $\mu_B$ physics constitute genuine holographic predictions. At $\mu_B=0$, the results we shall present for transport observables are also predictions, although the results for the equation of state and baryon susceptibility shown in Figs. \ref{fig2} and \ref{fig6}, respectively, are not predictions, but instead the results of fits to lattice data used to fix $V(\phi)$ and $f(\phi)$, as explained before.

\subsection{Model predictions for the equation of state at $\mu_B \neq 0$}
\label{sec3}

\hspace{5 mm} Let us begin this section by briefly reviewing the relevant thermodynamic relations that will be used to obtain the holographic equation of state at nonzero baryon chemical potential. The internal and free energy densities of the system at finite $\mu_B$ are given by, respectively
\begin{align}
\epsilon(s,\rho)&=Ts-p+\mu_B\rho,\\
\mathcal{F}(T,\mu_B)&=-p(T,\mu_B)=\epsilon(s,\rho)-Ts-\mu_B\rho,
\end{align}
from which follow the differential relations
\begin{align}
d\epsilon(s,\rho)&=Tds+\mu_B d\rho,\\
d\mathcal{F}(T,\mu_B)&=-dp(T,\mu_B)=-sdT-\rho d\mu_B.
\end{align}
Therefore, at fixed $\mu_B$, one finds
\begin{align}
dp(T,\textrm{fixed}\,\mu_B)=sdT\,,
\end{align}
such that the speed of sound squared at any fixed value of the chemical potential reads
\begin{align}
c_s^2(T,\mu_B)\equiv\frac{dp}{d\epsilon}\biggr|_{\mu_B}=
\left(\frac{T}{s}\frac{\partial s(T,\mu_B)}{\partial T}\biggr|_{\mu_B} +\frac{\mu_B}{s}\frac{\partial \rho(T,\mu_B)}{\partial T}\biggr|_{\mu_B}\right)^{-1}.
\label{cs2mu}
\end{align}
Also, for the trace anomaly at nonzero chemical potential one obtains
\begin{align}
I(T,\mu_B)\equiv\epsilon(T,\mu_B)-3p(T,\mu_B)=Ts(T,\mu_B)+\mu_B\rho(T,\mu_B)-4p(T,\mu_B).
\label{Imu}
\end{align}

\begin{figure}[h]
\begin{centering}
\includegraphics[width=0.46\textwidth]{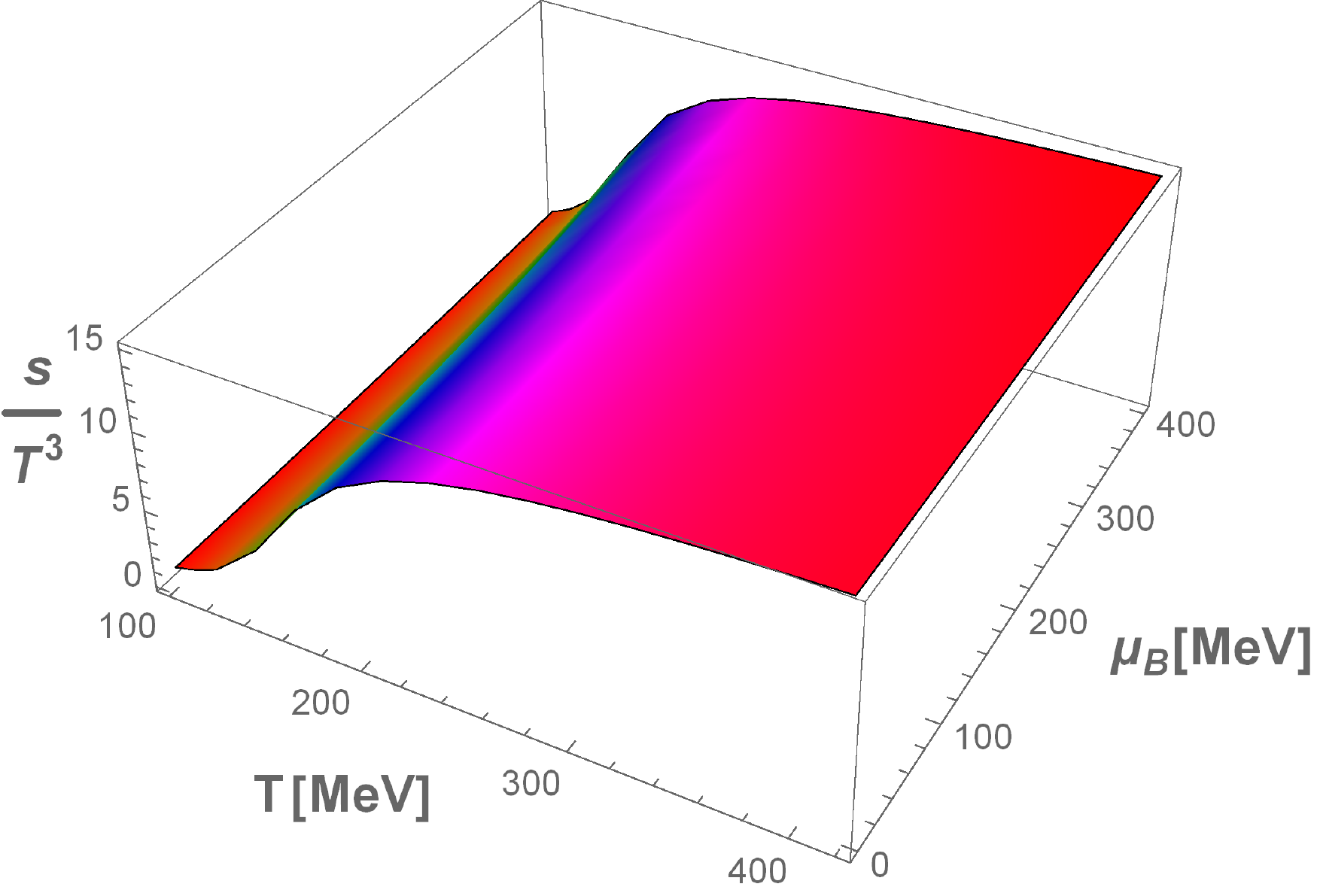}
\hskip0.03\textwidth
\includegraphics[width=0.5\textwidth]{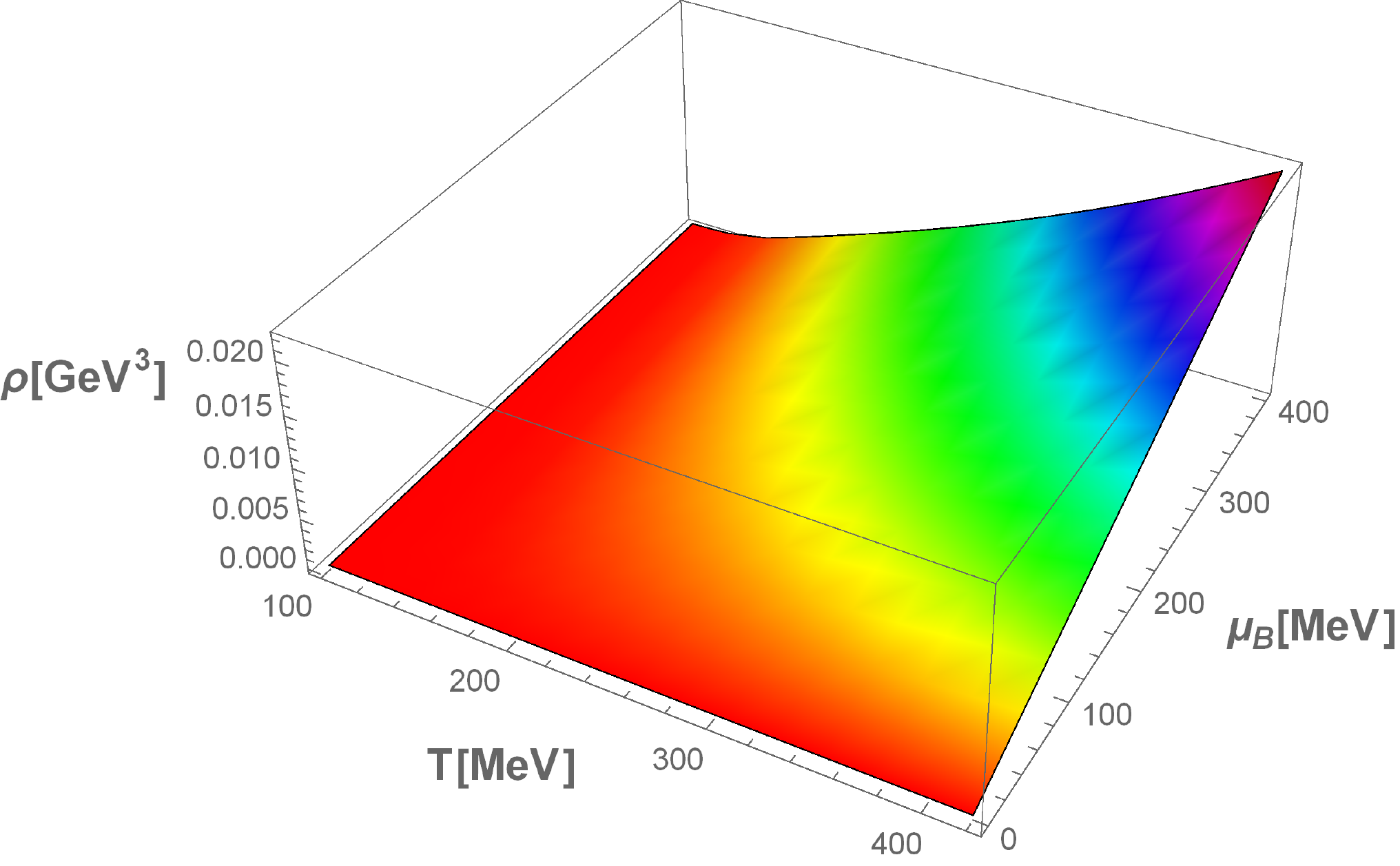}
\par\end{centering}
\caption{(Color online) Normalized entropy density (left) and baryon charge density (right) as functions of the temperature and baryon chemical potential with the color gradients depicting different values of these densities.
\label{fig7}}
\end{figure}

\begin{figure}[h]
\begin{tabular}{cc}
\includegraphics[width=0.46\textwidth]{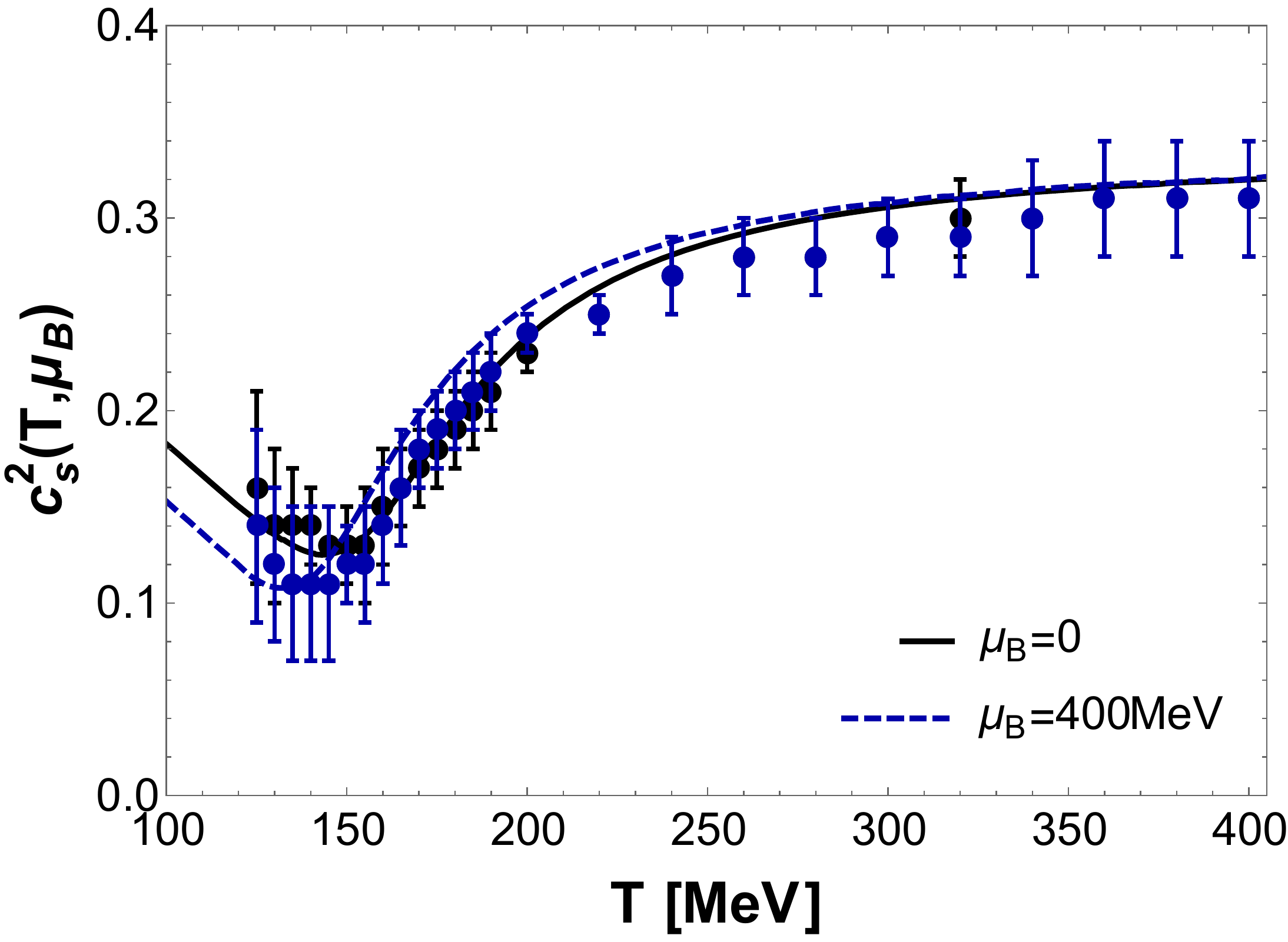} & \includegraphics[width=0.46\textwidth]{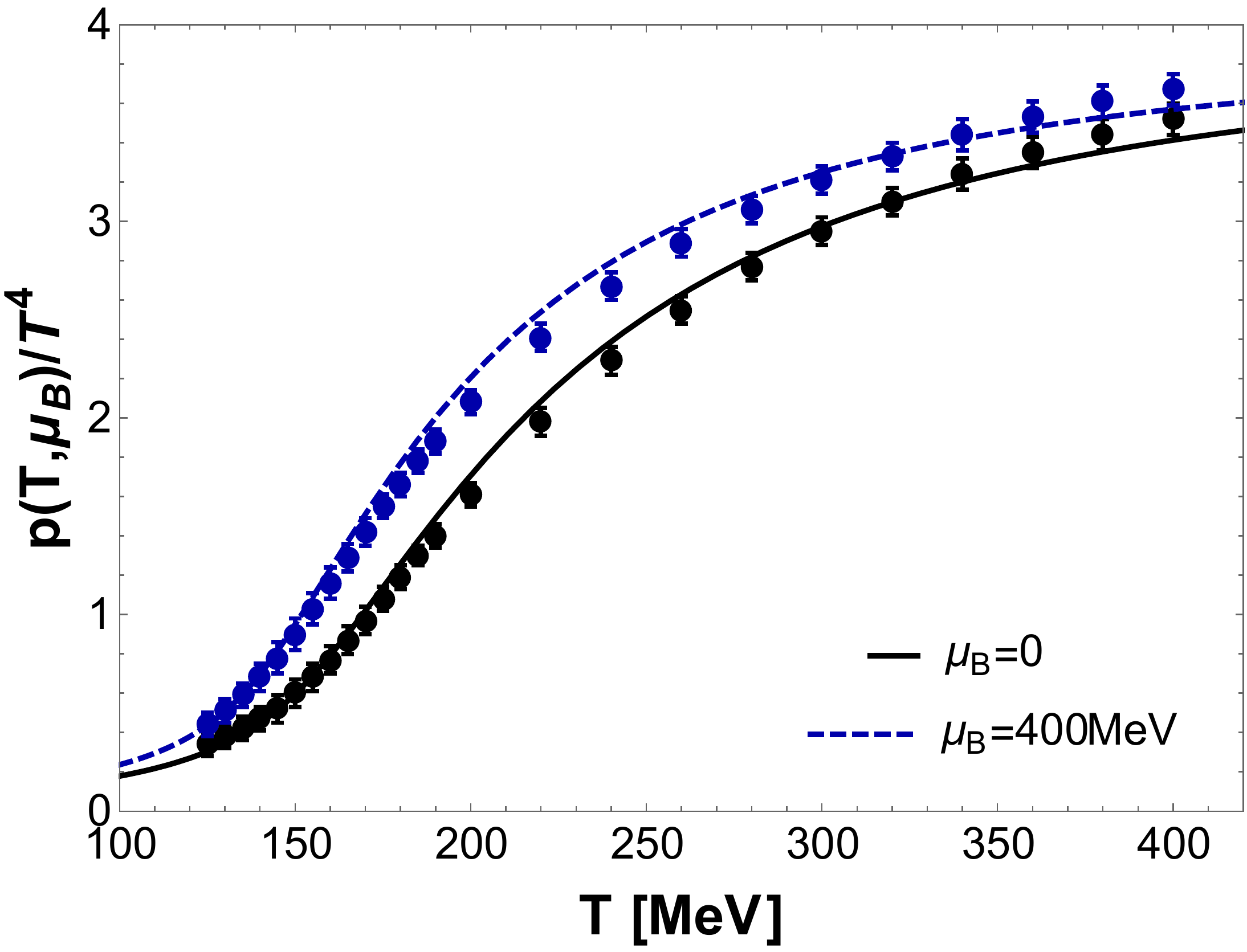} 
\end{tabular}
\begin{center}
\begin{tabular}{c}
\includegraphics[width=0.46\textwidth]{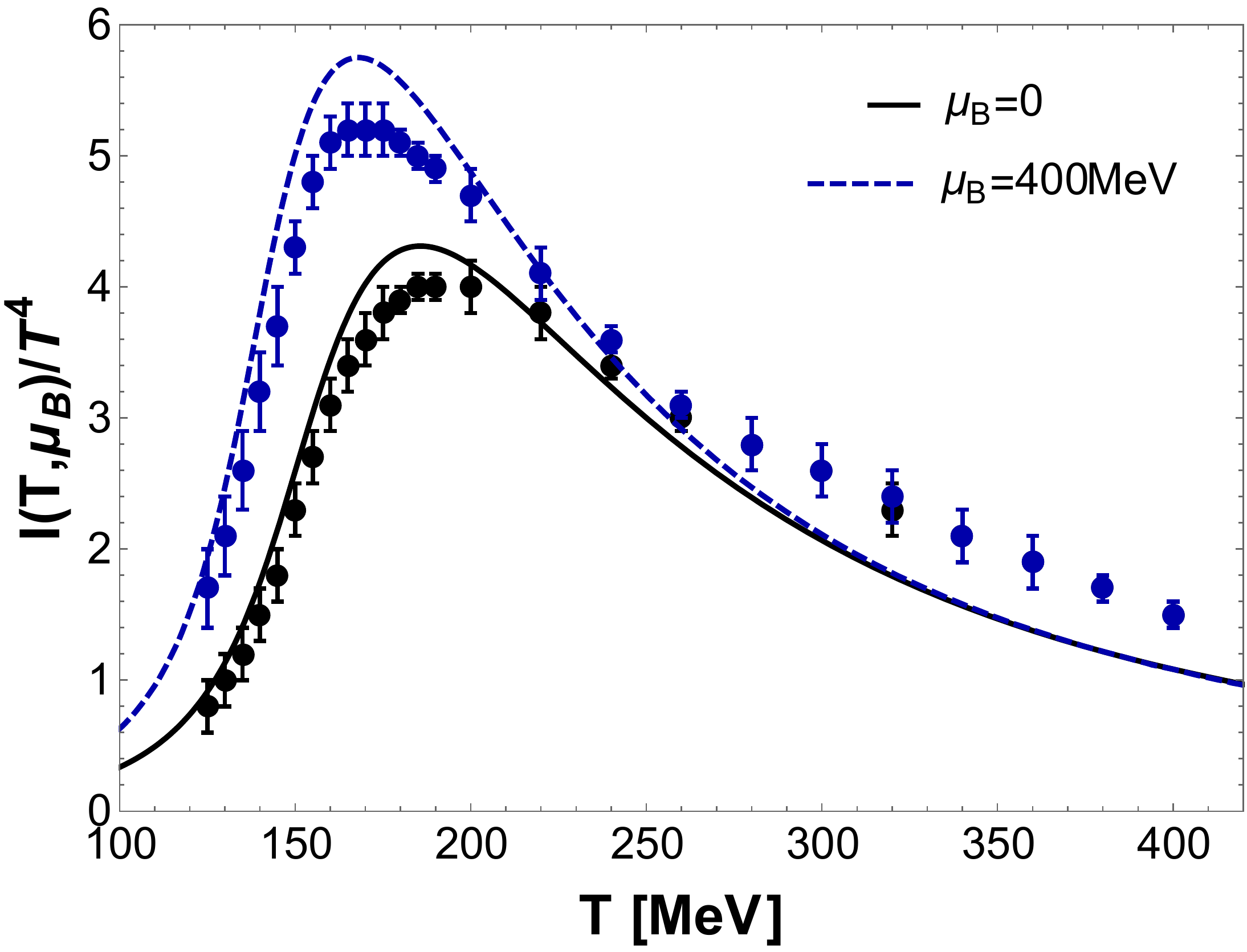} 
\end{tabular}
\end{center}
\caption{(Color online) Speed of sound squared (top left), normalized pressure (top right), and normalized trace anomaly (bottom) as functions of the temperature for different values of the baryon chemical potential, compared with the lattice data from \cite{fodor1}.
\label{fig9}}
\end{figure}

In order to determine the holographic equation of state at finite baryon density, we numerically generated a large number of $\sim 10^5$ charged black holes on a rectangular grid of initial conditions, with $\phi_0$ varying between 0.52 and 6.50 in equally spaced steps and $\frac{\Phi_1}{\Phi_1^{\textrm{max}}(\phi_0)}$ varying between 0 and 0.5 in equally spaced steps for each value of $\phi_0$. Our results are shown in Figs.\ \ref{fig7} and \ref{fig9}.

We first compute the entropy density using the Bekenstein-Hawking relation \eqref{2.22} and then integrate it with respect to the temperature, while keeping the chemical potential fixed to obtain the pressure differences. One can see in Fig.\ \ref{fig9} that the holographic result for the normalized pressure at finite chemical potential agrees quantitatively very well with the lattice data. We find also a good quantitative agreement between our holographic results and the lattice data for the speed of sound squared at nonzero baryon chemical potential while for the trace anomaly the disagreement found at large temperatures and zero density remains.

In Fig.\ \ref{fig7}, one notes from the behavior of the normalized entropy density that the change of the degrees of freedom of the system from a hadron gas to a QGP remains a smooth analytical crossover for the values of $T$ and $\mu_B$ considered here, as also seen in current lattice simulations. Therefore, as in \cite{gubser1}, a putative holographic critical point in our model could only appear at larger values of $\mu_B$, which are beyond the values probed experimentally via the low energy collisions at RHIC.

\begin{figure}[h]
\begin{centering}
\includegraphics[scale=0.5]{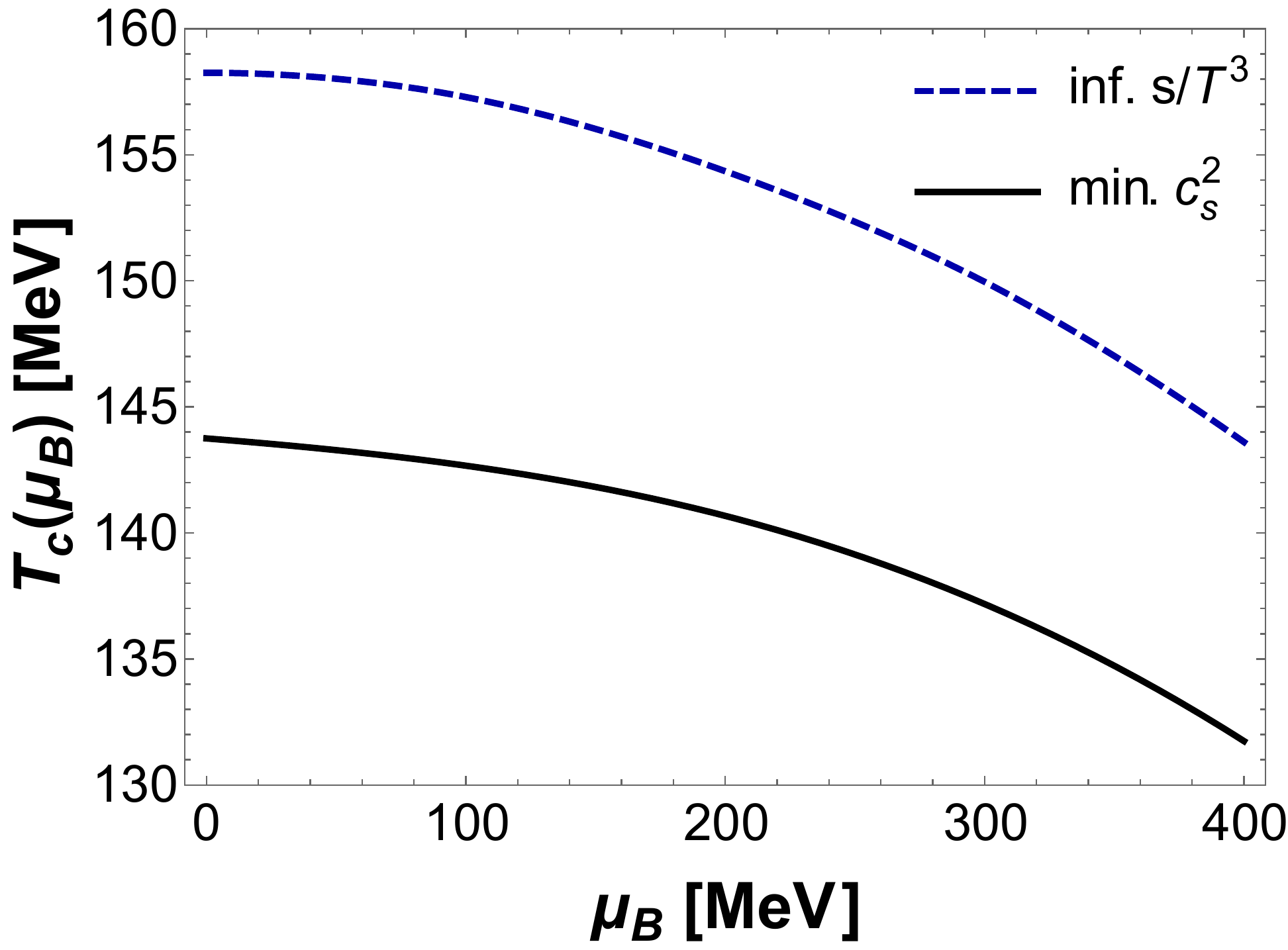}
\par\end{centering}
\caption{(Color online) Crossover band at finite baryon chemical potential in our bottom-up holographic model: the upper curve corresponds to the inflection point of the normalized entropy density while the lower curve refers to the minimum of the speed of sound squared. Within the region delimited by these curves, the degrees of freedom of the system are changing from a hadron gas to a QGP phase. One can see that the crossover temperature decreases with increasing baryon chemical potential.}
\label{figcross}
\end{figure}

In Fig.\ \ref{fig9}, we also note that the minimum of $c_s^2$ and the peak of the normalized trace anomaly are pushed towards lower temperatures with increasing chemical potential, which indicates a reduction in the crossover temperature as one increases the baryon chemical potential. In Fig.\ \ref{figcross} we defined the crossover region at nonzero baryon chemical potential in our holographic model using the inflection point of the normalized entropy density and the minimum of the speed of sound squared. One can see in Fig.\ \ref{figcross} that the crossover temperature decreases with increasing baryon chemical potential, which is the behavior also found in other (non-holographic) calculations (see, for instance, \cite{Fodor:2004nz,Fischer:2011mz,Du:2015psa,Bonati:2015bha,Bellwied:2015rza}). 

We follow \cite{Bonati:2015bha,Bellwied:2015rza} and fit our numerical results for the crossover lines displayed in Fig.\ \ref{figcross} to the series expansion
\begin{align}
\frac{T_c(\mu_B)}{T_c(0)}=1-\kappa\left(\frac{\mu_B}{T_c(0)}\right)^2+\lambda\left(\frac{\mu_B}{T_c(0)}\right)^4+\mathcal{O} (\mu_B^6).
\label{analyticalTcmu}
\end{align}
Combining the results for the curvature $\kappa$ of the crossover lines $T_c(\mu_B)$ calculated from the inflection point of $s/T^3$ and the minimum of $c_s^2$, we obtain the holographic estimate $\kappa\approx 0.013$, which is in remarkable agreement with very recent results obtained by different lattice collaborations: $\kappa_{\textrm{lattice}}^{\textrm{(I)}}=0.0135(20)$ \cite{Bonati:2015bha} and $\kappa_{\textrm{lattice}}^{\textrm{(II)}}=0.0149(21)$ \cite{Bellwied:2015rza}. Furthermore, we can also give a holographic prediction for the fourth order coefficient in \eqref{analyticalTcmu}, which could be tested in future lattice simulations: $\lambda\approx 5.4\times 10^{-5}$.

We should also point out that for high temperatures (above 300 MeV) recent results for thermodynamical quantities obtained from three-loop Hard Thermal Loop perturbation theory are in good agreement with the lattice data for baryon chemical potentials up to 400 MeV \cite{HTLvslattice}. In the near $T_c$ region, however, the weak coupling calculations start to deviate from the lattice data. Presumably, this region involves a somewhat large coupling and the holographic results obtained here provide a good quantitative description of the lattice data. This has motivated the detailed study of some non-equilibrium properties of the plasma presented in the next sections.

\section{Energy loss of heavy and light quarks}
\label{sec4}

\hspace{5 mm} The energy loss experienced by fast moving probes in the QGP \cite{Gyulassy:1990ye,Wang:1991xy} is of great interest to the heavy ion community (see \cite{Majumder:2010qh} for a review and \cite{Burke:2013yra} for a recent summary of the results from theoretical approaches defined at weak coupling). In this section we present our predictions for the heavy and light quark energy loss in the strongly coupled QGP near the crossover transition both at zero and nonzero baryon density. While the phenomenological relevance of hard probes in low energy heavy ion collisions is only marginal, it is interesting from a theoretical point of view to understand how nonzero baryon density effects may change the energy loss of heavy and light quarks moving through the baryon dense strongly interacting plasma.

The results to be discussed in this section make use of calculations involving classical strings in asymptotically $AdS_5$ backgrounds. Since our backgrounds support a nontrivial Maxwell field, one could consider in principle minimally coupling the string end points to the gauge field. In the case of infinitely heavy quarks, to be discussed in the calculations of the drag force and the Langevin diffusion coefficients in Sections \ref{sec4.1} and \ref{sec4.3}, respectively, the minimal coupling between the string and the gauge field is suppressed in the t' Hooft coupling relatively to the Nambu-Goto action, therefore, it does not contribute at leading order in the t' Hooft coupling. In the case of the jet quenching parameter to be discussed in Section \ref{sec4.2}, the minimal coupling does not play any role because the contributions coming from each string endpoint (located at the boundary) cancel each other out. In the case of the light quark energy loss in the shooting string setting to be discussed in Section \ref{sec4.4}, the minimal coupling may be relevant if one considers a flavor brane setup as the one discussed in \cite{Kiritsis:2011ha}, for instance. However, we postpone to a future work the generalization of the shooting string scenario with a minimal coupling to a Maxwell field.


\subsection{Heavy quark energy loss}
\label{sec4.1}

\hspace{5 mm} Here we compute the drag force experienced by heavy quarks in the plasma as a function of $T$ and $\mu_B$ using the holographic model constructed in the previous sections.

In order to calculate the energy loss that a heavy quark experiences in a strongly coupled medium, we will use the trailing string model of \cite{heavy1,heavy2} (see also Refs. \cite{heavy3,heavy4,heavy5}). In this model, a heavy quark probe moving with a constant velocity $v$ (say, in the spatial $x$ direction) in the strongly coupled thermal plasma is represented holographically by an open string with an endpoint moving with the same velocity at the boundary, while the rest of the string trails behind it in the bulk with the other endpoint being located at a black hole horizon formed on top of the string worldsheet. As the quark moves through a hot and dense medium, such as the one described by our holographic model, it loses energy and momentum through drag force $dp_x/dt$ which is holographically computed using the energy flow $dE/dx$ from the endpoint of the string at the boundary towards the worldsheet horizon.

A classical trailing string is described by the Nambu-Goto action, which in the Einstein frame has the following form
\begin{align}
S_{\textrm{NG}}=\frac{1}{2\pi\alpha'}\int d\tau d\sigma\, e^{\sqrt{\frac{2}{3}}\phi}\sqrt{-\det(\gamma_{ab})},
\label{4.1}
\end{align}
where we used the 5D non-critical string theory-inspired result to relate the metrics in the string and Einstein frames \cite{ihqcd-1,ihqcd-2}\footnote{We adopt the same conventions for the action \eqref{2.1} as in \cite{gubser1,gubser2}, which differ from the conventions in \cite{ihqcd-1,ihqcd-2} by a simple rescaling of the scalar field, $\breve{\phi}=\sqrt{\frac{3}{8}}\phi$, where $\breve{\phi}$ is the dilaton field from \cite{ihqcd-1,ihqcd-2} and $\phi$ is the dilaton field used here and in \cite{gubser1,gubser2}. The background metrics in string and Einstein frames are then related by $g_{\mu\nu}^{(S)}=\exp\left[\frac{4}{3}\breve{\phi}\right]g_{\mu\nu}=\exp\left[\sqrt{\frac{2}{3}}\phi\right]g_{\mu\nu}$.}. In \eqref{4.1}, $\alpha'=l_s^2$ is the square of the fundamental string length and
\begin{align}
\gamma_{ab}=g_{\mu\nu}\partial_a X^\mu\partial_b X^\nu,\,\,\,a,b\in\{\tau,\sigma\},
\label{4.2}
\end{align}
is the induced worldsheet metric with $X^\mu(\tau,\sigma)$ describing the worldsheet embedding with the background metric $g_{\mu\nu}$, which is considered here to be expressed in the standard coordinate system in Eq.\ \eqref{2.19}.

We choose the static gauge $(\sigma,\tau)=(\tilde{r},\tilde{t})$ for the worldsheet coordinates so that the worldsheet embedding for an endpoint moving in the $\tilde{x}$-direction is given by $X^\mu(\tilde{r},\tilde{t})=(\tilde{r},\tilde{t},\tilde{x}(\tilde{r},\tilde{t}),0,0)$. Following \cite{heavy1,heavy2}, we choose the stationary trailing string Ansatz $\tilde{x}(\tilde{r},\tilde{t})=v\tilde{t}+\xi(\tilde{r})$, with $\xi(\tilde r)$ describing the string radial profile. The Nambu-Goto action with this Ansatz has the following form
\begin{align}
S_{\textrm{NG}}=\int d\tilde{t} d\tilde{r}\, L_{\textrm{NG}},\,\,\,
L_{\textrm{NG}}=\frac{e^{\sqrt{\frac{2}{3}}\tilde{\phi}(\tilde{r})}}{2\pi\alpha'}\sqrt{-g_{tt}g_{rr}-g_{tt}g_{xx} \xi'(\tilde{r})^2 -g_{rr}g_{xx}v^2}.
\label{4.5}
\end{align}
The equation of motion for $\xi(\tilde r)$ implies that its conjugate radial momentum, $\Pi_\xi=\partial L_{\textrm{NG}}/ \partial\xi'$, is a constant of motion. Solving this relation for $\xi'(\tilde{r})$ one obtains
\begin{align}
\xi'(\tilde{r})=\sqrt{\frac{-g_{tt}g_{rr}-g_{rr}g_{xx}v^2}{g_{tt}g_{xx}\left[1+\frac{e^{\sqrt{8/3}\tilde{\phi}(\tilde{r})} g_{tt}g_{xx}}{(2\pi\alpha' \Pi_\xi)^2}\right]}},
\label{4.6}
\end{align}
where we chose the positive root which corresponds to the string trailing behind the endpoint. The numerator in Eq.\ \eqref{4.6} changes sign at some $\tilde{r}=\tilde{r}_\star$ given by the solution of the following equation
\begin{align}
g_{tt}(\tilde{r}_\star)+g_{xx}(\tilde{r}_\star)v^2=0.
\label{4.7}
\end{align}
As discussed in Refs.\ \cite{langevin1,langevin2,langevin3}, the induced worldsheet metric has the form of a 2D black hole geometry with a horizon at $\tilde{r}=\tilde{r}_\star$. The only way for $\xi(\tilde{r})$ to remain real for all $\tilde r$ is that the denominator in Eq. \eqref{4.6} also changes sign at the worldsheet horizon $\tilde{r}_\star$. This requirement fixes the value of the conjugate momentum 
\begin{align}
\Pi_\xi^2=-\frac{e^{\sqrt{\frac{8}{3}}\tilde{\phi}(\tilde{r}_\star)} g_{tt}(\tilde{r}_\star) g_{xx}(\tilde{r}_\star)}{(2\pi\alpha')^2}.
\label{4.8}
\end{align}

The heavy quark drag force is given by the flux of the momentum from the endpoint into the bulk of the string, which turns out to be given precisely by $\Pi_\xi$ \cite{heavy1,heavy2}
\begin{align}
\tilde{F}_{\textrm{drag}}=\frac{d\tilde{p}_x}{d\tilde{t}}=\frac{d\tilde{E}}{d\tilde{x}}=\Pi_\xi=- \frac{e^{\sqrt{\frac{2}{3}}\tilde{\phi}(\tilde{r}_\star)} \sqrt{-g_{tt}(\tilde{r}_\star) g_{xx}(\tilde{r}_\star)}}{2\pi\alpha'}.
\label{4.9}
\end{align}
Plugging in the Ansatz \eqref{2.19} for the background metric into Eq.\ \eqref{4.9} and expressing everything in terms of the numerical coordinates (as discussed in Section \ref{sec2.3}), one finally obtains the following dimensionless ratio (which is negative-definite)
\begin{align}
\frac{F_{\textrm{drag}}(T,\mu_B;v)}{\sqrt{\lambda_t}T^2}=-8\pi h_0^{\textrm{far}} v e^{\sqrt{\frac{2}{3}}\phi(r_\star)+ 2A(r_\star)},
\label{4.10}
\end{align}
where we defined the 't Hooft coupling $\lambda_t=(\alpha')^{-2}$. In the numerical coordinates, the worldsheet horizon defined by Eq. \eqref{4.7} is obtained by numerically solving the following equation
\begin{align}
\frac{h(r_\star)}{h_0^{\textrm{far}}}=v^2.
\label{4.11}
\end{align}

\begin{figure}[h]
\begin{centering}
\includegraphics[width=0.46\textwidth]{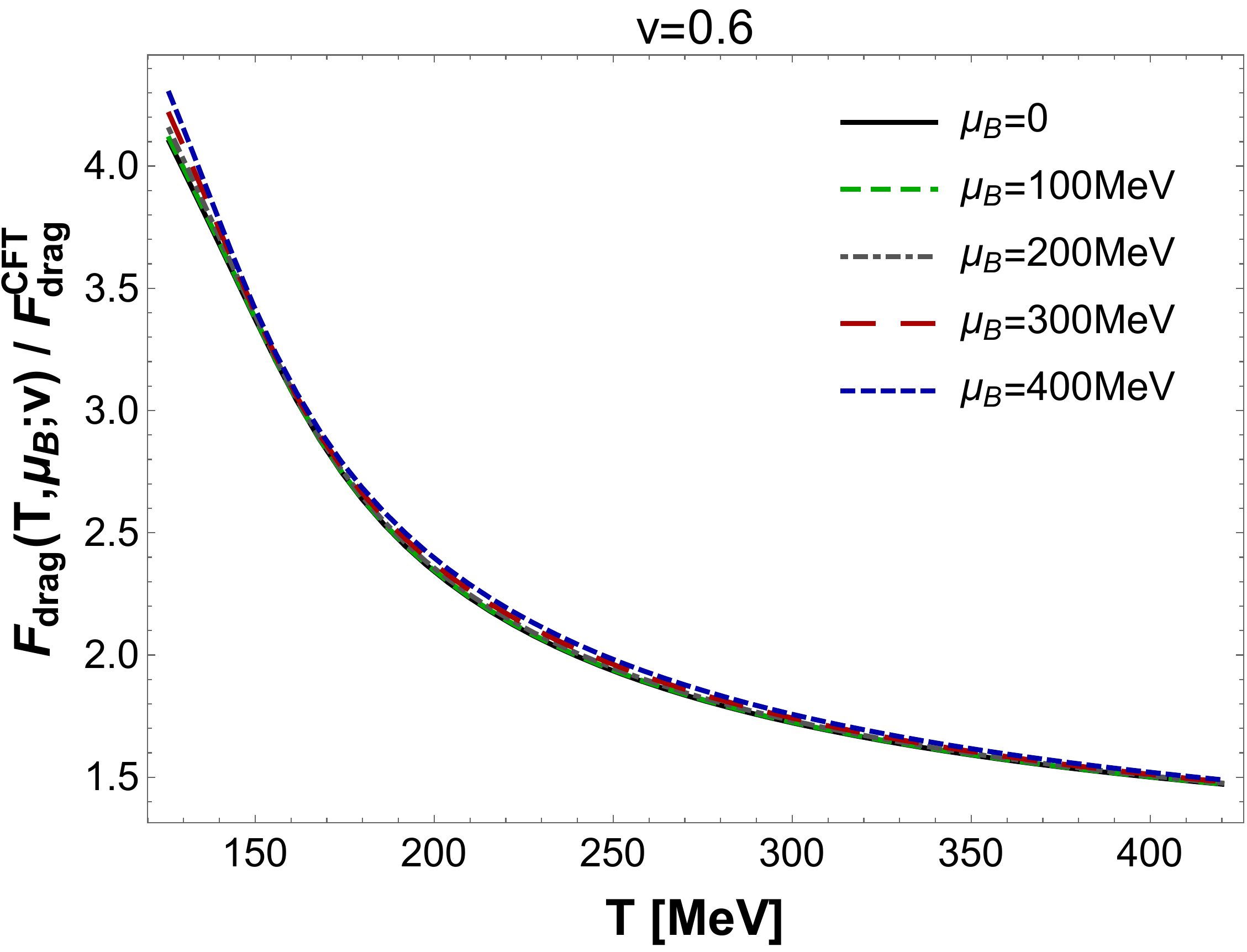}
\hskip0.07\textwidth
\includegraphics[width=0.46\textwidth]{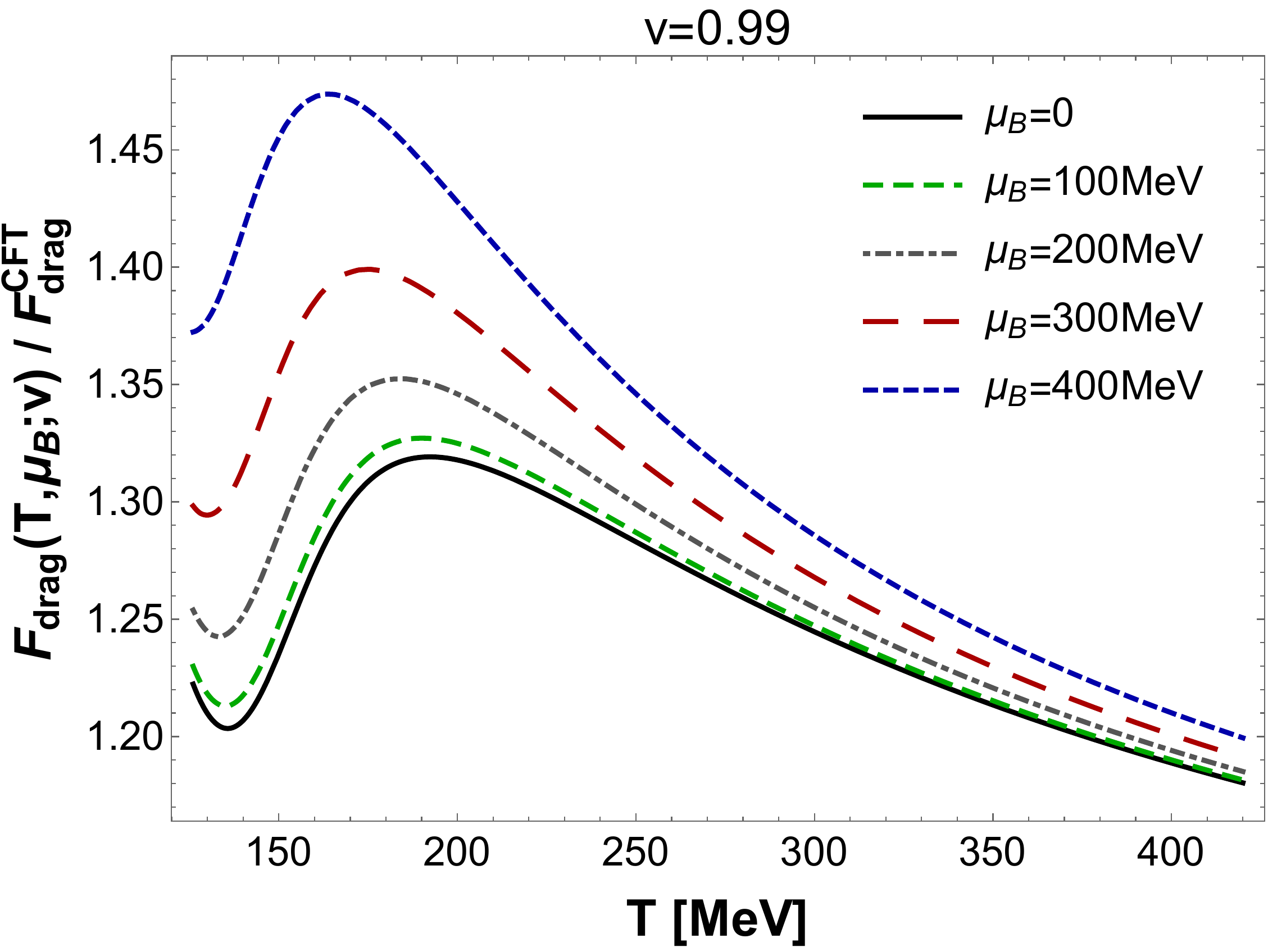}
\par\end{centering}
\caption{(Color online) Heavy quark drag force normalized by its conformal limit as a function of the temperature for different values of the baryon chemical potential and fixed quark velocities.
\label{fig11}}
\end{figure}

We show in Fig.\ \ref{fig11} our numerical results for the heavy quark drag force normalized by its conformal limit (which is only attained at high temperatures) as a function of $T$ for several values of $\mu_B$. The conformal result at $\mu_B=0$ is obtained by using the $AdS_5$-Schwarzschild background, which gives the analytical result $F_\textrm{drag}^\textrm{conformal}/(T^2 \sqrt\lambda_t) = -\pi \gamma(v) v/2 $, where $\gamma(v)=1/\sqrt{1-v^2}$ \cite{heavy1,heavy2}. In the left panel we set $v=0.6$ while in the right panel we take $v=0.99$. One can see that the value of the normalized drag force is very sensitive to the quark velocity and a nontrivial $\mu_B$ dependence and a non-monotonic $T$ dependence only appear at high velocities. This is because, for low velocities, $r_\star$ from (\ref{4.11}) is close to the background horizon where $\Phi(r)$ is small and the drag force is less sensitive to the finite density effects\footnote{This is opposite from the sensitivity to the nonconformal effects, which are most pronounced in the infrared (i.e. close to the horizon) and, as we approach the ultraviolet (boundary), the system becomes asymptotically conformal.}. At $v=0.99$, one can see that the normalized drag force peaks in the phase transition region $T\sim 150-200$ MeV. These results illustrate that the heavy quark energy loss is sensitive to the details of the phase transition and this should be taken into account in phenomenological strongly coupled models used in the calculation of the heavy quark nuclear modification factor \cite{Ficnar:2010rn,Ficnar:2011yj}. Moreover, the fact that a nontrivial baryon chemical potential dependence and a non-monotonic temperature dependence are only observed at large velocities suggests that charm quarks should be generally more sensitive to the medium properties in comparison to bottom quarks. Furthermore, one can see in Fig.\ \ref{fig11} that the peak produced at high velocities increases significantly when $\mu_B$ is raised to 400 MeV and so does the overall energy loss. It is also interesting to note that the peak in the energy loss is pushed towards lower temperatures with increasing chemical potentials, which is in agreement with the fact that in our model the crossover temperature decreases with increasing $\mu_B$.

\subsection{Heavy quark Langevin diffusion coefficients}
\label{sec4.3}

\hspace{5 mm} The holographic prescription for treating Langevin processes was studied in Refs.\ \cite{langevin1,langevin2} in the context of a strongly coupled $\mathcal{N}=4$ Supersymmetric Yang-Mills (SYM) plasma and the generalization to non-conformal holographic duals was put forward in Ref.\ \cite{langevin3}. Here we use the general formulas derived in \cite{langevin3} to compute the Langevin diffusion coefficients for heavy quarks in the zero-frequency limit, which corresponds to the long time behavior of the stochastic diffusion processes in the hot and dense medium described by our holographic model.  

In the trailing string model the effects of thermal fluctuations on the heavy quark propagation are not taken into account. Following Refs.\ \cite{langevin1,langevin2,langevin3}, we consider the Brownian motion of the heavy quark as it moves through the medium, which may be approximately modeled by a linearized local Langevin equation describing the fluctuations of the heavy quark trajectory with a constant velocity (see, e.g., \cite{langevin3} for a derivation and discussion of the approximations). Following the standard holographic dictionary, from the minimal coupling interaction for an external heavy quark, $\int dt \delta X^\mu(r\rightarrow\infty,t)\, \mathcal{F}_\mu(t)$, one identifies the boundary value of the fluctuation of the trailing string worldsheet embedding, $\delta X^\mu(r\rightarrow\infty,t)$, as the source for a boundary gauge invariant operator, $\mathcal{F}^\mu(t)$, which plays the role of a random force acting on the heavy quark. The momentum dependent Langevin diffusion coefficients can be calculated from the symmetrized real time 2-point correlation function of this boundary force operator and the Langevin diffusion constants are then obtained from the zero frequency limit of these coefficients. The standard way \cite{retarded1,retarded2,retarded3,retarded4} to obtain these correlation functions holographically is to solve the linearized bulk equations of motion for the trailing string fluctuations $\delta X^\mu(r,t)$ and compute the on-shell action. However, in the zero momentum limit of these correlation functions, an equivalent and simpler approach involves the membrane paradigm \cite{iqbal-liu} since in this case it is not necessary to explicitly solve the equations of motion. This was used in \cite{langevin3} to compute the parallel and perpendicular Langevin diffusion constants for a general 5-dimensional nonconformal holographic background,
\begin{align}
\kappa_\perp=\frac{T_\star}{\pi\alpha'}b^2_{(S)}(u_\star), \,\,\, \kappa_\parallel=\frac{16\pi T_\star^3}{\alpha'} \frac{b^2_{(S)}(u_\star)}{f'(u_\star)^2}, \,\,\, T_\star=\frac{v}{4\pi}\sqrt{f'(u_\star)\left[\frac{4b'_{(S)}(u_\star)}{b_{(S)}(u_\star)} +\frac{f'(u_\star)}{f(u_\star)}\right]}, 
\label{4.26}
\end{align}
where the Einstein-frame background metric is written in the conformal gauge (with boundary at $u=0$)
\begin{align}
ds^2=b^2(u)\left[-f(u)d\hat{t}^2+d\v{\hat{x}}^2+\frac{du^2}{f(u)}\right].
\label{4.27}
\end{align}
In \eqref{4.26}, $T_\star$ is the Hawking temperature of the worldsheet horizon defined by $u_\star$ computed via $f(u_\star)=v^2$, and $b_{(S)}^2(u)=e^{\sqrt{2/3}\hat{\phi}(u)}b^2(u)$ is the warping factor in the string frame\footnote{Note that here we used our normalization for the dilaton instead of the one used in \cite{langevin3}, as discussed in Section \ref{sec4.1}.}.

In order to relate our background \eqref{2.19} in the $\tilde{B}(\tilde{r})=0$ gauge to the background \eqref{4.27} in the conformal gauge, we equate \eqref{2.19} and \eqref{4.27} and require that $\hat{\phi}(u)=\tilde{\phi}(\tilde{r})$ and $(\hat{t},\v{\hat{x}})=(\tilde{t},\v{\tilde{x}})$, which gives
\begin{align}
b^2_{(S)}(u)=e^{\sqrt{\frac{2}{3}}\tilde{\phi}(\tilde{r})+2\tilde{A}(\tilde{r})},\,\,\, f(u)=\tilde{h}(\tilde{r}),
\label{4.28}
\end{align}
with the following radial coordinate transformation
\begin{align}
du=-e^{-\tilde{A}(\tilde{r})}d\tilde{r}.
\label{4.29}
\end{align}
We can now use this to write \eqref{4.26} in our standard coordinates and then express those equations in the numerical coordinates by following the discussion in Section \ref{sec2.3},
\begin{align}
\frac{\kappa_\perp(T,\mu_B;v)}{\sqrt{\lambda_t}T^3}=16\pi\, (h_0^{\textrm{far}})^{3/2} \,v \sqrt{h'(r_\star)\left[4A'(r_\star)+\sqrt{\frac{8}{3}}\phi'(r_\star)+\frac{h'(r_\star)}{h(r_\star)}\right]} e^{\sqrt{\frac{2}{3}}\phi(r_\star)+3A(r_\star)},
\label{4.31}
\end{align}
\begin{align}
\frac{\kappa_\parallel(T,\mu_B;v)}{\sqrt{\lambda_t}T^3}=16\pi(h_0^{\textrm{far}})^{5/2} \,v^3 \left(h'(r_\star)\left[4A'(r_\star)+\sqrt{\frac{8}{3}}\phi'(r_\star)+\frac{h'(r_\star)}{h(r_\star)}\right]\right)^{\frac{3}{2}} \frac{e^{\sqrt{\frac{2}{3}}\phi(r_\star)+3A(r_\star)}}{h'(r_\star)^2}.
\label{4.32}
\end{align}

\begin{figure}[h]
\begin{centering}
\includegraphics[width=0.46\textwidth]{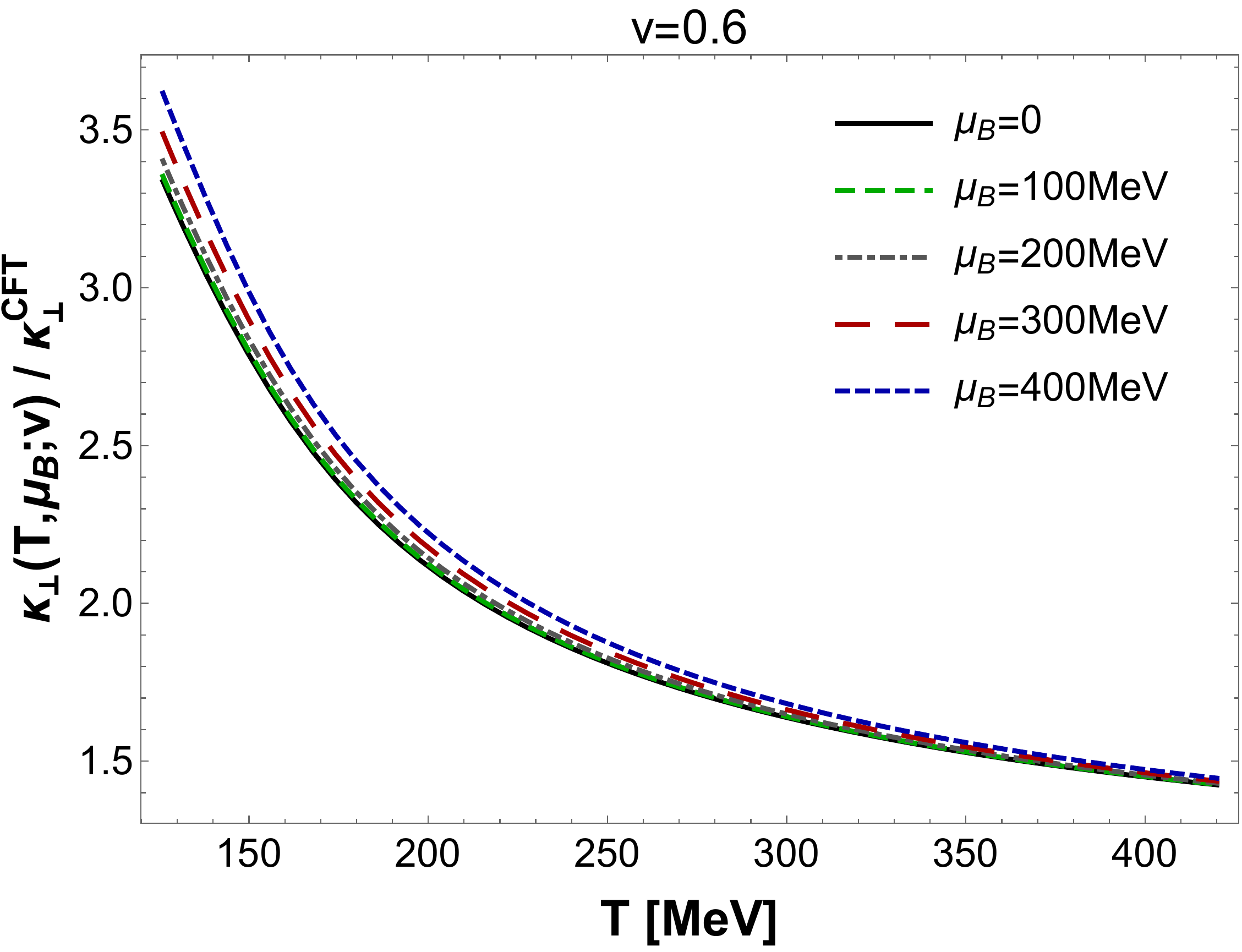}
\hskip0.07\textwidth
\includegraphics[width=0.46\textwidth]{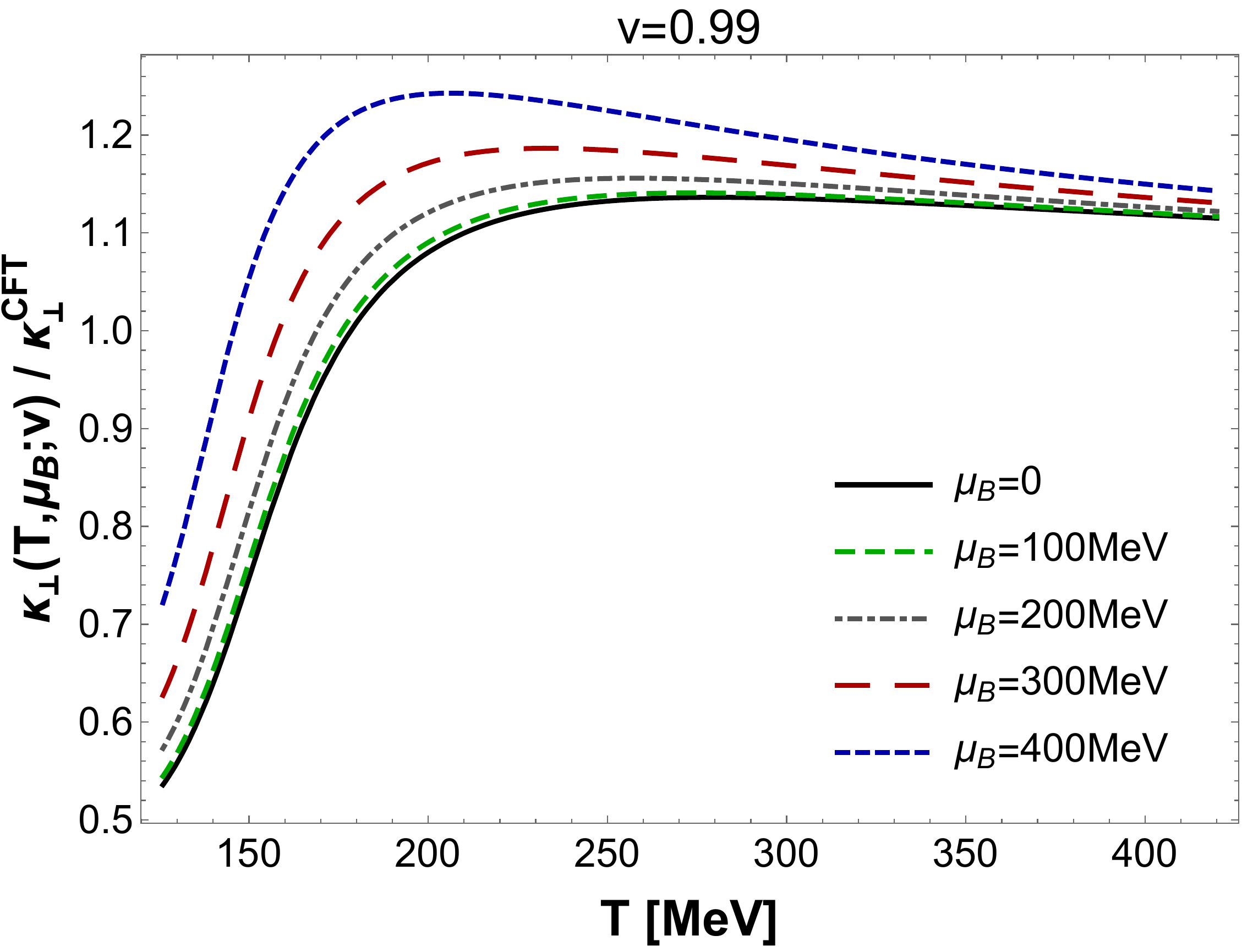}
\includegraphics[width=0.46\textwidth]{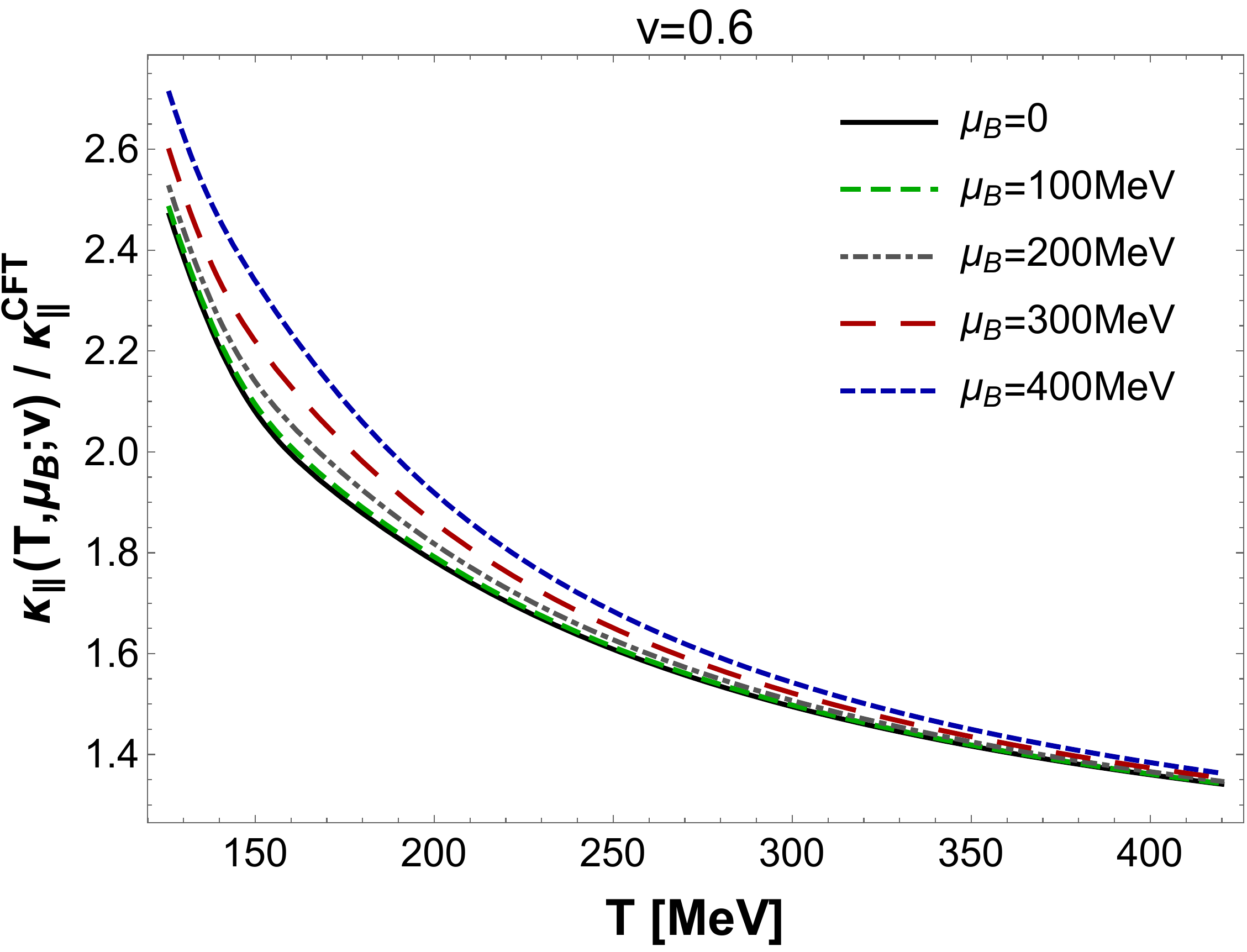}
\hskip0.07\textwidth
\includegraphics[width=0.46\textwidth]{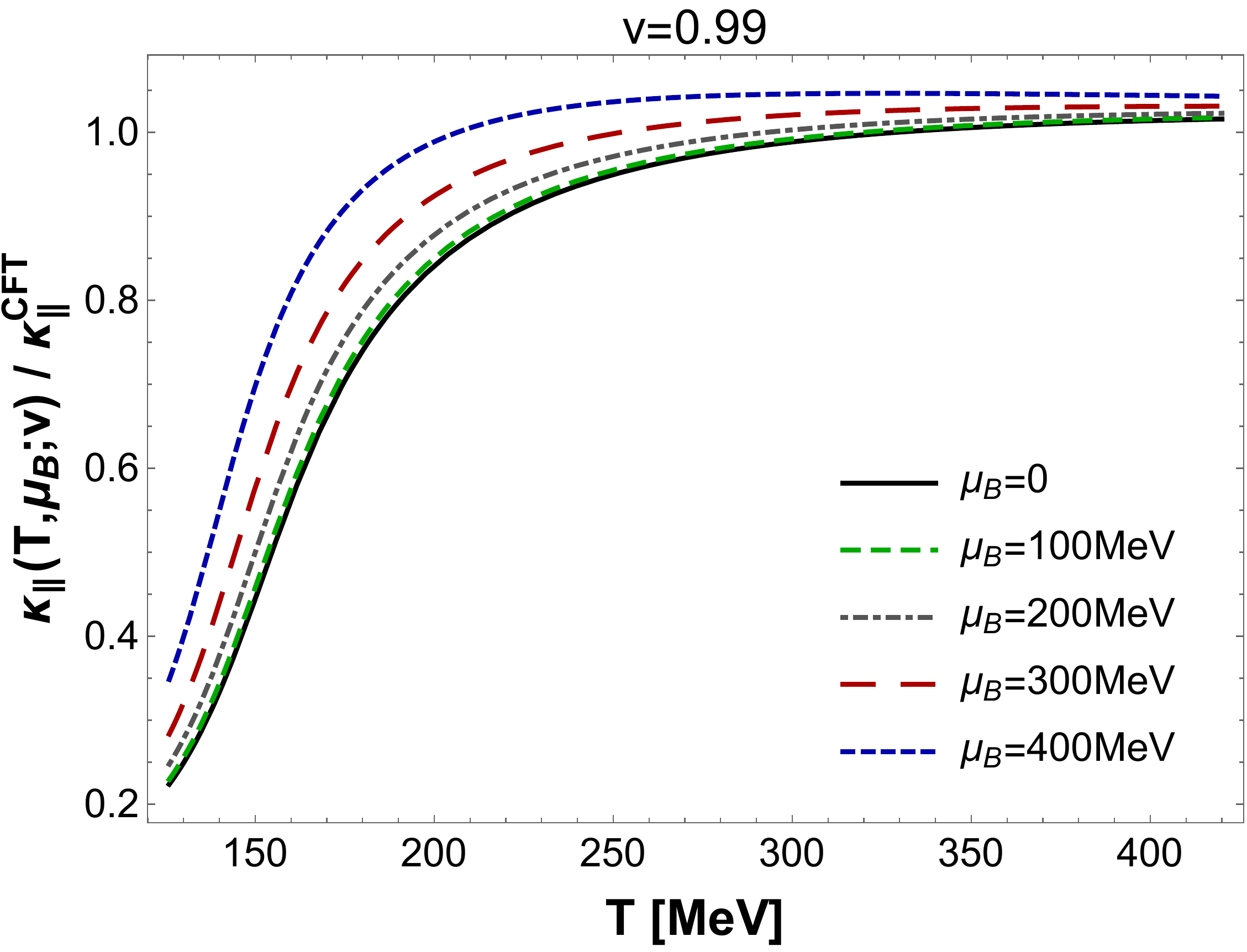}
\par\end{centering}
\caption{(Color online) Perpendicular (upper panel) and parallel (lower panel) Langevin diffusion constants normalized by their respective conformal limits as functions of the temperature for different values of the baryon chemical potential and fixed quark velocities.
\label{fig13}}
\end{figure}

Our numerical results for the perpendicular (upper panel) and parallel (lower panel) Langevin diffusion coefficients normalized by their respective conformal limits are shown in Fig.\ \ref{fig13}. The conformal values for these coefficients are $\kappa_\perp^\textrm{conformal}/(T^3 \sqrt{\lambda_t})=\pi\sqrt{\gamma(v)}$ and $\kappa_\parallel^\textrm{conformal}/(T^3 \sqrt{\lambda_t})=\pi\gamma^{5/2}(v)$ \cite{langevin1,langevin2}. In the left column we set $v=0.6$ while on the right we take $v=0.99$. Similarly to the curves for the heavy quark drag force in Fig.\ \ref{fig11}, one can see that these transport coefficients display a more appreciable $\mu_B$ dependence and a non-monotonic $T$ dependence when the heavy quark velocity is close to unity. This non-monotonic dependence with the temperature at $\mu_B=0$ may play an important role in phenomenological setups involving strongly coupled modeling of thermal fluctuations in heavy quark energy loss. For instance, it would be interesting to see how an improved treatment of the strongly coupled medium properties near the phase transition affects the study performed in Ref.\ \cite{Horowitz:2015dta}, which employed the conformal limit of the Langevin diffusion coefficients. Moreover, we find that these coefficients increase with $\mu_B$, which indicates that the thermal fluctuations of the trailing string become more relevant in a baryon dense medium.

\subsection{Light quark energy loss: the jet quenching parameter}
\label{sec4.2}

\hspace{5 mm} Even though weak coupling approaches to jet quenching phenomena seem to be consistent with heavy ion data \cite{Burke:2013yra}, one should keep in mind that the weak coupling description of the medium used in these approaches does not incorporate the defining property of the QGP, i.e., its nearly perfect fluidity. Energy loss in a hot and dense medium involves several different scales \cite{Majumder:2010qh} and, as discussed in \cite{qhat2,qhat3}, strong coupling approaches to the interaction between the jet and the medium at scales of order $\sim T$ may be called for. In this section we compute the jet quenching parameter $\hat{q}$ \cite{qhat1,qhat2} (see also \cite{heavy5,qhat3}) associated with light quark probes in our nonconformal backgrounds with nonzero baryon chemical potential. For recent lattice results at zero baryon chemical potential regarding this quantity see \cite{qhatlattice}.

The original formula for $\hat{q}$ in holography, derived in \cite{wiedemann}, was inspired by the dipole approximation \cite{Zakharov:1997uu} for the perturbative QCD jet quenching parameter $\hat q$ \cite{Baier:1996sk}
\begin{equation}
\langle W^{\textrm{(adjoint)}}_{L\times L^-}\rangle \approx \exp \left[-\frac{1}{4 \sqrt{2}} \hat{q} L^- L^2\right],
\end{equation}
where $L$ and $L^-$ (with $L\ll L^-$) are the sides of an adjoint light-like rectangular Wilson loop described by a quark-antiquark pair. The authors of \cite{qhat1,qhat2} used this to \emph{define} a non-perturbative version of $\hat{q}$ that would be valid at strong coupling
\begin{align}\label{aa1}
\hat{q}\equiv -\frac{4\sqrt{2}}{L^- L^2}\times \ln\left(\langle W^{\textrm{(adjoint)}}_{L\times L^-}\rangle\right).
\end{align}
According to the standard holographic dictionary \cite{wilsonloops1} (see also Refs. \cite{wilsonloops2,wilsonloops3,wilsonloops4}) the expectation value of a Wilson loop in the fundamental representation in a strongly coupled $SU(N_c)$ gauge theory in the large $N_c$ limit is given by the exponential of minus the on-shell Nambu-Goto action for a string worldsheet whose boundary coincides with that Wilson loop. Hence, we have from \eqref{aa1}
\begin{equation}
\hat{q} = \frac{8\sqrt{2}}{L^- L^2} S_{\textrm{NG}}^{\textrm{on-shell}},
\label{4.12}
\end{equation}
where the extra factor of 2 comes from the fact that at large $N_c$ the adjoint Wilson loop is twice the value of the fundamental loop.

Since we are interested in a light-like Wilson loop, we define light-cone coordinates
\begin{align}
\tilde{x}_\pm=\frac{\tilde{x}\pm\tilde{t}}{\sqrt{2}},
\label{4.13}
\end{align}
leaving $\tilde{r}$, $\tilde{y}$, and $\tilde{z}$ unchanged. In these coordinates, the background metric \eqref{2.19} has the following form
\begin{align}
ds^2=e^{2\tilde{A}(\tilde{r})}\left[ d\tilde{y}^2+d\tilde{z}^2+\frac{1}{2}(1-\tilde{h}(\tilde{r}))(d\tilde{x}_+^2+ d\tilde{x}_-^2) + (1+\tilde{h}(\tilde{r}))d\tilde{x}_+d\tilde{x}_- \right] + \frac{d\tilde{r}^2}{\tilde{h}(\tilde{r})}.
\label{4.14}
\end{align}
The geometric configuration considered here corresponds to a string attached to a quark-antiquark pair at the boundary, separated by a distance $L$ and moving through a much longer (light-like) distance $L^-$, tracing out a rectangle at the boundary. The string sags into the bulk describing an U-shaped string worldsheet whose tip turns out to lie at the radial position of the bulk horizon, as we will see in a moment. It is convenient to use the static gauge $(\tau,\sigma)=(\tilde{x}_-,\tilde{y})$ such that the worldsheet embedding reads $X^\mu(\tilde{x}_-,\tilde{y})=(\tilde{r}(\tilde{x}_-,\tilde{y}),0,\tilde{x}_-,\tilde{y},0)$. Since we are taking the $L^-\gg L $ limit, there is translational invariance in the $\tilde{x}_-$-direction and the worldsheet embedding reduces to
\begin{align}
X^\mu(\tilde{x}_-,\tilde{y})=(\tilde{r}(\tilde{y}),0,\tilde{x}_-,\tilde{y},0).
\label{4.15}
\end{align}
The Nambu-Goto action for this string configuration is given by
\begin{align}
S_{\textrm{NG}}=\frac{L^-}{2\pi\alpha'}\int_{-L/2}^{L/2}d\tilde{y}\,e^{\sqrt{\frac{2}{3}}\tilde{\phi}(\tilde{r})+2\tilde{A} (\tilde{r})}\sqrt{\frac{1-\tilde{h}(\tilde{r})}{2}\left(1+\frac{e^{-2\tilde{A}(\tilde{r})}\tilde{r}'(\tilde{y})^2}{ \tilde{h}(\tilde{r})}\right)}.
\label{4.17}
\end{align}
Since the integrand in Eq.\ \eqref{4.17} does not depend explicitly on $\tilde{y}$, we can identify the following constant of motion
\begin{align}
H_{\textrm{NG}}=\tilde{r}'\frac{\partial L_{\textrm{NG}}}{\partial\tilde{r}'}-L_{\textrm{NG}}
=-e^{\sqrt{\frac{2}{3}}\tilde{\phi}(\tilde{r})+2\tilde{A}(\tilde{r})}\sqrt{ \frac{\tilde{h}(\tilde{r})(1-\tilde{h}(\tilde{r}))}{2(\tilde{h}(\tilde{r})+e^{-2\tilde{A}(\tilde{r})}\tilde{r}'(\tilde{y})^2)} }\equiv\frac{C_0}{\sqrt{2}}.
\label{4.18}
\end{align}
Solving Eq.\ \eqref{4.18} for $\tilde{r}'(\tilde{y})$, we obtain the following on-shell relation
\begin{align}
\frac{d\tilde{r}}{d\tilde{y}}=\sqrt{\frac{e^{2\tilde{A}(\tilde{r})}\tilde{h}(\tilde{r})}{C_0^2} \left[ e^{\sqrt{\frac{8}{3}}\tilde{\phi}(\tilde{r})+4\tilde{A}(\tilde{r})} (1-\tilde{h}(\tilde{r})) - C_0^2 \right]}.
\label{4.19}
\end{align}
Note that for sufficiently small $C_0$ the term in the brackets in \eqref{4.19} is always positive and $\tilde{r}'(\tilde{y})$ vanishes only when $\tilde{h}(\tilde{r})$ vanishes, which happens at the bulk horizon $\tilde{r}=\tilde{r}_H$ and it corresponds to the tip of the string worldsheet. Since we place the origin of the $\tilde{y}$-axis at the center of the $L$-side of the rectangular Wilson loop, the U-shaped profile implies the symmetry $\tilde{r}(\tilde{y})=\tilde{r}(-\tilde{y})$, and, consequently, the turning point must correspond to $\tilde{r}_H=\tilde{r}(\tilde{y}=0)$. Then, after integrating Eq.\ \eqref{4.19} we have
\begin{align}
\frac{L}{2}=C_0\int_{\tilde{r}_H}^{\tilde{r}_{\textrm{max}}} \frac{d\tilde{r}}{\sqrt{e^{2\tilde{A}(\tilde{r})}\tilde{h}(\tilde{r}) \left[ e^{\sqrt{8/3}\tilde{\phi}(\tilde{r})+4\tilde{A}(\tilde{r})}(1-\tilde{h}(\tilde{r})) - C_0^2 \right]}},
\label{4.20}
\end{align}
where, formally, $\tilde{r}_{\textrm{max}}\rightarrow\infty$. For small values of $C_0$ one may expand the integrand of \eqref{4.20} in powers of $C_0$ to obtain at leading order
\begin{align}
C_0 \approx \frac{L}{2\int_{\tilde{r}_H}^{\tilde{r}_{\textrm{max}}}d\tilde{r} \frac{e^{-\sqrt{2/3}\tilde{\phi}(\tilde{r})-3\tilde{A}(\tilde{r})}}{\sqrt{\tilde{h}(\tilde{r}) (1-\tilde{h}(\tilde{r}))}}}.
\label{4.21}
\end{align}
This shows that a small value of $C_0$ corresponds to a small $L$ (when compared to $L^-$). Plugging the solution \eqref{4.19} in Eq.\ \eqref{4.17} we obtain the on-shell Nambu-Goto action
\begin{align}
S_{\textrm{NG}}=\frac{L^-}{\pi\alpha'} \int_{\tilde{r}_H}^{\tilde{r}_{\textrm{max}}}d\tilde{r} \frac{e^{\sqrt{8/3}\tilde{\phi}(\tilde{r})+3\tilde{A}(\tilde{r})} (1-\tilde{h}(\tilde{r}))}{\sqrt{2\tilde{h}(\tilde{r}) \left[ e^{\sqrt{8/3}\tilde{\phi}(\tilde{r})+4\tilde{A}(\tilde{r})}(1-\tilde{h}(\tilde{r})) - C_0^2 \right]}},
\label{4.22}
\end{align}
where we used \eqref{4.20} and $\tilde{r}(\tilde{y})=\tilde{r}(-\tilde{y})$ symmetry. Following \cite{qhat1}, we now compute the disconnected contribution for this correlator corresponding to the action of two straight strings in the bulk. In order to compute these disconnected string configurations, it is convenient to define another static gauge for the worldsheet parameters, namely, $(\tau,\sigma)=(\tilde{x}_-,\tilde{r})$, such that the corresponding worldsheet embeddings are of the form $X^\mu(\tilde{x}_-,\tilde{r})=(\tilde{r},0,\tilde{x}_-,0,0)$. The sum of the actions for these two disconnected string configurations is then given by
\begin{align}
S_{\textrm{NG}}^{\textrm{straight}}=\frac{L^-}{\pi\alpha'} \int_{\tilde{r}_H}^{\tilde{r}_{\textrm{max}}}d\tilde{r} e^{\sqrt{\frac{2}{3}}\tilde{\phi}(\tilde{r})}\sqrt{g_{--}g_{rr}} = \frac{L^-}{\pi\alpha'} \int_{\tilde{r}_H}^{\tilde{r}_{\textrm{max}}}d\tilde{r} \,e^{\sqrt{\frac{2}{3}}\tilde{\phi}(\tilde{r})+\tilde{A}(\tilde{r})} \sqrt{\frac{1-\tilde{h}(\tilde{r})}{2\tilde{h}(\tilde{r})}}.
\label{4.23}
\end{align}
Subtracting the term \eqref{4.23} from Eq.\ \eqref{4.22} expanded up to $\mathcal{O}(C_0^2)$ and using Eq.\ \eqref{4.21}, one obtains the final on-shell (subtracted) Nambu-Goto action at leading order in $L / L^-$ describing the $L$-dependent interaction between the quark-antiquark pair moving through the hot and dense plasma
\begin{align}
S_{\textrm{NG}}^{\textrm{on-shell}}=S_{\textrm{NG}}-S_{\textrm{NG}}^{\textrm{straight}} \approx\frac{1}{2\pi\alpha'}\frac{L^- L^2}{4\sqrt{2}} \frac{1}{\int_{\tilde{r}_H}^{\tilde{r}_{\textrm{max}}}d\tilde{r} \frac{e^{-\sqrt{2/3}\tilde{\phi}(\tilde{r})-3\tilde{A}(\tilde{r})}}{\sqrt{\tilde{h}(\tilde{r})(1-\tilde{h}(\tilde{r}))}}}.
\label{4.24}
\end{align}
Finally, plugging \eqref{4.24} into \eqref{4.12} and going back to the numerical coordinates, as discussed in Section \ref{sec2.3}, we arrive at the formula for the jet quenching parameter in our background
\begin{align}
\frac{\hat{q}(T,\mu_B)}{\sqrt{\lambda_t}T^3} = \frac{64\pi^2 h_0^{\textrm{far}}}{\int_{r_{\textrm{start}}}^{r_{\textrm{max}}}dr \,\frac{e^{-\sqrt{2/3}\phi(r)-3A(r)}}{\sqrt{h(r)(h_0^{\textrm{far}}-h(r))}}}.
\label{4.25}
\end{align}

\begin{figure}[h]
\begin{centering}
\includegraphics[scale=0.50]{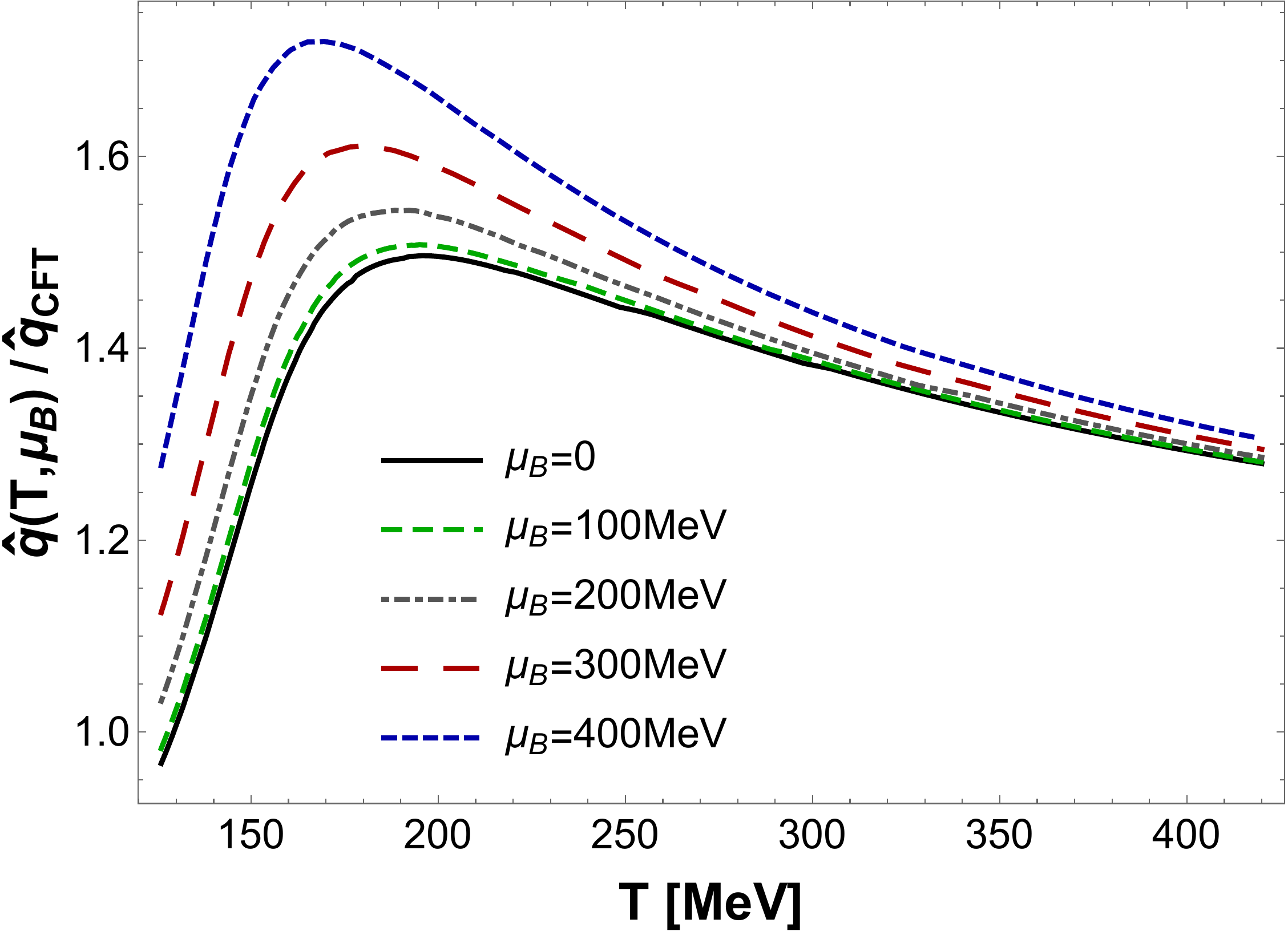}
\par\end{centering}
\caption{(Color online) Jet quenching parameter normalized by its conformal limit as a function of the temperature for different values of the baryon chemical potential.
\label{fig12}}
\end{figure}

Our numerical results for the $\hat{q}$ parameter \eqref{4.25} normalized by its conformal limit, $\hat{q}_{\textrm{CFT}}/\sqrt{\lambda_t}T^3=\pi^{3/2}\Gamma(3/4)/\Gamma(5/4)$ \cite{qhat1,qhat2}, are shown in Fig.\ \ref{fig12}. The jet quenching parameter displays a peak in the crossover region and its overall value increases with increasing baryon chemical potential. As observed in Ref.\ \cite{xingling}, the behavior of the curves for the jet quenching parameter are qualitatively similar to the ones found for the normalized trace anomaly (Fig.\ \ref{fig9}) and, consequently, the peak of this observable may be also employed to characterize the crossover temperature. The phenomenological relevance of a peak in the energy loss near the phase transition was pointed out in \cite{Liao:2008dk} and, more recently, in \cite{Xu:2014tda}.

Our results show that the energy loss of light probes in this plasma is not only sensitive to $\mu_B$ but also to the rapid change in degrees of freedom that occurs at the phase transition. The qualitative temperature dependence of $\hat{q}$ for $\mu_B=0$ found here is consistent with the analysis performed in \cite{Burke:2013yra}, though the overall magnitude of our result is larger than the values extracted via weak coupling calculations in \cite{Burke:2013yra}. Also, note that Fig.\ \ref{fig12} is consistent with our calculations for the energy loss of rapidly moving heavy quarks presented in the previous sections. More importantly, even though this strongly coupled medium cannot be understood in terms of quasiparticles, $\hat{q}$ is overall enhanced in a baryon dense system. 

Let us make a brief comparison between our result illustrated in Fig.\ \ref{fig12} at $\mu_B=0$ (solid curve) and the lattice result recently obtained in Ref.\ \cite{qhatlattice}. There, by taking the gauge coupling $g_{\textrm{QCD}}^2\approx 2.6$ at $T\approx 398$ MeV, it was estimated that $\hat{q}$ is around $6$ GeV$^2/$fm for QCD with two light flavors at this temperature (we note that this value is larger than the phenomenologically extracted result of \cite{Burke:2013yra}). From our result computed at $\mu_B=0$ one obtains $\hat{q}\approx 9.7\sqrt{\lambda_t}T^3$ at $T\approx 398$ MeV and, taking $\lambda_t=g_{\textrm{QCD}}^2N_c\approx 2.6\times 3=7.8$, one finds the holographic estimate $\hat{q}(T\approx 398\,\textrm{MeV},\mu_B=0) \approx 8.6 \,\textrm{GeV}^2/\textrm{fm}$. This is quite close to the lattice result even though our holographic model was constructed to mimic the thermodynamics of $(2+1)$-flavor QCD instead of QCD with only two light flavors considered in Ref.\ \cite{qhatlattice}.

Finally, let us note that it is also possible (as discussed, for instance, in \cite{langevin3}) to define a Langevin jet quenching parameter $\hat{q}_\perp$ (associated with the transverse momentum broadening of heavy quarks in the Langevin processes) from the perpendicular Langevin diffusion coefficient calculated in Section \ref{sec4.3}: $\hat{q}_\perp=2\kappa_\perp/v$. This parameter is different than the jet quenching parameter $\hat{q}$ \cite{qhat1} since the latter is strictly valid for light-like trajectories (see, for instance, the discussion in \cite{langevin2}). However, as one can see in Fig.\ \ref{fig13}, this Langevin $\hat{q}_\perp$ (which is trivially related to $\kappa_\perp$) at high velocities displays a qualitative behavior similar to the one seen in the jet quenching parameter $\hat{q}$ in Fig.\ \ref{fig12}.

\subsection{Light quark energy loss: the shooting string}
\label{sec4.4}

\hspace{5 mm} In this section we will use the holographic model of ``shooting strings'' proposed in \cite{light} to calculate the light quark energy loss in our nonconformal backgrounds. The shooting string is the simplest phenomenological implementation of the finite endpoint momentum approach \cite{FEM} to express the energy loss of light quarks in a way that can be used to calculate observables such as the nuclear modification factor $R_{AA}$ at RHIC and LHC\footnote{Other holographic models of light quark energy loss include the holographic version of the brick problem of \cite{PaulKrishna} and its hybrid phenomenological implementation in \cite{Hybrid}, as well as the recent work of \cite{WillRazieh} where different holographic definitions of light quark jets were explored.}. As shown in \cite{FEM}, adding finite momentum at the endpoints of classical strings allows them to travel further in the $AdS_5$-Schwarzschild background than the previous falling string configurations \cite{Chesler-Jets,Chesler-Stop}. At the same time, it was argued that strings with finite momentum at their endpoints may provide a more natural description of energetic light quarks plowing through a hot, strongly coupled plasma. These results were reassuring as they had the potential to reconcile the apparent quantitative inconsistencies of the first attempts at holographic light quark energy loss of \cite{FicnarSolo,Light-Old} and the LHC suppression data for light hadrons \cite{CMS-Light}. 

When finite momentum is added to the endpoint, it turns out that its motion simplifies \cite{FEM}: the endpoint must follow null geodesics, while the rest of the string sags behind it. In this approach, the quark energy loss is naturally described as the flow of energy from the endpoint into the bulk of the string, as the latter represents the color field generated by the quark. This quantity has a simple analytical expression \cite{light}:
\begin{equation}\label{s1}
\frac{d\tilde E}{d\tilde x}=-\frac{|R|}{2\pi\alpha'}\tilde G_{xx}^{(S)}(\tilde r)\,,
\end{equation}
where we assumed that the endpoint is moving in the $\tilde x$-direction, and that the string frame metric $\tilde G_{\mu\nu}^{(S)}$ is diagonal in $(\tilde t,\vec{\tilde x},\tilde r)$ coordinates and depends only on the radial coordinate $\tilde r$, as is the case in our background (\ref{2.19}). The constant $R$ parametrizes the geodesic that the endpoint is following, and is equal to the ratio of energy and momentum that a free particle following this geodesic would have conserved. 

In order to cast (\ref{s1}) into a phenomenologically usable form, we need to relate $R$ to some observable quantity and express $\tilde r$ in terms of $\tilde x$, which involves inverting the geodesic trajectory $\tilde x_{\rm geo}(\tilde r)$. A simple way to do this was proposed in \cite{light} where it was shown that, for phenomenological reasons, it makes sense to consider the endpoints that are initially close to the horizon. In this case, the endpoint is ``shot'' towards the boundary, and, for moderately large maximum $\tilde r$ that the endpoint can reach, the energy loss can be well approximated by using the critical geodesic with $R=1$. This is called the \emph{shooting string limit} and for an $\mathcal{N}=4$ SYM plasma (corresponding to the $AdS_5$-Schwarzschild background) it leads to the following expression for the energy loss
\begin{equation}\label{s2}
\frac{d\tilde E_{\rm conformal}}{d\tilde x}=-\frac{\pi}{2}\sqrt{\lambda_t}T^2\left(1+\pi T \tilde x\right)^2\,.
\end{equation}

To obtain the analog of (\ref{s2}) in our numerical backgrounds, we first need to find the critical geodesics numerically. To do that, we will parametrize the geodesics with time $\tilde t$ and seek to obtain $\tilde r_{\rm geo}(\tilde t)$ and $\tilde x_{\rm geo}(\tilde t)$. Along any geodesic in background (\ref{2.19}), the energy and the momentum (of a test particle) are conserved:
\begin{eqnarray}\label{s3}
\tilde E_{\rm geo} &=& \frac{1}{\eta}e^{2\tilde A}\tilde h \,,\\ \label{s4}
\tilde p_{x,{\rm geo}} &=& \frac{1}{\eta}e^{2\tilde A}\frac{d\tilde x_{\rm geo}}{d\tilde t}\,,
\end{eqnarray}
where $\eta(\tilde t)$ is an auxiliary field (the ``metric'' on the worldline). Defining $R\equiv \tilde E_{\rm geo}/\tilde p_{x,{\rm geo}}$, we can solve for $d\tilde x_{\rm geo}/d\tilde t$
\begin{equation}\label{s5}
\frac{d\tilde x_{\rm geo}}{d\tilde t}=\frac{\tilde h(\tilde r_{\rm geo}(\tilde t))}{R}\,.
\end{equation}
Plugging this in the null condition $d\tilde s^2=0$ allows us to solve for $d\tilde r_{\rm geo}/d\tilde t$:
\begin{equation}\label{s6}
\frac{d\tilde r_{\rm geo}}{d\tilde t}=\frac{1}{R}e^{\tilde A(\tilde r_{\rm geo}(\tilde t))}\tilde h(\tilde r_{\rm geo}(\tilde t))\sqrt{R^2-\tilde h(\tilde r_{\rm geo}(\tilde t))}\,.
\end{equation}
We can now in principle solve for null geodesics by dividing (\ref{s5}) and (\ref{s6}) and integrating the resulting $d\tilde x_{\rm geo}(\tilde r)/d\tilde r$. However, close to the maximum $\tilde r$ that the geodesic reaches (which we call $\tilde r_*$ and is given by $R^2=\tilde h(\tilde r_*)$), the quantity $d\tilde x_{\rm geo}/d\tilde r$ diverges and the numerical procedure is unstable. This is precisely the reason why we chose to parametrize the geodesics with coordinate $\tilde t$. The first order equation (\ref{s6}) still vanishes at $\tilde r=\tilde r_*$, resulting in ``stiffness'' of the numerical system: in order to avoid this, we will use the second-order equation for $\tilde r_{\rm geo}(\tilde t)$, which is just the geodesic equation (with an additional term that corrects for the fact that $\tilde t$ is not an affine parameter). 

\begin{figure}[h]
\begin{centering}
\includegraphics[width=0.46\textwidth]{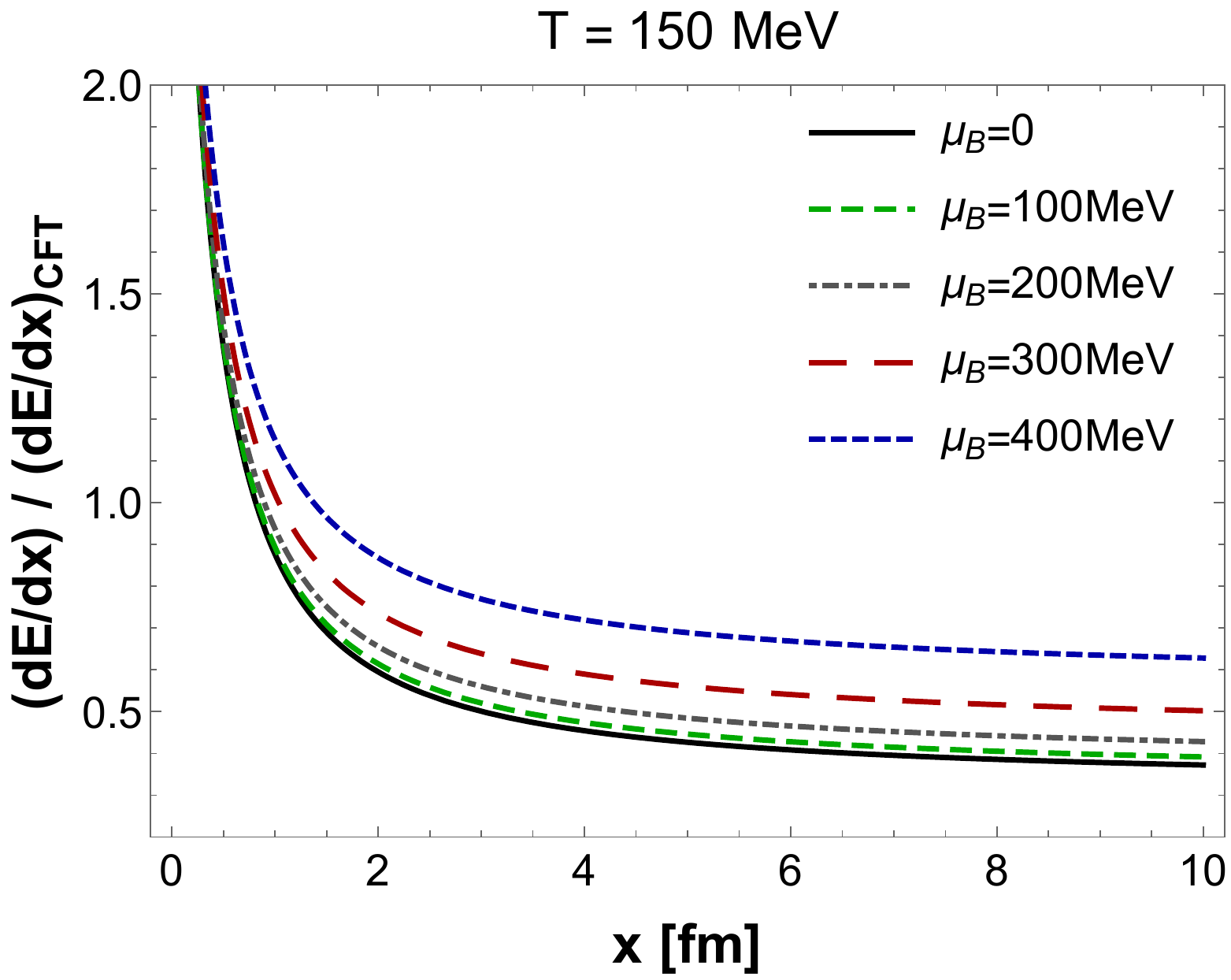}
\hskip0.07\textwidth
\includegraphics[width=0.46\textwidth]{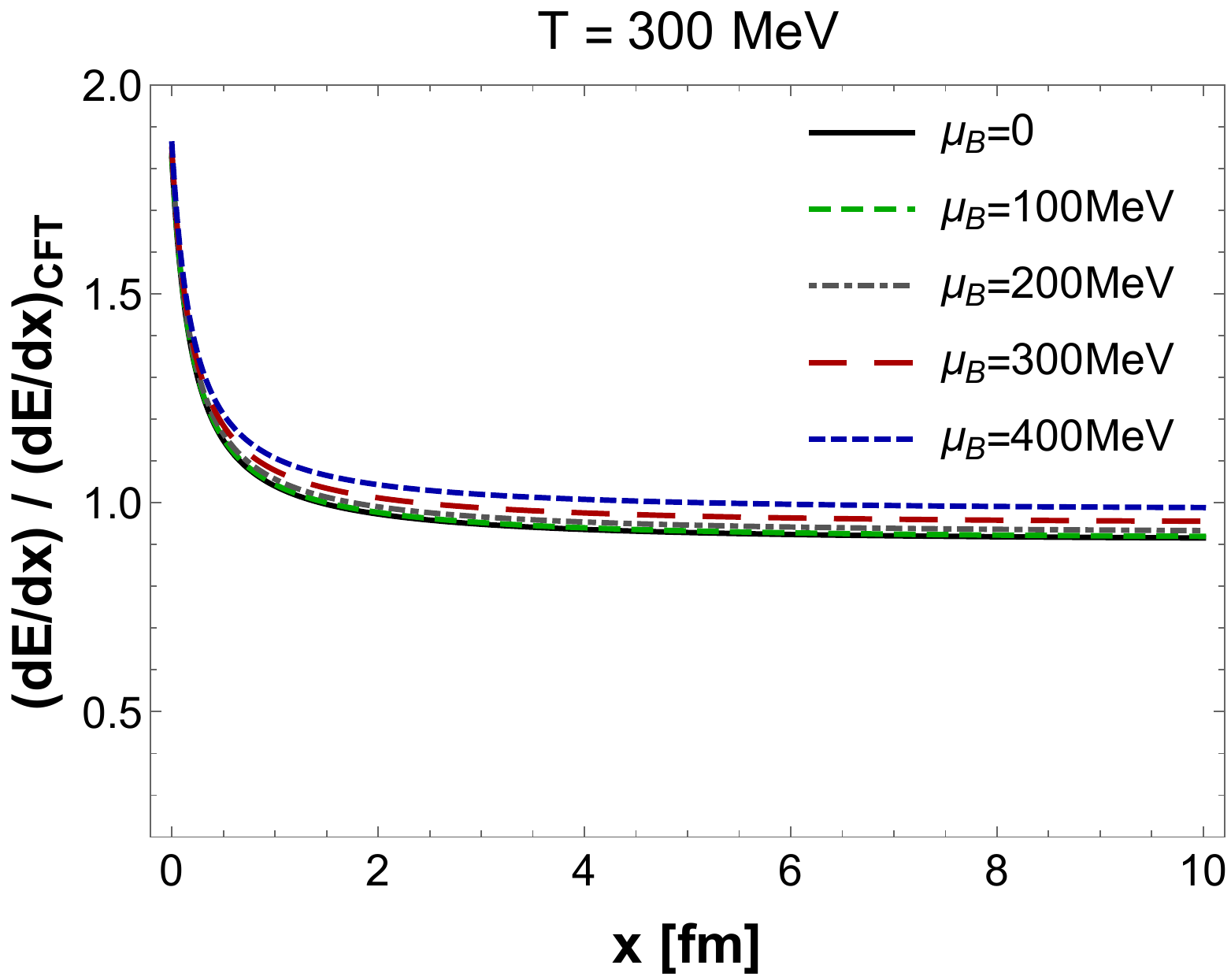}
\par\end{centering}
\caption{(Color online) Ratio of the shooting string energy loss in our model (\ref{s7}) and its conformal limit (\ref{s2}) as a function of distance $\tilde x$ computed at two fixed temperatures and different values of the baryon chemical potential. \label{dEdxShootingFig}}
\end{figure}

Finally, in order to take the shooting limit, we choose as large $\tilde r_*$ as the numerical code allows, solve for the null geodesic trajectory with initial conditions $\tilde r_{\rm geo}(0)=\tilde r_H$ and $\tilde x_{\rm geo}(0)=0$, plug the solution $\tilde r_{\rm geo}(\tilde t)$ in (\ref{s1}), which explicitly looks like
\begin{equation}\label{s7}
\frac{d\tilde E}{d\tilde x}=-\frac{1}{2\pi\alpha'}e^{2\tilde A+\sqrt{2/3}\tilde \phi}\,,
\end{equation}
and check that changing $\tilde r_*$ around that value does not affect $d\tilde E/d\tilde x$ within some phenomenologically relevant range of $\tilde x$. The results for a range of different chemical potentials and temperatures are shown in Fig.\ \ref{dEdxShootingFig}.

One can see in Fig.\ \ref{dEdxShootingFig} that the greatest deviations from the conformal limit occur at early times (small $\tilde x$), which is a novel feature of the shooting string setup: at small $\tilde x$ the endpoint is close to the horizon and hence sees the strongest effects of conformal symmetry breaking. We also see that, at early times, the energy loss in this model is larger than its conformal limit, while at later times, it becomes smaller and asymptotes to the same functional $\tilde x$-dependence as the conformal one (which is, again, expected, since in the shooting limit, large $\tilde x$ means that the endpoint is closer to the boundary). Finally, the higher the temperature, the closer the ratios in Fig. \ref{dEdxShootingFig} are to unity, correctly recovering the conformal limit. Furthermore, the energy loss increases with increasing $\mu_B$, which is consistent with expectations based on simple density arguments (though this strongly coupled dense medium does not admit a quasiparticle description).

Another interesting observation about Fig.\ \ref{dEdxShootingFig} is that, in the phenomenologically relevant range of $\tilde{x} \sim 0-1$ fm, the shooting string energy loss is larger relatively to its conformal limit at lower temperatures than it is at higher temperatures. This has potentially interesting phenomenological implications for the problematic simultaneous match of the nuclear suppression factors $R_{AA}$ of light hadrons at RHIC and LHC \cite{SurprisingTransparency}. As shown in \cite{light}, if one tries to fix the parameters of a conformal model by matching, say, the $R_{AA}$ data at the LHC and then use the same set of parameters at RHIC, the RHIC $R_{AA}$ ends up being above the data indicating that the predicted energy loss is simply too small. However, in this more realistic nonconformal model that can assess the rapid change in degrees of freedom in the phase transition, we see that the energy loss at RHIC (which is, on average, colder than LHC) is higher than the conformal expectation and, thus, this will result in a better match with the data and perhaps even completely close the ``gap'' between RHIC and LHC. We hope to come back to this issue in the near future.

\begin{figure}[h]
\begin{center}
\includegraphics[width=0.46\textwidth]{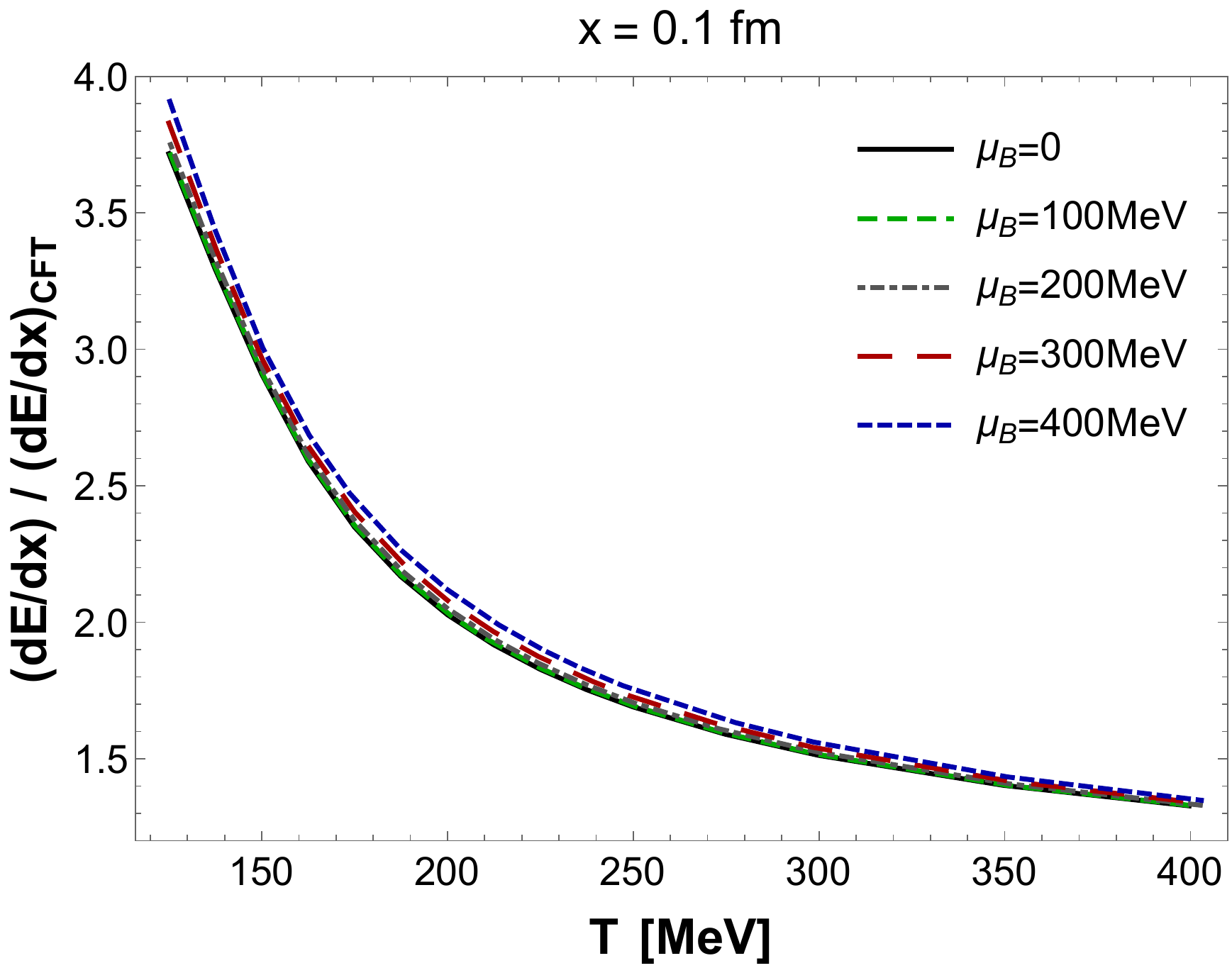}
\includegraphics[width=0.46\textwidth]{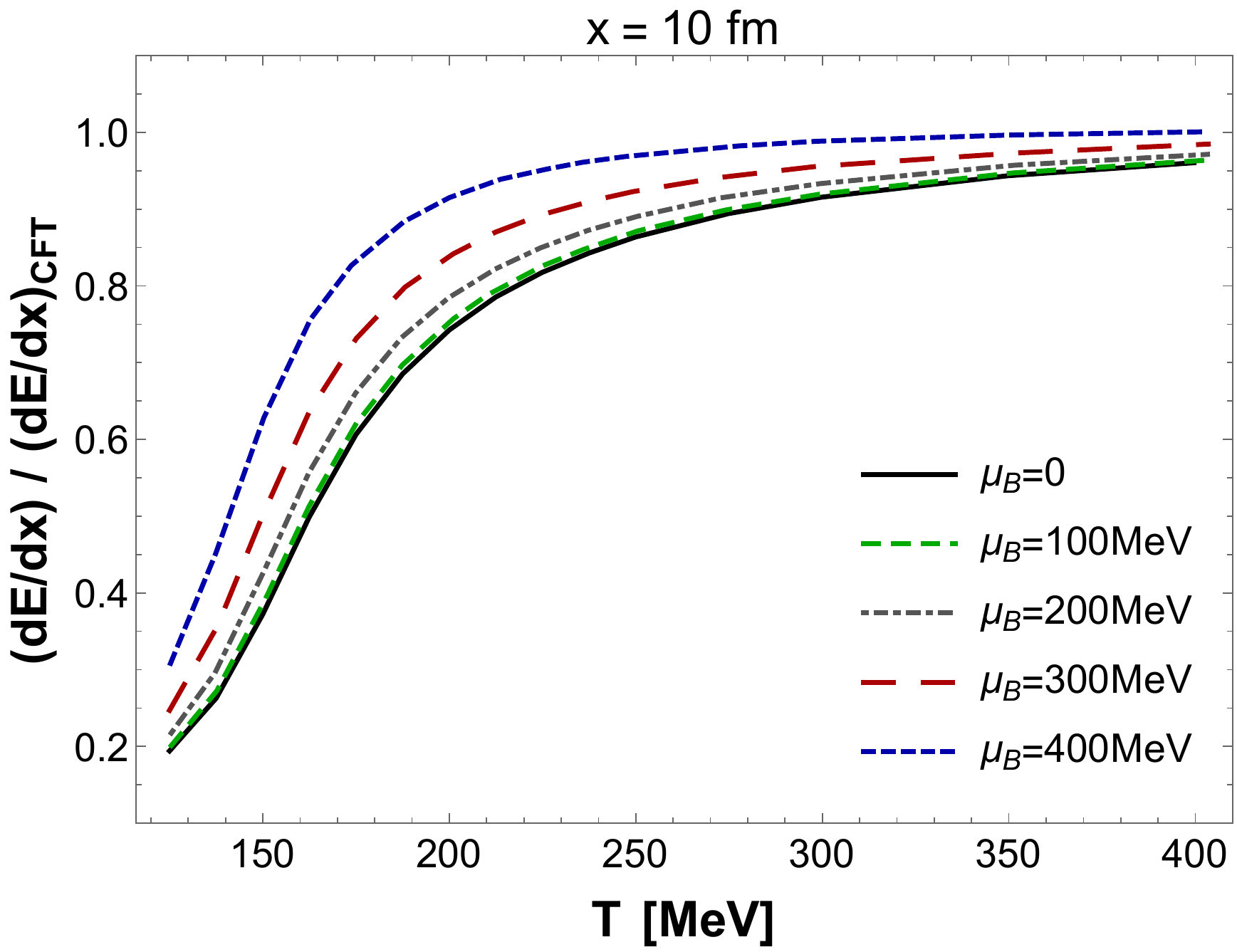}
\end{center}
\caption{(Color online) Ratio of the shooting string energy loss in our model (\ref{s7}) and its conformal limit (\ref{s2}) as a function of temperature for fixed distances and different values of the baryon chemical potential. \label{dEdxShootingTDep}}
\end{figure}

It is also interesting to consider energy loss as a function of temperature at some fixed distance $\tilde x$, as shown in Fig.\ \ref{dEdxShootingTDep}. We see that, at early times (left plot), the light quark energy loss (normalized by its conformal limit) approaches unity from above and looks in fact more similar to the heavy quark drag force at low velocities (Fig. \ref{fig11}) than to $\hat q$ (Fig. \ref{fig12}). At late times (right plot), this behavior changes qualitatively and this ratio approaches unity from below, developing features qualitatively similar to $\hat q$. This nontrivial behavior is quite specific to the shooting string setup and it would be interesting to see how this is translated into observables such as $R_{AA}$ and especially the elliptic flow parameter $v_2$, which is very sensitive to the specific path dependence of energy loss.

One should also note how, at early times, the energy loss is almost insensitive to the baryon chemical potential while at later times it becomes more sensitive to it. This is again a reflection of the shooting string setup in which, at early times, the endpoint is close to the horizon where $\tilde \Phi(\tilde r)=0$. A similar behavior has been observed before in the drag force (Fig. \ref{fig11}) where at low velocities one is more sensitive to the metric close to the horizon.

In summary, our holographic model allows for the calculation of quantities that characterize the energy loss of heavy quarks (see Figs.\ \ref{fig11} and \ref{fig13}) and light quarks (see Figs.\ \ref{fig12}, \ref{dEdxShootingFig} and \ref{dEdxShootingTDep}) in a hot and baryon dense strongly interacting plasma that not only behaves as a nearly perfect fluid but also displays thermodynamic properties that are in quantitative agreement with current lattice QCD calculations in the crossover region. Energy loss is found to overall increase with $\mu_B$ while also displaying a nontrivial, fast-varying behavior as a function of the temperature near the crossover. These results can be readily implemented in phenomenological studies of the nuclear modification factor in the QGP such as, for instance, \cite{Cao:2015hia}.

\section{Equilibration time of a baryon rich strongly coupled plasma}
\label{QNM}

\hspace{5 mm} In this section we will explore some generic aspects concerning the equilibration properties of our strongly coupled plasma at finite temperature and baryon chemical potential by computing the lowest quasinormal mode (QNM) associated with an external massless scalar perturbation in the bulk. The main idea behind this is that non-hydrodynamical degrees of freedom\footnote{A non-hydrodynamical mode is defined by the following condition for its dispersion relation: $\omega(k=0)\neq 0$.} play an important role in the transition from a non-equilibrium state to an equilibrated plasma\footnote{This can be easily understood in kinetic theory where the timescale associated with the lowest quasinormal mode enters in the effective hydrodynamic description via the relaxation time coefficient associated with the shear stress tensor \cite{Denicol:2011fa}. This detail is important in studies of the applicability of different hydrodynamic theories when compared to exact kinetic theory solutions, such as in \cite{Florkowski:2013lza,Denicol:2014xca,Denicol:2014tha}.} and these modes in a strongly coupled gauge theory correspond, via the holographic duality, to the quasinormal modes of a black hole background in the bulk \cite{HorowitzHubeny,SonStarinets}. The motivation for studying this subject comes from the interest in the thermalization properties of the strongly coupled quark-gluon plasma produced in heavy ion collisions. Here we will focus on the lowest quasinormal modes as these represent the dominant, longest-lived non-hydrodynamical degrees of freedom that can be used to provide an upper bound on the equilibration time in the dual plasma.

Quasinormal modes for nonconformal holographic duals have been recently investigated in \cite{JanikQNM} and \cite{BuchelQNM1,BuchelQNM2}, motivated by the potential effects that the breaking of conformal invariance might have on the equilibration of strongly coupled plasmas that are qualitatively more similar to the actual quark-gluon plasma formed in ultrarelativistic heavy ion collisions. In \cite{JanikQNM} the authors looked at a class of bottom-up nonconformal Einstein-Dilaton models with the dilaton potential tuned to reproduce the QCD equation of state at zero baryon chemical potential\footnote{One of the dilaton potentials used in \cite{JanikQNM} was the same we used here in \eqref{2.53} and, previously, in \cite{transport}.}, while in \cite{BuchelQNM1} and \cite{BuchelQNM2} the authors studied the rate of equilibration in two nonconformal top-down models, the $\mathcal{N}=2^*$ plasma \cite{Buchel:2007vy} and the Klebanov-Strassler model \cite{Klebanov:2000hb}, respectively. These studies have found that even when the deviation from conformal behavior was large, as it happens for instance around the crossover phase transition, the imaginary parts of the lowest quasinormal mode frequencies differed by roughly a factor of two from their conformal limit. Although this does not represent a qualitative change, a factor of two difference in the exponential damping of these degrees of freedom close to the phase transition may be of phenomenological importance. 

In fact, as originally pointed out in \cite{Denicol:2011fa,Noronha:2011fi}, the fact that the non-hydrodynamical modes in spatially isotropic and uniform strongly coupled models possess a nonzero real part \cite{Kovtun:2005ev} implies that the way these systems approach their universal hydrodynamic limit is radically different than what is observed in the case of dilute gases described by the relativistic Boltzmann equation. In the latter, spatially isotropic and uniform dilute gases have purely imaginary non-hydrodynamic modes, which lead to a characteristic exponential decay of these excitations. However, the nonzero real part found for these modes at strong coupling leads to an oscillatory component for the dissipative currents (such as the shear stress tensor) that seems to be a novel feature revealed by the holographic correspondence. This property has not yet been incorporated in the hydrodynamical modeling of the QGP \cite{reviewQGP1} though it is clearly seen in real time, holographic calculations \cite{JanikLocality}.      

Here we investigate the effect of a nonzero baryon chemical potential on the lowest QNM of an external massless scalar field in the limit of zero spatial momentum, $k\to 0$, as in \cite{JanikQNM,BuchelQNM1,BuchelQNM2}. In this limit, as discussed in \cite{JanikQNM}, all the relevant information about the QNM's of the model may be extracted from the study of an external massless scalar perturbation. We will also check how the QNM frequency varies with $k$ and see that it shows only a very mild dependence up to $k\approx 2\pi T$, in accordance with the so called ``ultralocality'' property first pointed out in \cite{JanikLocality} for a conformal plasma and later found to hold also in nonconformal plasmas at zero chemical potential \cite{JanikQNM}.

We start by writing our background \eqref{2.2} in the in-falling Eddington-Finkelstein coordinates since in those coordinates the condition that the mode is ingoing at the horizon translates into simple regularity there. We define the Eddington-Finkelstein time $v$, as follows
\begin{equation}\label{qnm1}
dv=dt+e^{-A}\frac{dr}{h}\,.
\end{equation}
With this transformation, the metric becomes
\begin{equation}\label{qnm2}
ds^2=e^{2A}\left(-hdv^2+d\vec x^2\right)+2e^{A}dvdr\,.
\end{equation}
The equation of motion of an external, minimally coupled massless scalar field\footnote{As discussed in \cite{JanikQNM}, it is important to distinguish this external scalar perturbation from the dilaton perturbation since the latter mixes with the perturbations for the metric and the vector field.} in this geometry is just the Klein-Gordon equation, 
\begin{equation}\label{qnm3}
\nabla_\mu\nabla^\mu \Psi=0\,,
\end{equation}
and we are considering the (spatially isotropic) harmonic solution in the boundary directions with a nontrivial dependence on the radial coordinate: $\Psi=e^{-i\omega v +ikx}\psi(r)$. With this Ansatz, the equation of motion has the following form
\begin{equation}\label{qnm4}
he^A\psi''+\left(e^Ah'+4e^AhA'-2i\omega\right)\psi'-\left(k^2e^{-A}+3i\omega A'\right)\psi=0\,.
\end{equation}
This equation needs to be solved with the condition that $\psi$ is regular at the horizon while it obeys a Dirichlet condition at the boundary, which means that this equation for a given background will admit solutions only for discretely many complex frequencies $\omega$. We searched for these frequencies numerically using a simple shooting method where one solves \eqref{qnm4} on a grid of complex $\omega$, specifying some initial value for $\psi$ at the horizon and its derivative as given by the near horizon expansion of \eqref{qnm4}:
\begin{equation}\label{qnm5}
\psi(0)=\psi_0,\qquad \psi'(0)=-\frac{3\omega}{i+2\omega}A_1\psi_0\,,
\end{equation}
where we set $k=0$. We note that here we are working with the numerical coordinates and we used the near-horizon expansions of the metric from Section \ref{sec2.2} (in particular, $A_1$ is given by \eqref{2.26}). Once the numerical solution is obtained, we look for the QNM frequencies by demanding that the value of $\psi$ at some large $r$ near the boundary vanishes\footnote{In practice, one could start with some rough grid in complex $\omega$ centered around the conformal value and then find clusters of small values of $|\psi(r\to\infty)|$. These points can then be used as centers for a finer grid and one can keep repeating this until the required accuracy is reached.}.

\begin{figure}[h]
\begin{center}
\includegraphics[width=0.49\textwidth]{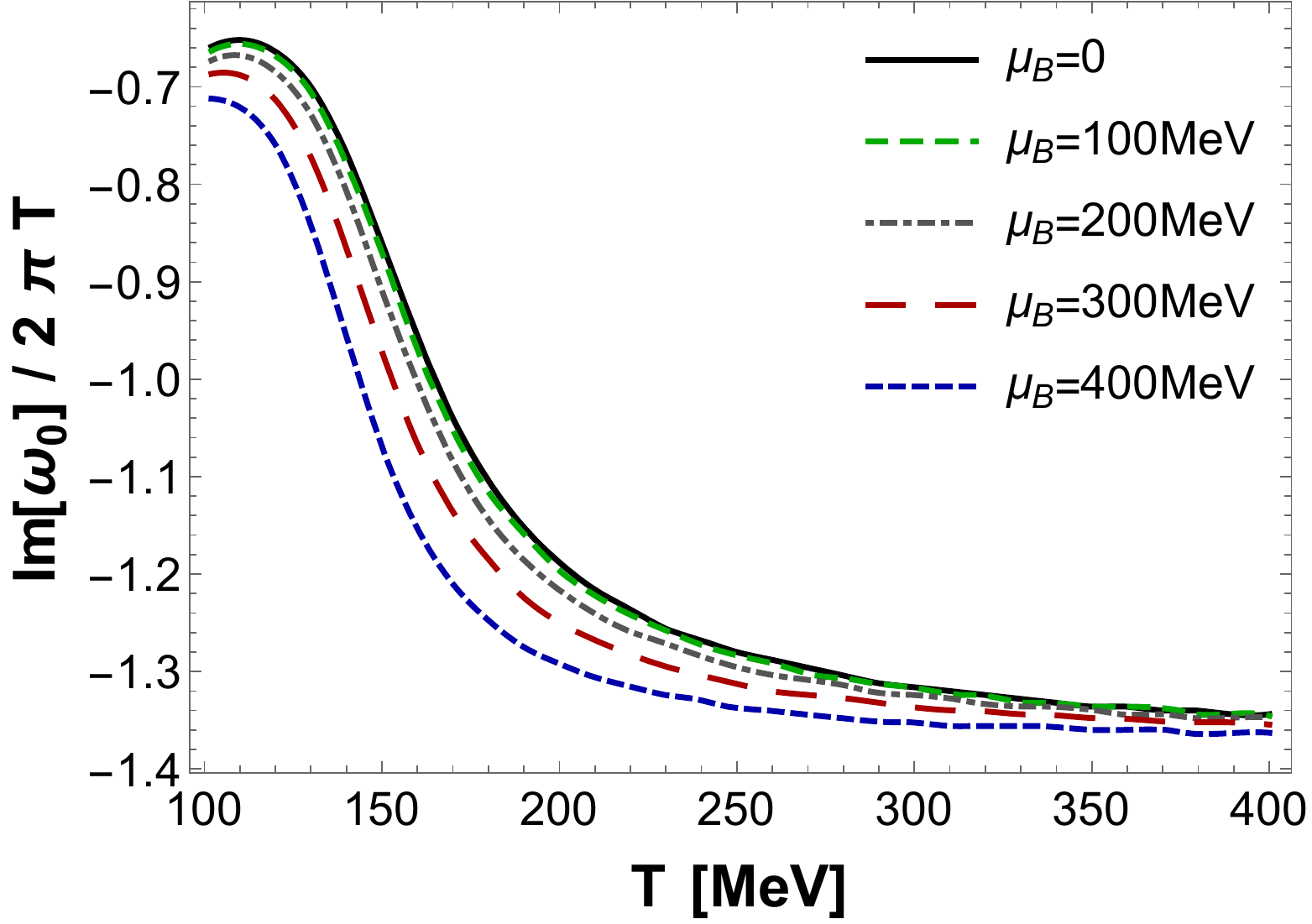}
\includegraphics[width=0.49\textwidth]{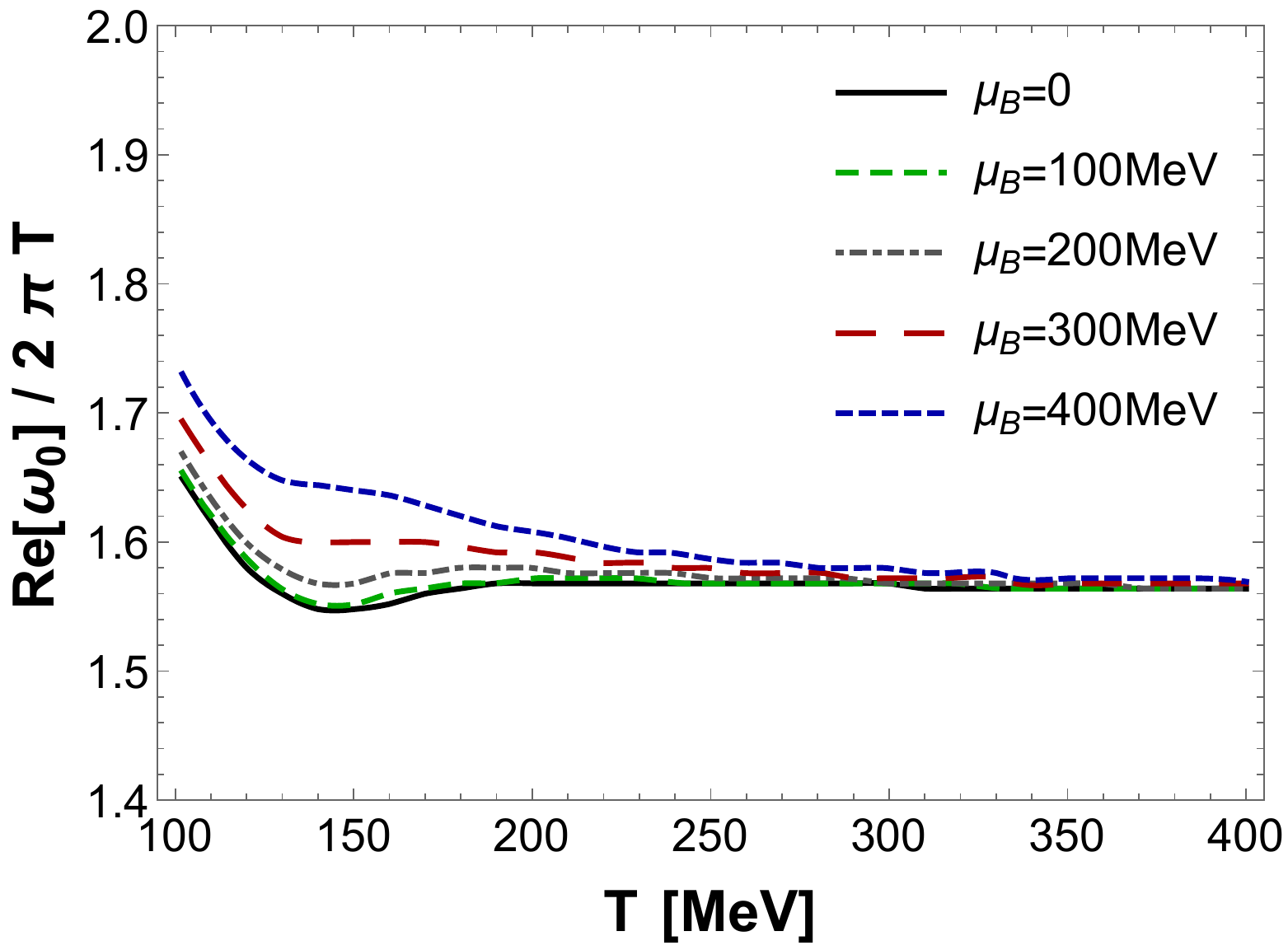}
\end{center}
\caption{(Color online) Imaginary (left panel) and real (right panel) parts of the lowest quasinormal mode of a massless scalar with spatial momentum $k=0$ as functions of the temperature for different values of the baryon chemical potential. \label{ImReFig}}
\end{figure}

Our results for the lowest QNM mode $\omega_0$ with vanishing momentum are shown in Fig.\ \ref{ImReFig}, where one can see how the temperature dependence of the real and imaginary parts of $\omega_0$ change as we vary the baryon chemical potential. Our result for the imaginary part of $\omega_0$ at zero chemical potential is very similar to the one found in \cite{JanikQNM} as it exhibits a factor of two decrease (in absolute value) close to the phase transition while returning to the conformal limit of $\frac{{\rm Im}[\omega_{0,{\rm CFT}}]}{2\pi T}=-1.373$ at high temperatures. This indicates that, generically, the typical equilibration times close to the phase transition may in fact be a factor of two larger than what one would expect from the conformal results. In the same plot we can also see the effects of the finite chemical potential on ${\rm Im}[\omega_0]$, which show that, by increasing $\mu_B$, one generally gets smaller equilibration times. The real part of $\omega_0$, shown in the right plot, is showing very little (relative) deviation from the conformal limit as one varies both the temperature and the baryon chemical potential. If a strong coupling description based on holography is appropriate for the QGP formed in low energy heavy ion collisions at RHIC, our results show that this plasma should still be a nearly perfect liquid\footnote{As discussed in \cite{Liao:2009gb}, at nonzero density $\eta T/(\epsilon+p)$ may be a more appropriate measure of fluidity than $\eta/s$. In our model, $\eta/s=1/4\pi$ and the baryon density $\rho$ is positive, which implies that $\eta T/(\epsilon+p) < 1/4\pi$.} in which the equilibration process towards hydrodynamic behavior is similar to that found at zero baryon density \cite{JanikLocality}, which is not included in the current numerical hydrodynamic codes. 

\begin{figure}[h]
\begin{center}
\includegraphics[width=0.70\textwidth]{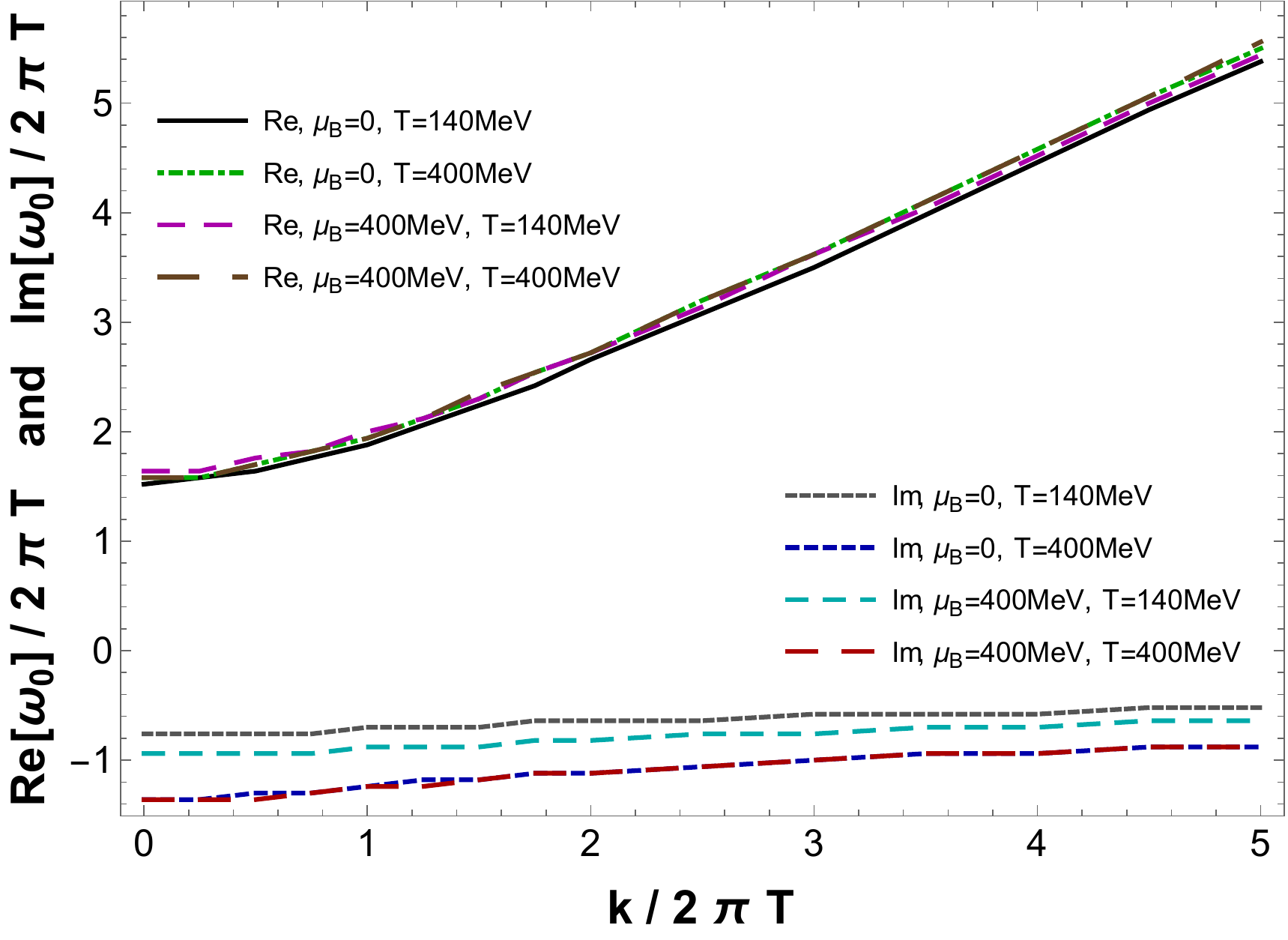}
\end{center}
\caption{(Color online) Imaginary and real parts of the lowest quasinormal mode of a massless scalar as functions of the momentum for four backgrounds corresponding to different values of temperature and baryon chemical potential. \label{NonZeroK}}
\end{figure}

In Fig.\ \ref{NonZeroK} we show the $k$-dependence of the imaginary and real parts of the frequency of the lowest QNM for four different backgrounds that span our phase space with temperatures between $140$ and $400$ MeV and baryon chemical potentials between $0$ and $400$ MeV. We see that the $k$-dependence of the imaginary parts is quite mild while the real parts quickly attain the conformal linear dispersion.



\section{Concluding remarks and perspectives}
\label{conclusion}

\hspace{5 mm} In this work, we used a 5-dimensional Einstein-Maxwell-Dilaton bottom-up holographic model \cite{gubser1,gubser2} that provides a comprehensive framework to investigate both the equilibrium and out-of-equilibrium properties of a baryon rich strongly coupled QGP. The parameters of the model were determined using current lattice data for $(2+1)$-flavor QCD thermodynamics at zero baryon chemical potential. Once all the parameters were fixed by $\mu_B=0$ lattice data, we proceeded to compare the thermodynamics of our holographic model with $\mu_B\neq 0$ lattice results obtaining a good quantitative agreement for the pressure and the speed of sound for baryon chemical potentials up to 400 MeV. We also determined the crossover region in our holographic model for this range of chemical potentials and we found that the crossover temperature is lowered as one increases the value of the baryon chemical potential. The curvature of the crossover band in our model is in remarkable agreement with recent lattice results. Therefore, this holographic model was shown to adequately describe many of the equilibrium properties of the hot and dense QGP near the crossover transition, besides also displaying the nearly perfect fluidity property inherent to holographic approaches \cite{Kovtun:2004de,Buchel:2003tz}.

This model was used to investigate two important aspects involving non-equilibrium processes that can take place in the QGP: the energy loss of light and heavy quarks and the equilibration time associated with non-hydrodynamic modes. We found that the energy loss of both light and heavy flavors near the phase transition is enhanced once finite baryon density effects are taken into account. This justifies the expectation that highly energetic probes should lose more energy in denser media at a given temperature even in the absence of a quasiparticle description. We also found that the inclusion of nonzero baryon density effects in the nonconformal plasma does not change the qualitative behavior displayed by the QNM's of a massless scalar excitation: the spectrum of QNM's still possesses both real and imaginary parts at zero spatial momentum and their values are reduced by a factor of nearly two at the crossover transition \cite{JanikQNM}, though we found that the equilibration time associated with the lowest lying QNM becomes shorter in a dense medium.

The holographic model used in the present work may be also employed to obtain predictions for many other physical observables that are relevant for the study of the QGP formed in heavy ion collisions. Some projects we hope to pursue in the near future include the generalization of the recent calculations performed in Ref.\ \cite{transport} involving the second order hydrodynamic transport coefficients by taking into account a nonzero $\mu_B$, which would entail in the calculation of new transport coefficients associated with baryon diffusion in the dense plasma. In this regard, by using the very same Einstein-Maxwell-Dilaton holographic model constructed in the present work, some of us have recently obtained in Ref. \cite{Rougemont:2015ona} holographic predictions for the $T$ and $\mu_B$ dependence of some of the transport coefficients associated with the baryon charge sector, namely, the baryon and thermal conductivities, and also the baryon diffusion constant. In Ref. \cite{Finazzo:2015xwa}, the same model was used to obtain predictions for the electric charge sector of the hot and baryon dense QGP, with calculations of the $T$ and $\mu_B$ dependence of the electric conductivity and electric diffusion constant, and also the thermal photon and dilepton production rates. It would also be interesting to see how the holographic Polyakov loop \cite{Noronha:2009ud,Noronha:2010hb} and the imaginary part of the rectangular, time-like Wilson loop \cite{Noronha:2009da,Finazzo:2013rqy} are affected by a finite baryon density.

\acknowledgments

We thank R.~Critelli for insightful comments on the manuscript. R.~R. acknowledges financial support by the S\~{a}o Paulo Research Foundation (FAPESP) under FAPESP grant number 2013/04036-0. A.~F. was supported by the European Research Council (ERC) under the European Union's Seventh Framework Programme (ERC Grant agreement 307955). S.~I.~F. was successively supported by a FAPESP PhD fellowship, under FAPESP grant number 2011/21691-6, and by a FAPESP postdoctoral fellowship, granted under a FAPESP and Coordena\c{c}\~ao de Aperfei\c{c}oamen\-to de Pessoal de N\'{i}vel Superior (CAPES) agreement, under FAPESP grant number 2015/00240-7. J.~N. acknowledges financial support by FAPESP and Conselho Nacional de Desenvolvimento Cient\'ifico e Tecnol\'ogico (CNPq) and thanks the Physics Department at Columbia University for its hospitality.

\appendix
\section{Some technical aspects of the model}
\label{apa}

\hspace{5 mm} In this Appendix we discuss some technical aspects of our holographic model concerning mainly the forms of $V(\phi)$ and $f(\phi)$ fixed in Eqs.\ \eqref{2.53} and \eqref{2.54}, respectively, and their relations with the thermodynamic stability of our black hole solutions.

\begin{figure}[h]
\centering
\includegraphics[width=0.48\textwidth]{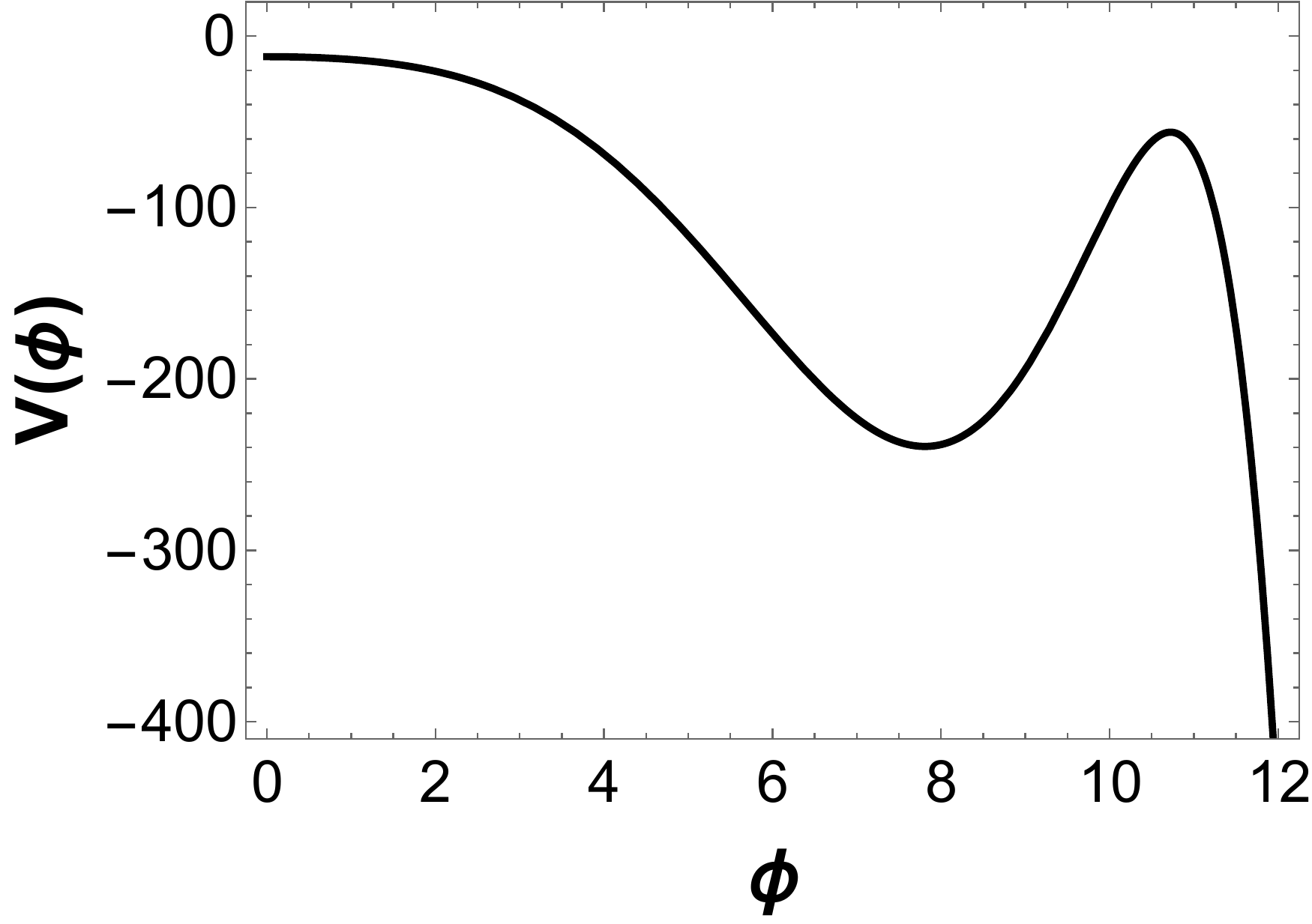}
\includegraphics[width=0.48\textwidth]{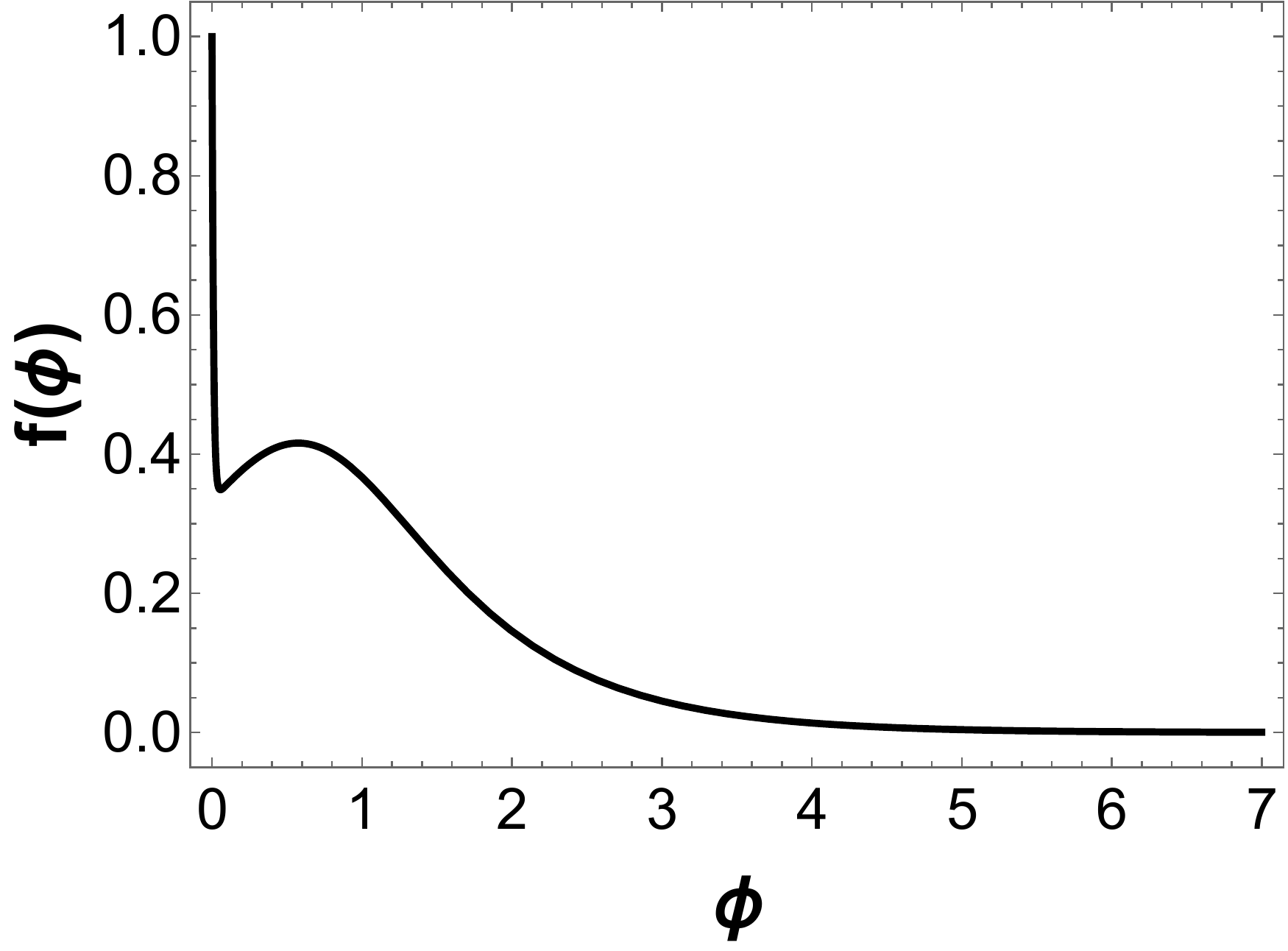}
\caption{Dilaton potential and Maxwell-Dilaton coupling used in the present work for $\phi\ge 0$. \label{Vandf}}
\end{figure}

In the present work, as done originally in \cite{gubser1,gubser2}, we restricted our calculations to positive values of the initial condition $\phi_0$ corresponding to the value of the dilaton field at the horizon. In this case, we checked that all the numerical solutions we produced have $\phi(r)\ge 0$. We show in Fig.\ \ref{Vandf} the plots for the dilaton potential $V(\phi)$ and the Maxwell-Dilaton coupling $f(\phi)$ fixed in Eqs.\ \eqref{2.53} and \eqref{2.54}.

By following \cite{Charmousis:2010zz,Gouteraux:2012yr}, we identify the fixed points of our model:
\begin{itemize}
\item Ultraviolet fixed points: these are given by the maxima of the potential at $\phi=0$ and $\phi\approx\pm 10.72$;
\item Standard infrared fixed points: these are given by the local minima of the potential at $\phi\approx\pm 7.81$;
\item Hyperscaling violating infrared fixed points: these are given by $\phi\rightarrow\pm\infty$;
\item $AdS_2$ infrared fixed points: these are given by the vanishing of $V'/V+f'/f$ and are located at $\phi\approx 0.06,\,0.70,\,10.98,\,13.45$.
\end{itemize}

We remark that the ultraviolet fixed point reached by our solutions corresponds to $AdS_5$ geometries with a vanishing dilaton field at the boundary.

Note that we may replace the non-monotonic part of our potential by a monotonic piece smoothly connecting the deep infrared at large $\phi$ to the part of the potential we effectively used in our present calculations and the results obtained in the region of the phase diagram explored in our work, 100 MeV $\le T\le$ 400 MeV and 0 $\le \mu_B\le$ 400 MeV would not change since the replaced piece of the potential has been always hidden behind the horizon in our current calculations. Such replacement, however, would modify the results in other regions of the phase diagram, such as the physics at very low $T$, and it would also modify (part of) the structure of fixed points of the model. However, as discussed before, the infrared region of our potential, corresponding to large values of $\phi$ or low values of $T$, was not constrained by lattice data and, therefore, we have no interest in this region in the present work. In fact, as discussed in \cite{gubser1}, by adequately adjusting the large $\phi$ piece of $V(\phi)$, one could in principle have a variety of different behaviors at low $T$.

We also note that the non-monotonicity of our dilaton potential poses some technical complications if one intends to explore larger regions of the plane of initial conditions $(\phi_0,\Phi_1)$ and the corresponding generated region in the $(T,\mu_B)$ phase diagram. This is because each local extrema of the dilaton potential corresponds to a singular point of the Einstein-Maxwell-Dilaton equations of motion. The numerical integration of these equations of motion in cases when extra singular points between the boundary and the horizon are present is complicated and we have not considered such cases in the present work\footnote{The singular points corresponding to the boundary and the horizon have been properly dealt with in our numerical calculations, as discussed in Section \ref{sec2.2}.}. This, in turn, limits the range of initial conditions one considers in practice, because if one wants to avoid the complicated task of numerically integrating the equations of motion with extra singular points between the boundary and the horizon, then the local extrema of $V(\phi)$ must be always hidden behind the horizon (as we have done in all of our current calculations). In fact, if we take $\phi_0$ above 7.81, then we stumble in an extra singular point for any value of $\Phi_1$. The same happens for $\phi_0<7.81$ if we take high enough values of $\Phi_1$ (even if it satisfies the physical bound $\Phi_1<\Phi_1^{\textrm{max}}(\phi_0)$) and the dilaton field $\phi$ becomes a non-monotonic function of $r$ eventually probing the region where $\phi(r)>7.81$. We have excluded all such initial conditions from our calculations by restricting the ranges of $\phi_0$ and $\Phi_1$ explored in the present work. However, we remark that the range of initial conditions we employed is enough to cover the relevant region of the phase diagram which we are interested in, 100 MeV $\le T\le$ 400 MeV and 0 $\le \mu_B\le$ 400 MeV, and in fact, a fairly broader region than that.

\begin{figure}[h]
\centering
\includegraphics[width=0.48\textwidth]{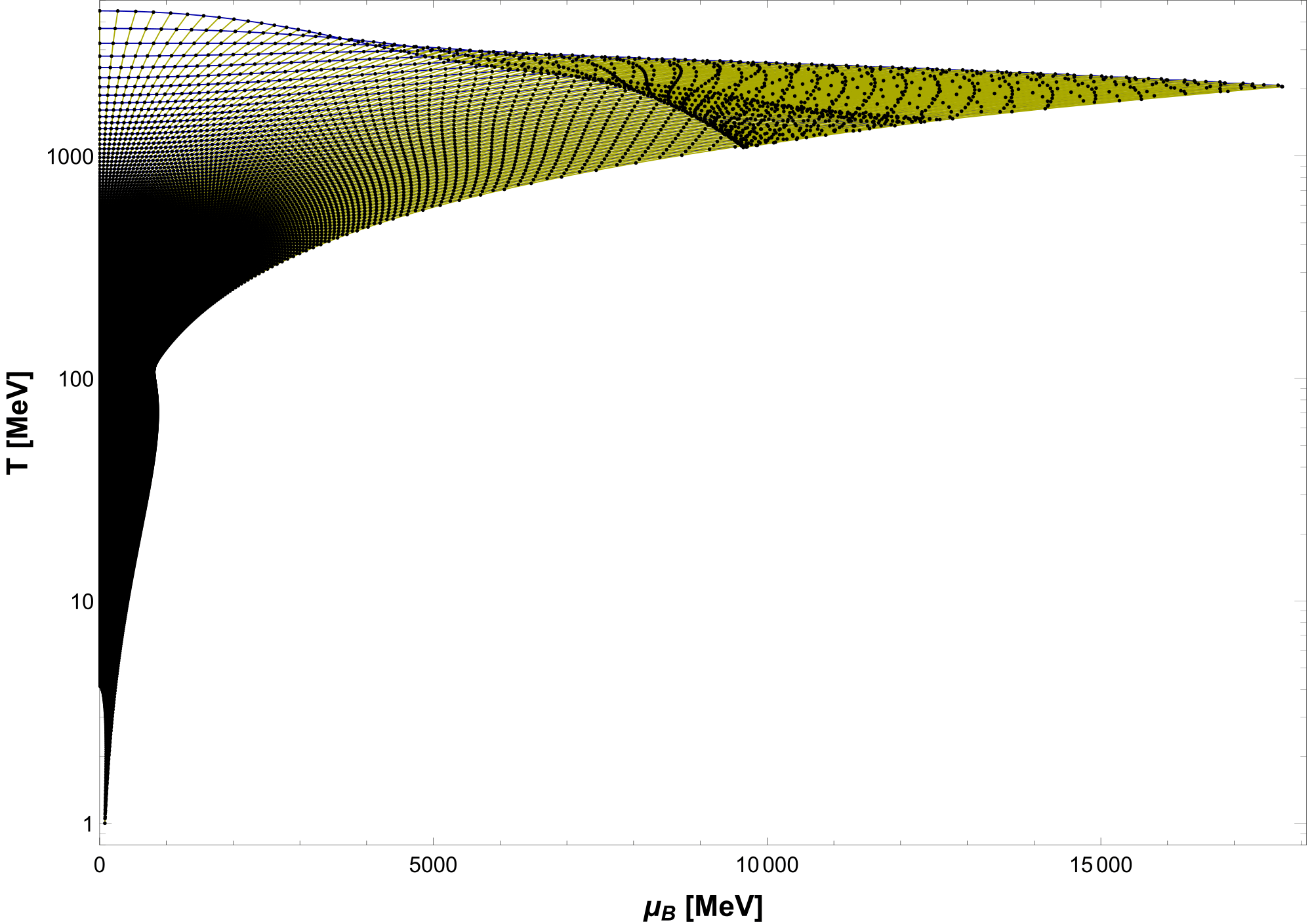}
\includegraphics[width=0.48\textwidth]{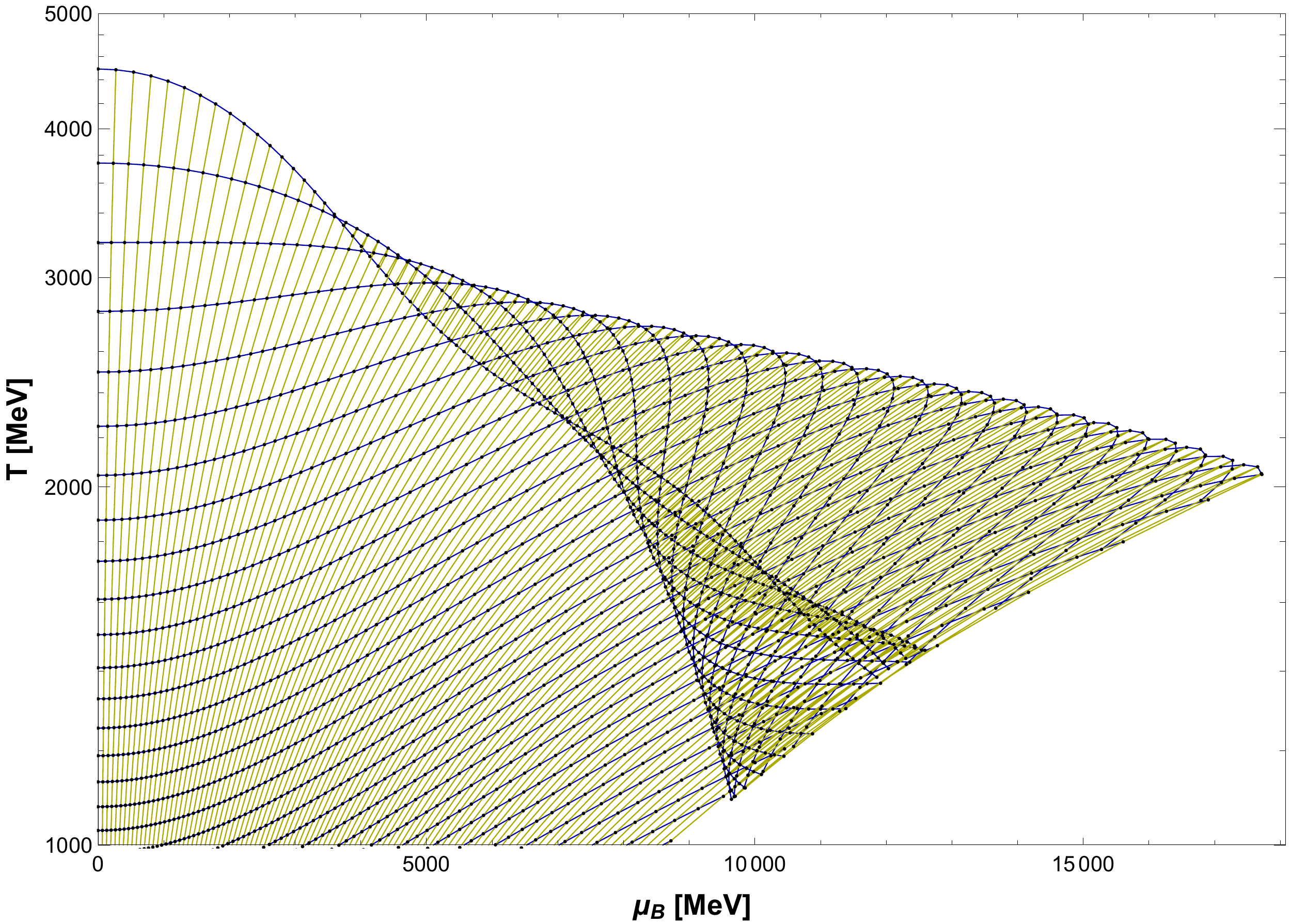}
\includegraphics[width=0.56\textwidth]{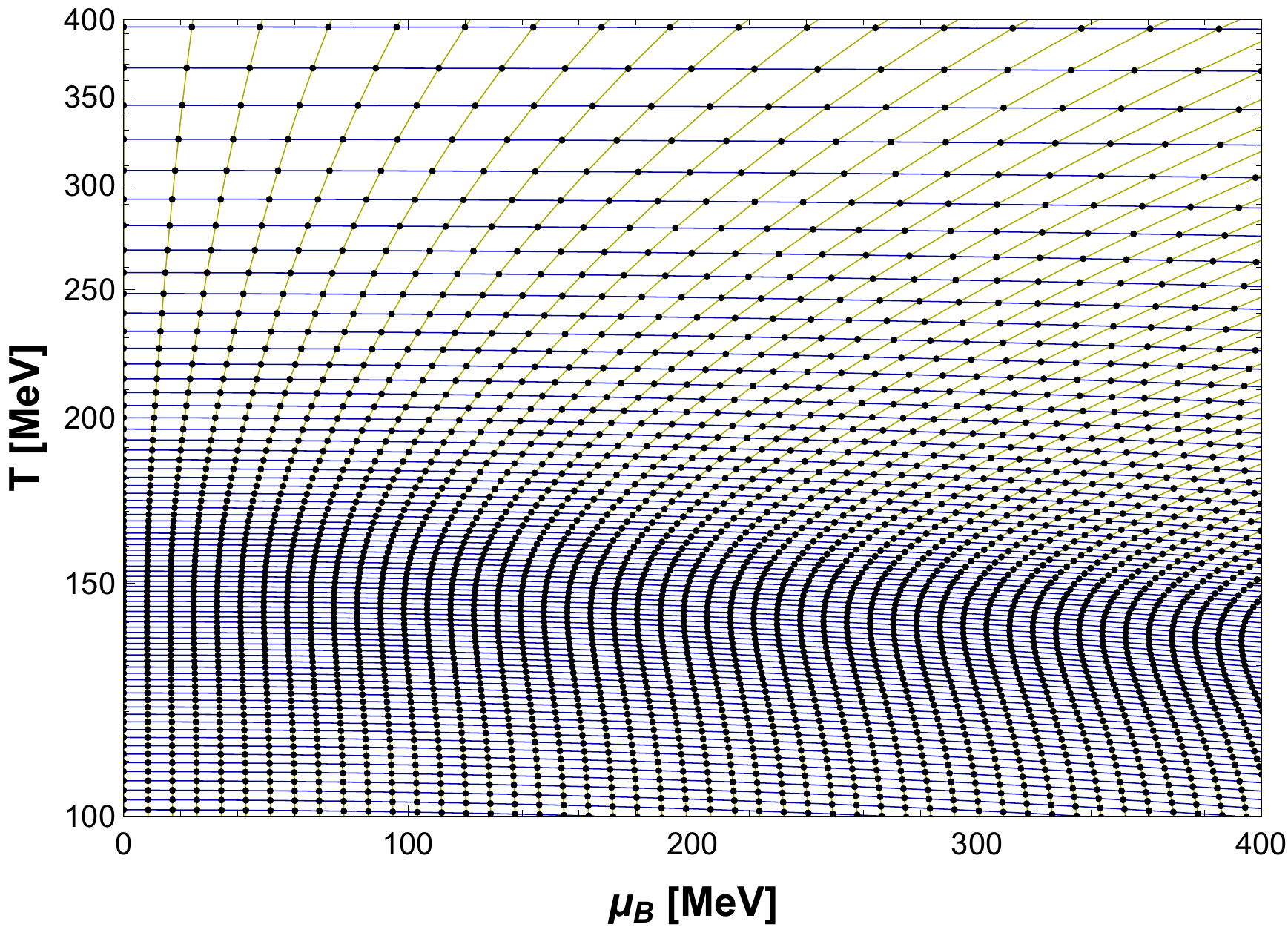}
\caption{{\it Top left:} grid shape in the $(T,\mu_B)$ phase diagram generated by the initial conditions $\phi_0\in[0.05,7.3]$ and $\Phi_1 / \Phi_1^{\rm max}(\phi_0)\in[0,0.5]$. Solutions with the same value of $\phi_0$ are connected by blue lines, while solutions with the same value of $\Phi_1 / \Phi_1^{\rm max}(\phi_0)$ are connected by golden lines. {\it Top right:} detail of the region of the phase diagram with competing black hole solutions. {\it Bottom:} region of the phase diagram explored in our work, where there are no competing black hole solutions. \label{ICPlot}}
\end{figure}

Next, we give evidence that the black hole solutions we generated in the region of interest of the phase diagram are unique. In order to do this, we examine the mapping between the initial conditions and the relevant thermodynamic state variables, i.e., the mapping $(\phi_0,\Phi_1)\to (T,\mu_B)$, exploring the (extended) range of initial conditions comprised by $\phi_0 \in [0.05,7.3]$ and $\Phi_1 \in [0,0.5\,\Phi_1^{\rm max}(\phi_0)]$. These solutions are displayed in Fig. \ref{ICPlot}. As one can see in the bottom plot in Fig. \ref{ICPlot}, for the range of initial conditions analyzed here, there are no competing black hole solutions in the region of the phase diagram which we are interested in our work since no two solutions with different $(\phi_0,\Phi_1)$ give the same $(T,\mu_B)$ inside this region (note that each point is only crossed by one blue line and one golden line). Moreover, as one can see from the top plots in Fig. \ref{ICPlot}, extending the region covered in the plane of initial conditions just add points in the $(T,\mu_B)$ plane which are more and more distant from the phenomenologically relevant region studied in our work (for very high values of $T$ and $\mu_B$, which were not explored in the present work, one notes that there are indeed competing black hole solutions). This analysis constitutes evidence that in the region of the phase diagram we worked with in our manuscript the black hole solutions are, in fact, unique.

In spite of the above argument, let us now assume the unlikely possibility that, if we consider some initial conditions outside the region of the plane of initial conditions we covered, then a hypothetical competing branch of black hole solutions could fall within the region of the phase diagram which we are interested in the present work. This could only be the case for solutions where $\phi(r)$ probes values above 7.81, since in our scanning we have excluded this possibility for $\phi(r)<7.81$. In this case, since the region where $\phi(r)$ probes values above 7.81 was always hidden behind the horizon in our calculations, then we could in principle just modify the form of the potential in that region such as to remove any possible competing branch of black hole solutions hypothetically falling inside the relevant region of the phase diagram investigated in the present work, while maintaining the solutions we used in our calculations completely unmodified. Consequently, all of our current results would remain exactly the same (which is indeed desirable, given the quantitative agreement between our holographic thermodynamics and corresponding lattice data even at nonzero $\mu_B$).

The last question we address in this Appendix regards the stability of the black hole solutions used in our calculations. Since we argued before that it is unlikely that there are competing branches of black hole solutions in the relevant region of the phase diagram covered in the present work\footnote{And that, even if a hypothetical extra competing branch were indeed present, we could in principle simply remove it by deforming the large $\phi$ part of $V(\phi)$, which has been always hidden behind the horizon in the present work without modifying our current results.}, one still has to show that the black hole solutions we derived have larger pressure than the thermal gas solution, which is just the vacuum solution with the imaginary Euclidean time direction compactified over a circle with circumference $\beta=1/T$.

In the present model, it is numerically difficult to obtain very low temperature geometries, and in particular the vacuum solution, because one would need to integrate the equations of motion with the extra singular points corresponding to the local minimum and the local maximum of $V(\phi)$. Here we present an argument of plausibility for why the black hole solutions we obtained should be expected to be thermodynamically preferred, i.e. to possess a larger pressure than the thermal gas solution. We divide our argument in two parts:
\begin{enumerate}
\item First, we calculated the discrete version of the Jacobian $J=\partial(s,\rho)/\partial(T,\mu_B)$ defined in \cite{gubser1} for all the black hole solutions generated within the region of interest of the phase diagram and checked that $J>0$ for all of them, which shows that there are no thermodynamically unstable black hole solutions inside the region of the phase diagram investigated in our work;
\item Second, since there are no unstable black hole solutions, these solutions could be at least metastable, if they had smaller pressure than the thermal gas solution. But this is not plausible, because we can go as low as $T\approx 0.005$ MeV at $\mu_B=0$ (this geometry is generated by the initial conditions $(\phi_0,\Phi_1)=(7.8,0)$) and check that the pressure of the black hole solutions monotonically increases with increasing $T$. Since the pressure of the thermal gas solution must be just a constant, which is usually matched with the black hole pressure in the extremal limit of vanishing horizon \cite{He:2013qq}, it cannot be larger than the pressures we calculated using black hole solutions in the region of the phase diagram explored in our work, where $T\gtrsim 100$ MeV.
\end{enumerate}
Therefore, the black hole solutions we derived should not only be thermodynamically stable but they should also be thermodynamically preferred over the thermal gas solution in the entire region of the phase diagram studied in our work. Furthermore, at nonzero $\mu_B$ this conclusion also holds since, as we have shown before, the pressure of the black hole solutions increases with increasing $\mu_B$ (what is quantitatively confirmed by current lattice data up to $\mu_B=400$ MeV), while the thermal gas solution does not depend on $\mu_B$. Then, the numerical black hole solutions we used to compute the physical observables in the present paper are indeed expected to be the dominant saddle points of the model.

In the future, it would be interesting to use a purely monotonic dilaton potential that can describe the lattice data for the QCD equation of state at $\mu_B=0$ since in this case the technical complications we commented upon in this Appendix would not be present. An example of such a potential can be found in Ref.\ \cite{Yaresko:2015ysa}.

\bibliography{Bibliography}{}
\bibliographystyle{JHEP}

\end{document}